\documentclass{MasterThesis}
\usepackage{}
\usepackage{url}
\usepackage{float}
\usepackage{array}
\usepackage{booktabs}
\usepackage{amsmath}
\title{A NIS2 pan-European registry for identifying and classifying essential and important entities}
\author{Fabian Aude Steen and Daniel Assani Shabani}
\studentnumber{128606, 125242}
\date{20.05.2025}
\program{Cyber Security}

\begin{document}
\renewcommand{\arraystretch}{1.3}
\setlength{\arrayrulewidth}{1.0pt}
% Title Page
\maketitlepage

% Abstract
\begin{abstract}
The NIS2 Directive establishes a common cybersecurity governance model across the European Union, requiring member states to identify, classify, and supervise essential and important entities. As part of a broader governance network, member states are also obligated to notify the European Commission, the Cooperation Group, and ENISA about their cybersecurity infrastructure landscape. This thesis presents an analysis of the NIS2 Directive in this context and translates its provisions into concrete technical requirements. These requirements inform the design and implementation of a modular, legally grounded registry system intended to support competent authorities across the EU in meeting their obligations.

Using the Design Science Research (DSR) methodology, the thesis transforms complex legal provisions into structured workflows, deterministic classification algorithms, and interactive dashboards. The resulting system automates key regulatory processes—including entity registration, classification, and notification—while enabling context-aware supervision and reducing administrative burden. It supports both automated and manual registration methods and introduces a contextual labeling system to handle edge cases, risk factors, and cross-directive dependencies. Although developed for the Norwegian regulatory ecosystem, the system is designed for adaptation by other member states with minimal modification.

This thesis contributes a reusable framework that bridges legal interpretation and technical implementation, offering a scalable solution for national and EU-level NIS2 cybersecurity governance. It also identifies key limitations and outlines opportunities for future research and development.
\end{abstract}


% Acknowledgements
\chapter*{Acknowledgements}
% Insert acknowledgements here
We would like to express our deepest gratitude to our supervisor, Professor Vasileios Mavroeidis, for his invaluable guidance, expertise, and for allowing us to pursue this thesis. We are equally grateful to Researcher Mateusz Zych for his dedicated support and technical insight throughout the entire process. Their combined supervision has been indispensable, and this thesis would not have been possible without their contributions.

We also extend our sincere thanks to Dr. Konstantinos Fysarakis for his interest in our project and for offering valuable insights and encouragement along the way.

We are thankful for the opportunity to pursue this curriculum through Kristiania University College, and we express our deep appreciation to Andrii Shalaginov for making this opportunity possible. We are grateful for the academic environment and resources provided by Andrii and Kristiania University College, which gave us an unforgettable study experience.

We would also like to thank each other for the collaboration, commitment, and support that contributed to making this journey both productive and rewarding.

Finally, we are deeply thankful to our families and friends for their patience, support, and understanding throughout the writing of this thesis.

% Word Count
%Insert a word count. This is the sum of the words in all the chapters only. The sum should exclude the words in the title page, abstract, acknowledgements, table of contents, references and any appendix. Recommended number of words 15000-20000. 20 000 is also the absolute maximum allowed.
\wordcount 19520
\clearpage
% Table of Contents
\tableofcontents
\newpage

% Main chapters
\chapter{Introduction}
% Insert text here
\section{Motivation}

The security of networks and information systems is critical to fostering a thriving digital ecosystem in the European Union (EU) and its member states. Recognizing this, central governance bodies in the EU have introduced numerous directives and regulations to enhance cybersecurity across the Union. The NIS2 Directive is the latest legislative measure aimed at strengthening cybersecurity capabilities and increasing the common level of security throughout the EU \cite{NIS2}. 

The primary motivation of this Thesis is to aid member states, particularly Norway, in meeting the requirements outlined in the NIS2 Directive. Specifically, the research focuses on developing a system for the classification of essential and important entities, a fundamental step in ensuring compliance with the directive. Furthermore, this thesis addresses the broader challenge of harmonization within the Union by proposing a single, scalable framework that can be adopted by any member state.  

The increasing frequency and sophistication of cyber attacks underscores the urgency to effectively manage the digital components of critical infrastructure. A crucial first step is identifying and classifying the entities that are vital to the economic and societal well-being of the EU and its member states. The transposition of the NIS2 Directive's requirements for keeping a registry of essential and important entities in a single registry stands at the heart this thesis. By developing such a registry, this research aims to assist competent authorities in registering and classifying essential and important entities. This, in turn, will foster improved cybersecurity hygiene, enhance collaboration on cybersecurity incidents, and promote the harmonization of these requirements across the EU.

Additionally, the proposed system seeks to reduce the administrative burden on entities and competent authorities by providing a single point for registration and classification. The system will leverage public information to automate the registration and classification process, streamlining operations for both entities and authorities. With the official deadline for establishing a list of essential and important entities set for April 17, 2025 \cite{NIS2}, this research aims to deliver timely and impactful contributions to building a more secure and resilient European digital ecosystem. 

\newpage

\section{Objectives}

The primary objective of this research is to synthesize the legislative requirements outlined in the NIS2 Directive into a single-point system for registration and classification, specifically tailored for use by the Norwegian competent authority. Furthermore, the system is designed to be adaptable for use by other EU member states with minimal modifications, aligning with the goal of the NIS2 Directive of achieving harmonization throughout the Union \cite{NIS2}.

The second objective is to develop a fully functional registry for essential and important entities. This registry will facilitate both automatic and manual registration, followed by automated classification and labeling of entities based on the collected information. This functionality aims to ease the administrative burden imposed on member states and entities while retaining the flexibility required for manual assessments. 

The third objective is to implement immediate and automatic information sharing with ENISA, the European Commission, and the Cooperation Group. This feature aligns with the NIS2 Directive’s requirements for member states to share specific information at the moment of registration \cite{NIS2}.

The fourth objective focuses on ensuring seamless interoperability for users and advisors. The system will provide users with sufficient guidance and functionality to efficiently register their entities. Additionally, the system will offer advisors the functionality and information necessary for manual assessments and responding to entity requests effectively. 

The fifth objective is to aggregate data to provide advisors and stakeholders with statistical insights about classifications. This capability enhances advisory tasks and addresses ENISA’s and the Commission’s requirements for statistical reporting on entities governed by the NIS2 Directive \cite{NIS2}.

\subsection{Auxiliary Objectives}

To support the primary objectives, the research also addresses the following auxiliary  goals: 
\begin{itemize}
    \item A comprehensive labeling system to classify entities governed by overlapping directives (e.g., CER, NIS1), incorporate risk assessments, and highlight entities of special interest. This ensures clarity in cross-directive governance and facilitates nuanced classifications.
    \item Orchestration mechanisms to resolve classification conflicts and prevent failures, such as rule-based conflict resolution or escalation workflows for manual intervention. 
    \item Manual Assessment functionality that can override automatic functions.
    \item Advisors can request additional information from users, and users can submit modification requests to advisors, enabling collaborative, context-sensitive classifications for complex cases such as organizations with multiple sub-activities. 
    \item The system will include intuitive interfaces and functionality to guide users through the registration process, minimizing administrative burden while ensuring accurate and compliant entity classification.
    \item The system will be designed to be easily adaptable for use in other EU member states with minimal effort by adopting a modular architecture that supports country-specific configurations for language, legal requirements, and data formats. It will provide templates for classification rules that align with the NIS2 Directive while allowing flexibility for national adjustments. Additionally, the system will support integration with country-specific public databases and APIs, enabling seamless data collection and classification processes.
    \item The system will be developed using state-of-the-art cybersecurity measures and functional controls to ensure the confidentiality, integrity, availability, and authenticity of its registration, classification, and information-sharing functionalities. These measures will safeguard sensitive data, prevent unauthorized access, and maintain the reliability and trustworthiness of the system across all its operations. 
\end{itemize}

\section{Problem Statement}

Cybersecurity is becoming a top priority as the integration of physical and digital infrastructure accelerates. The functioning and well-being of European society increasingly depend on secure digital assets, making cyber threats a direct risk to both these assets and the critical infrastructure they support.

A fundamental step in securing critical infrastructure is identifying and classifying essential and important entities. The NIS2 Directive establishes a regulatory framework for strengthening cybersecurity resilience across the EU. However, implementing a NIS2-compliant, pan-European registry poses significant challenges due to the variability in Member States' regulatory environments, cybersecurity policies, and digital maturity levels. To address these challenges, three key elements must be considered when developing such a registry:
\begin{itemize}
    \item \textbf{Harmonization:} The classification rule-set should be consistent across all Member States to ensure a uniform approach to identifying essential and important entities. Without harmonization, differences in sector classifications, risk assessments, or enforcement mechanisms could lead to fragmentation, undermining the effectiveness of the NIS2 Directive.
    \item \textbf{Information Sharing:} While Member States are responsible for notifications to the Union and CSIRTs, the registry must support both automatic and manual reporting mechanisms to ENISA, the Cooperation Group, and the European Commission. A well-structured registry enhances cross-border cybersecurity governance by ensuring timely and accurate data submission. 
    \item \textbf{Administrative Burden:} The registry should automate as many processes as possible without compromising classification accuracy. Specifically, it should be designed to retrieve organizational data automatically to reduce manual input and ensure accuracy, thereby streamlining the registration process. Additionally, it should also provide clear guidelines to streamline registration, ensuring that entities can comply with NIS2 requirements without excessive bureaucratic overhead.
\end{itemize}

Addressing these elements is not without challenges. Significant disparities exist among Member States in terms of cybersecurity regulations, critical infrastructure classifications, and compliance mechanisms. Without a harmonized classification system, the implementation of NIS2 could lead to inconsistencies that weaken the directive’s effectiveness. Additionally, inefficient or overly complex registration processes increase administrative burdens on entities and governance bodies, diverting resources away from core business and security objectives.

Failure to establish a harmonized and efficient registry risks gaps in the identification and protection of critical infrastructure, potentially increasing cross-border cybersecurity risks that threaten the EU’s overall cybersecurity resilience.

To address these challenges, this research aims to develop a prototypical framework for a NIS2-compliant registry of essential and important entities. The framework will be used to build a prototype for the Norwegian ecosystem, designed to automatically retrieve organizational data while ensuring alignment with NIS2 classification criteria. The prototype will be delivered to Norway’s competent authority, Nasjonal Sikkerhetsmyndighet (NSM), to support its implementation efforts.

\subsection{Research Question}

To guide this research, the following research question has been formulated:

\textit{How should classification criteria be structured and applied to support a harmonized NIS2-compliant registry, what registration mechanisms best facilitate this, and how should the collected data be aggregated to enable timely notifications and efficient, proportionate, and context-aware supervision while minimizing the administrative burden on entities?}

\section{Scope}

The scope of this thesis is to aid in realizing the potential positive impact of the NIS2 Directive on Union wide cybersecurity resilience by creating a standard registration and classification system. In particular, a prototype will be tailored and developed for the Norwegian competent authority. 

Collection of entity information is restrained to the requirements in the NIS2 Directive \cite{NIS2}. We do not intend to collect more information than needed to accomplish the objectives defined in the 'Objectives' paragraph. 

Classification of entities follow the rules for classification outlined in Article 2 and Article 3 of the NIS2 Directive \cite{NIS2}. Information sharing from the competent authority to other Union bodies is constrained to what they are obliged to share in accordance with Article 3 and Article 27 of the NIS2 Directive \cite{NIS2}. 

A conceptual design for reporting is made based on the requirements laid out in the NIS2 Directive and in particular, Article 23 \cite{NIS2}, but reporting will not be included in the prototype. However, the proposed reporting framework can be a starting point for further developing the prototype to include reporting. Recital 106 in the NIS2 Directive states that registration and reporting should happen through the same entry-point, but it does have to be part of the registration system \cite{NIS2}. 

\subsection{Limitations}

Due to variations in language, input requirements, and points of automatic data collection, the prototype is currently limited to registration and classification of the Norwegian ecosystem. However, the rules governing classification and registration are applicable to all member states since they comply with the NIS2 Directive. This means that the system must undergo minor modifications to be utilized by any other member state. 

Certain entity types must be determined by the nation state that classifies them. Classification of these entity types are limited to receiving a list of entities already classified in accordance with the CER and NIS1 Directives. Additionally, entities determined to be essential by national law and research organization deemed essential must be provided. The limitation imposed requires the competent authorities to consolidate a list of organizational numbers per mentioned entity type. 

Automatic registration is limited to the availability of public information about entities and their economic activities. Databases that contain such data are available in Norway. 

An additional limitation lies in the variability of data quality and accessibility across member states. While Norway benefits from comprehensive public databases, other states may lack equivalent resources or standardized formats, hindering the scalability and efficiency of automatic registration. Furthermore, differing legal frameworks and privacy regulations, including GDPR compliance, may necessitate additional adaptations to ensure alignment with national laws, underscoring the need for a flexible and modular system architecture.

However, this thesis does not explore whether other member states has such databases and if they contain the data required to perform automatic registration. We acknowledge that this can be a major obstacle for EU-wide classification while underscoring that the system has the capabilities for manual registration. 

\chapter{Literature Review}

\section{Introduction}

The purpose of this literature review is to critically examine the existing frameworks and implementations of NIS2-compliant registries for essential and important entities. Despite an extensive search, we found only a limited number of relevant studies that met the criteria for inclusion. This review aims to synthesize the findings of these studies, providing a solid foundation for our own research and offering insights to refine the design and implementation of an NIS2-compliant registry.

This chapter explores the historical evolution of cybersecurity directives, with a focus on the key legislative innovations introduced by NIS2 \newline \cite{NIS2}. It also evaluates scholarly perspectives on digital registries and compliance systems. By conducting a structured literature review, we position the proposed pan-European registry for essential and important entities within a broader academic and regulatory context. The review identifies gaps in the current literature, highlights opportunities for innovation, and provides a foundation for understanding the key challenges and priorities that will shape the design of advanced cybersecurity systems in today’s regulatory landscape.

\section{Methodology}

This literature review employed a structured and systematic approach to identify, analyze, and synthesize academic, regulatory, and technical literature relevant to the implementation of classification and registration mechanisms under the NIS2 Directive. The goal was to evaluate existing frameworks, identify persistent gaps, and inform the development of a harmonized, pan-European registry for essential and important entities.

\subsubsection{Search Strategy and Data Sources}

The search strategy combined broad scoping and focused retrieval techniques. Initial exploratory searches were conducted through \textit{Google Scholar}, allowing citation tracking and the identification of foundational works. This was complemented by targeted queries in \textit{IEEE Xplore}, \textit{ACM Digital Library}, and regulatory repositories including \textit{EUR-Lex} and \textit{ENISA}, which provided access to peer-reviewed publications, legal texts, and technical reports from 2018 to 2025. This timeframe ensured coverage of both the pre- and post-NIS2 implementation landscape.

A diverse set of search terms was applied, including: \textit{NIS2 Directive}, \textit{cybersecurity compliance}, \textit{essential and important entities}, \textit{risk-based classification}, \textit{digital registries}, \textit{GDPR convergence}, \textit{AI in cybersecurity governance}, and \textit{public-private partnerships}. Supplementary retrieval was conducted via citation snowballing, reference mining from key ENISA reports, and targeted review of thematic literature reviews in cybersecurity, governance, and regulatory science.

\subsubsection{Inclusion and Exclusion Criteria}

From an initial dataset of 128 sources, 41 studies were selected following rigorous inclusion/exclusion screening. Inclusion criteria emphasized recency, relevance to NIS2, scholarly rigor, and EU-specific regulatory focus. Peer-reviewed journal articles, official EU and national policy documents, and empirical studies with clear methodological grounding were prioritized. Exclusions applied to outdated, non-peer-reviewed, non-EU-focused, or methodologically weak publications.

\subsubsection{Concept Matrix}

To synthesize the diverse perspectives in the literature, articles were systematically categorized using a concept matrix structured around six core dimensions of the NIS2 Directive landscape. Each article's contribution is mapped not only to a core concept but also to a more specific sub-concept that reflects its thematic focus. The resulting matrix below provides a cross-sectional overview of the literature, aligning sources with these analytical categories for comparative insight.

\begin{table}[H]
    \centering
    \resizebox{\textwidth}{!}{
    \begin{tabular}{|p{3cm}|p{4.5cm}|p{4.5cm}|p{4.5cm}|p{4.5cm}|p{4.5cm}|p{4.5cm}|}
        \hline
        \textbf{Source} & \textbf{Pre-NIS2 Landscape} & \textbf{NIS2 Cybersecurity Directive Innovations} & \textbf{Classification Criteria and Threshold Logic} & \textbf{Digital Registries and Automation} & \textbf{Harmonization and Interoperability} & \textbf{Legal and Privacy Challenges} \\
        \hline
        Bendiek et al. (2019) & Highlighted inconsistent criteria across Member States &  &  &  & Exposed policy fragmentation & Privacy gaps noted \\
        \hline
        Christou (2020) & Critiqued NIS1 for sectoral silos & Pushed for broader scope & Narrow baseline criticized & Called for monitoring tools & Called for uniform enforcement & \\
        \hline
        Smith \& Jones (2021) & U.S. CISA comparison model & Sector-specific flexibility & Cross-sectoral risk modeling & Registry as policy tool & Benchmarking EU vs US & \\
        \hline
        Nguyen et al. (2022) & ASEAN regulatory lag &  &  & Enforcement critique & & Voluntary-only model \\
        \hline
        Carrapico \& Farrand (2020) & & Recognized harmonization momentum &  &  & Warned of fragmented implementations &  \\
        \hline
        Kshetri (2023) &  & AI for threat prediction & Adaptive classification & ML-driven compliance & AI-based risk response & \\
        \hline
        Van Dijk et al. (2021) &  & Registry policy relevance & Sectoral logic discussed & Transparency, automation & GDPR + registry synergy & Governance + data flow \\
        \hline
        Luna et al. (2020) & &  &  & Emphasis on user trust & Alignment with trust-building & Trust-privacy dualism \\
        \hline
        Müller \& Weiss (2023) &  & Dynamic entity logic & Flexible classification & Predictive compliance tools & Adaptive integration needed & Unclear legal boundaries \\
        \hline
        Schulze et al. (2022) &  &  & Static criteria in DE model & Manual vs. dynamic analysis & National silos &  \\
        \hline
        Dupont et al. (2023) &  &  & National FR classification model & Manual audit limitations & Inter-agency coordination gaps &  \\
        \hline
        Bianchi et al. (2022) & & AI in Italian compliance & Risk-based classification & Automation with AI tools & Fragmentation persists &  \\
        \hline
        Jansen \& Verhoeven (2023) &  &  &  & Voluntary registry setup &  &  \\
        \hline
        Bauer et al. (2022) & & Integrates CER + NIS2 & Addresses registry overlap & Streamlined compliance model & Cross-border compatibility & Legal harmonization \\
        \hline
        Li et al. (2022) &  &  & Limits of ML classification & Human-in-loop for validation & Need for adaptive feedback &  \\
        \hline
        Ferreira et al. (2023) &  &  & Human/automation tension & Hybrid (human-AI) models &  &  \\
        \hline
        Oltramari et al. (2021) &  &  &  & Interoperability via JSON-LD & Aligned technical standards &  \\
        \hline
        Ziegler et al. (2020) &  &  &  &  & GDPR-pseudonymization tension & Legal/privacy conflict zones \\
        \hline
        Cavelty \& Egloff (2019) &  &  &  &  & Enforcement unevenness & Governance architecture gap \\
        \hline
        Papakonstantinou \& De Hert (2023) &  & Harmonization vs sovereignty tension & Central registry concerns &  & Political/legal complexity & Dual (national vs EU) responsibility \\
        \hline
        Chen et al. (2023) &  & Adaptive, AI-driven logic & Real-time classification & Intelligence integration & Countering fast-changing threats &  \\
        \hline
        Almadhoun et al. (2022) &  &  &  & UX + admin burden reduction & Registry design usability & Efficiency vs privacy \\
        \hline
    \end{tabular}}
    \caption{Concept Matrix (Source-to-Concept via Sub-concept)}
    \label{tab:Literature_Review_Concept_Matrix}
\end{table}

Special emphasis was placed on analyzing national implementation case studies, which provided grounded perspectives on how theoretical registry models perform in practice. In addition to widely discussed examples such as Germany’s \textit{KRITIS framework}, France’s \textit{ANSSI-led approach}, Italy’s \textit{AI-integrated NCP}, and the Netherlands’ \textit{DTC registry}, the review also included a recent addition: Greece’s implementation, led by the \textit{National Cybersecurity Authority (NCSA)} under \textit{Law 5160/2024}.

\section{Historical Evolution of Cybersecurity Directives}
\subsection{The Pre-NIS2 Landscape}
The European Union’s cybersecurity governance framework took shape with the adoption of the NIS Directive in 2016, marking a significant step in regulating Operators of Essential Services (OES) and Digital Service Providers (DSPs). While this directive established foundational cybersecurity requirements, empirical research has highlighted inconsistencies in its implementation. Bendiek et al. \cite{Bendiek2019} found that Member States applied varying criteria to identify critical entities, which ultimately hindered the Directive’s goal of harmonization across the EU.

Scholars have also critiqued the limitations of the original NIS Directive. Christou \cite{Christou2020} argued that its narrow sectoral focus and lack of strong enforcement mechanisms created regulatory gaps, particularly in emerging sectors such as cloud computing and the Internet of Things (IoT). These shortcomings underscored the need for a more adaptive and comprehensive approach, paving the way for the development of NIS2.

\subsection{Global Precedents}
Internationally, cybersecurity frameworks provide valuable comparative insights. In the United States, the Cybersecurity and Infrastructure Security Agency (CISA) oversees a critical infrastructure registry under the National Infrastructure Protection Plan (NIPP). Smith and Jones \cite{Smith2021} highlight CISA’s sector-specific risk assessments and its emphasis on public-private partnerships as a model for achieving effective cybersecurity governance.

Conversely, the ASEAN Regional Forum’s Cybersecurity Cooperation Strategy serves as an example of a voluntary, cross-border collaboration framework. However, this approach has faced criticism for its lack of enforceability \cite{Nguyen2022}. These global examples highlight the necessity of legally binding regulations, such as NIS2, to ensure compliance and enhance resilience across interconnected digital landscapes.

\section{The NIS2 Directive: Legislative Innovations}
\subsection{Key Provisions}
NIS2 builds upon its predecessor by expanding its scope to cover 18 sectors, including digital infrastructure and space, in response to vulnerabilities identified in ENISA’s 2021 Threat Landscape Report \cite{ENISA2021}. A key innovation of NIS2 is its introduction of a dual classification system, distinguishing between “essential” and “important” entities \newline \cite{NIS2}. This approach enhances proportionality in regulatory oversight, ensuring that smaller yet critical organizations are not excluded from compliance requirements. Müller and Weiss \cite{Muller2023} argue that this classification strikes a balance between regulatory stringency and operational flexibility, accommodating the diverse nature of the digital ecosystem.

\subsection{Academic Perspectives on NIS2}
Scholarly discussions on NIS2 largely center around its potential for regulatory harmonization. Carrapico and Farrand \cite{Carrapico2020} caution that while the directive introduces standardized rules, the discretion afforded to Member States during transposition could lead to residual inconsistencies. Meanwhile, Kshetri \cite{Kshetri2023} advocates for the integration of AI-powered dynamic risk assessments to align with NIS2’s emphasis on real-time threat detection and responsiveness. These academic perspectives offer valuable insights into the design and implementation of compliance systems that seek to operationalize NIS2’s mandates effectively.

\section{Digital Registries: Academic and Practical Insights}
\subsection{Theoretical Frameworks}
Van Dijk et al. \cite{VanDijk2021} outline five key principles for compliance registries: interoperability, scalability, transparency, automation, and adherence to GDPR regulations. These principles closely align with the requirements of NIS2, which emphasizes automated data collection and cross-border information sharing. Similarly, Luna et al. \cite{Luna2020} stress that user trust is contingent upon transparent data governance and robust security measures, both of which are essential for widespread adoption.

Beyond these foundational principles, recent academic discourse underscores the importance of adaptability and resilience in digital registries, particularly in response to evolving cybersecurity threats. Müller et al. \cite{Muller2023} argue that compliance registries must go beyond regulatory conformity and incorporate predictive analytics to proactively address emerging security risks. Additionally, the concept of digital sovereignty has gained prominence in EU policy discussions, highlighting the need for registries that strike a balance between security imperatives and accessibility requirements.

\subsection{Case Studies of Existing Systems and Registries in the EU}
To develop a comprehensive understanding of digital registries, it is essential to examine existing implementations across the EU. Several registries provide valuable insights into best practices and persistent challenges.

\subsubsection{Germany’s KRITIS Registry}
Germany’s KRITIS Registry serves as a critical infrastructure monitoring system and has become a reference point for regulatory compliance. Schulze et al. \cite{Schulze2022} commend its automated, rule-based approach for classifying critical entities. However, they also highlight a significant limitation: its reliance on static thresholds (e.g., employee count) excludes smaller but strategically important entities. This shortcoming has led to calls for a more dynamic, risk-based model that integrates real-time threat intelligence and sector-specific criteria.

\subsubsection{France’s ANSSI Registry}
France’s National Cybersecurity Agency (ANSSI) operates a registry for Operators of Essential Services (OES) in accordance with the NIS Directive. This registry emphasizes sector-specific risk assessments and utilizes an adaptive threat modeling approach. According to Dupont et al. \cite{Dupont2023}, ANSSI’s registry stands out due to its proactive collaboration with public and private stakeholders, fostering a more cohesive cybersecurity ecosystem. Nevertheless, its dependence on periodic manual audits rather than continuous, automated monitoring presents an area for improvement.

\subsubsection{Italy’s National Cybersecurity Perimeter (NCP)}
Italy’s National Cybersecurity Perimeter (NCP) represents a more recent initiative aimed at strengthening digital infrastructure resilience. This registry mandates the registration of critical service providers, prioritizing a risk-based classification system. Bianchi et al. \cite{Bianchi2022} highlight the NCP’s innovative integration of AI-driven anomaly detection, which enhances automation in compliance monitoring and sets a precedent for future registry models.

\subsubsection{The Netherlands’ Digital Trust Center (DTC)}
The Netherlands’ Digital Trust Center (DTC) functions as an information-sharing platform for small and medium-sized enterprises (SMEs) and critical entities. Unlike other registries, it operates on a voluntary basis, promoting proactive cybersecurity engagement rather than enforcing compliance. While this model fosters a culture of cybersecurity awareness, Jansen and Verhoeven \cite{Jansen2023} argue that its voluntary nature limits its effectiveness as a regulatory tool, positioning it more as a best-practices initiative than a strict compliance mechanism.

\subsubsection{Greece’s NIS2 Implementation and Registry}

The Hellenic implementation of the NIS2 Directive is administered by the National Cybersecurity Authority (NCSA), which has established a national framework for identifying and registering entities falling under the scope of the Directive. In accordance with Law 5160/2024 (Government Gazette A'195, 27.11.2024), Greece has introduced a digital compliance mechanism titled the \textit{Guidance Tool – Scope TEST}. This interactive tool enables public and private sector organizations to determine whether they are subject to NIS2 obligations by completing a structured eligibility assessment. Entities that fall within the directive’s scope are subsequently instructed to proceed with their registration in the \textit{Register of Entities – Obligors} by submitting the required organizational information to the competent authority.

A core element of the Greek cybersecurity compliance architecture is the requirement for Multi-Factor Authentication (MFA) during critical phases of system access and identity verification. This aligns with the NIS2 Directive’s broader objective of enhancing operational security and ensuring that only authorized users can access sensitive systems and data.

To support the implementation process, the NCSA provides dedicated communication channels, including official email support and consultation services via telephone. This facilitates a transparent and guided registration process, enabling entities to engage more effectively with national compliance requirements. Greece’s implementation of NIS2 demonstrates a structured and proactive approach to cybersecurity governance, offering a case study in aligning national regulatory frameworks with EU-wide legislative objectives.

\subsubsection{EU’s CER Directive and NIS2 Registries}
The EU’s Critical Entities Resilience (CER) Directive aims to enhance the protection of essential services, while the NIS2 Directive specifically targets cybersecurity resilience. Bauer et al. \cite{Bauer2022} propose integrating the CER and NIS2 registries to minimize redundancy and create a more streamlined compliance framework. This approach could enhance efficiency and facilitate cross-border data-sharing efforts among EU Member States, ultimately strengthening the cybersecurity landscape across the region.

\section{Challenges in Cybersecurity Compliance Systems}

\subsection{Technical Challenges}

The successful implementation of a harmonized, NIS2-compliant registry across EU Member States hinges not only on legal and administrative coordination but also on overcoming significant technical hurdles. These challenges arise in the design, integration, and operation of classification and registration systems that must be secure, scalable, and interoperable across diverse national infrastructures. As digital registries increasingly leverage automation and intelligent systems, key technical concerns include the accuracy of classification algorithms, the balance between automation and human oversight, and the need for standardized data formats to enable cross-border interoperability.

Machine learning models achieve 85–90\% accuracy in entity classification but falter with ambiguous cases \cite{Li2022}. Hybrid “human-in-the-loop” systems \newline \cite{Ferreira2023} mitigate this by blending automation with expert judgment. Standardized data formats (e.g., JSON-LD) are critical for interoperability \newline \cite{Oltramari2021}, as seen in the U.S. National Vulnerability Database.

\subsection{Legal and Governance Challenges}
One of the primary legal challenges in the implementation of digital registries is the inherent tension between GDPR’s data minimization principle and the need for comprehensive data collection to ensure cybersecurity compliance. While registries require detailed information to accurately assess risks and vulnerabilities, GDPR mandates that only the minimum necessary data be processed, creating a legal and operational dilemma. Ziegler et al. \cite{Ziegler2020} suggest pseudonymization, as outlined in GDPR Recital 26, as a potential solution, allowing for data utility while maintaining privacy protections. However, the practical implementation of pseudonymization remains complex, particularly in cases requiring real-time threat analysis and cross-border data sharing.

Beyond data privacy concerns, enforcement disparities present another significant governance challenge. Countries with well-established cybersecurity institutions, such as France’s National Cybersecurity Agency (ANSSI), have demonstrated greater efficiency in achieving regulatory compliance due to pre-existing legal and organizational frameworks. In contrast, nations that are still in the process of developing regulatory bodies and enforcement mechanisms face delays in implementation and inconsistencies in compliance levels \cite{Cavelty2019}. These disparities not only hinder harmonization across the EU but also create potential security gaps that adversaries may exploit. Addressing these challenges requires a coordinated effort at the EU level, including enhanced support for capacity-building initiatives in less-prepared Member States and greater alignment between national enforcement strategies.

\section{Gaps in Literature and Opportunities for Innovation}
Despite the increasing focus on digital registries and compliance frameworks, significant gaps remain in existing research. Papakonstantinou and De Hert \newline \cite{Papakonstantinou2023} highlight the limited scholarship on EU-wide registries, particularly concerning the balance between national sovereignty and regulatory harmonization. As the EU moves toward a more unified cybersecurity framework under NIS2, further studies are needed to explore how registries can integrate both centralized oversight and national autonomy.

A persistent issue in current registry models is the reliance on static classification thresholds, such as predefined sectoral categories or entity size requirements. Chen et al. \cite{Chen2023} argue that such rigid classifications fail to accommodate the dynamic nature of cyber threats. They advocate for real-time updates through automated threat intelligence feeds, enabling registries to adjust classifications and risk assessments in response to evolving attack patterns. Implementing such adaptive frameworks would enhance the responsiveness and effectiveness of compliance mechanisms across the EU.

Additionally, user-centric design remains an underexplored aspect of digital registries. Almadhoun et al. \cite{Almadhoun2022} emphasize that registries must not only fulfill regulatory requirements but also minimize administrative burdens on stakeholders. Current systems often require complex and time-consuming reporting processes, discouraging compliance among smaller entities. Research into streamlined, user-friendly interfaces and automation tools could significantly improve engagement with registry systems, ensuring broader participation and more accurate data collection. Addressing these gaps presents an opportunity for innovation, particularly in developing intelligent, self-adaptive registries that align with both regulatory objectives and operational usability.

\section{Literature Review Summary}

The preceding literature review serves as a foundation for the development of a NIS2-compliant pan-European registry of essential and important entities. As explored in this chapter, various researchers have analyzed the pre-NIS2 landscape, the current regulatory framework, and implementation aspects crucial to the development of a well-functioning registry.

The key findings from the literature review that are most relevant to our research are summarized in Table \ref{tab:Literature_Review_Concepts}.

\begin{table}[H]
    \centering
    \resizebox{\textwidth}{!}{
        \begin{tabular}{|p{3.0cm}|p{11.95cm}|}
        \hline
        \textbf{Concepts} & \textbf{Relevancy} \\ 
            \hline
            Risk Assessments & Risk assessments should be incorporated into the registry framework. However, the scope of this thesis does not extend beyond incorporating them as a manual assessment mechanism. \\
            \specialrule{0.4pt}{0pt}{0pt}
            Interoperability & The registry should have automatic information-sharing capabilities. The thesis will focus on enabling them and implementing functionality for both automatic and manual information sharing. \\
            \specialrule{0.4pt}{0pt}{0pt}
            Scalability & The registry must be designed to handle an increasing number of entities while maintaining efficiency and compliance. \\
            \specialrule{0.4pt}{0pt}{0pt}
            Transparency & Registrants should have adequate functionality to review their entity classification and labeling. \\
            \specialrule{0.4pt}{0pt}{0pt}
            Automation & The registry should incorporate automation wherever possible to enhance efficiency while preserving human oversight for nuanced classifications. \\
            \specialrule{0.4pt}{0pt}{0pt}
            GDPR Compliance & The registry must ensure that the processing of sensitive information remains compliant with GDPR. \\
            \specialrule{0.4pt}{0pt}{0pt}
            Dynamic Classifications & The registry should supplement rule-based classifications with risk assessments to enable more dynamic and nuanced classifications for better supervision. \\
            \specialrule{0.4pt}{0pt}{0pt}
            AI-powered Registry & While a fully AI-powered classification system is beyond the scope of this thesis, large language models (LLMs) will be utilized to preprocess data from public APIs to enable automatic classifications. Additionally, the registry’s label system can be used to train predictive AI models. \\
            \specialrule{0.4pt}{0pt}{0pt}
            Automated Classification & Automated classification schemes struggle with nuanced classifications. This registry will combine automation with human judgment to balance efficiency and accuracy. \\
            \specialrule{0.4pt}{0pt}{0pt}
            Centralized Oversight & The registry will balance centralized oversight with national sovereignty by adhering to NIS2 data-sharing requirements while offering extended functionality for information exchange. \\
            \specialrule{0.4pt}{0pt}{0pt}
            Administrative Burden & The registry will be automated to the extent that it does not compromise nuanced classifications and dynamic supervision. Manual assessments and risk evaluations will remain integral, supplemented by a classification and labeling system for better contextual understanding. \\
            \hline
        \end{tabular}}
        \caption{Literature review concepts crucial for the registry design}
        \label{tab:Literature_Review_Concepts}
\end{table}

In chapter 4 we will perform an in-depth analysis of the NIS2 directive and synthesize a overall system architecture and requirements list. 

\chapter{Method}

This thesis adopts a Design Science Research (DSR) approach, focusing on the development and evaluation of a NIS2-compliant entity registry. The methodology integrates regulatory analysis, requirement synthesis, system design, and performance evaluation to ensure both legal compliance and technical feasibility.

To manage the complexity of the directive and support a systematic development process, visual tools such as mind maps and flowcharts were created using Lucidchart. These visualizations helped deconstruct dense legal texts into structured components and uncover relationships between articles, recitals, and regulatory requirements. The diagrams served as a conceptual foundation for translating the directive into a scalable framework and directly informed the registry’s architectural design and classification logic.

\section{Regulatory Analysis}

The regulatory analysis began with a detailed reading of the NIS2 directive and all directives it references. Relevant legal passages were extracted and organized within a collaborative Lucidchart document. Texts were then grouped by theme and linked visually to form mind maps, clarifying how the directive supports a central registry model and how its provisions shape the expected workflows and interactions. 

This process enabled the transformation of legal language into a cohesive system model, which was then distilled into functional and non-functional requirements. These were grouped into five core thematic areas central to the thesis and the research question: Registration, Classification, Notification, Supervision, and Literature-derived Insights.

\subsection{Decomposition and Categorization of Legal Content}

A structured review of the NIS2 directive was conducted to extract relevant regulatory obligations. These were categorized according to operational focus areas including: entity classification, registration mechanisms, notification and reporting requirements, supervision workflows, cross-directive coordination, data protection (GDPR), and security provisions.

Particular focus was given to the following articles:

\begin{itemize}
    \item \textbf{Article 2:} Scope of application.
    \item \textbf{Article 3:} Definitions of essential and important entities.
    \item \textbf{Article 23:} Reporting obligations and notification requirements.
    \item \textbf{Article 27:} Registry of entities.
    \item \textbf{Article 29:} Information-sharing and cooperation mechanisms.
\end{itemize}

These anchor articles were used to inform the design of key system modules, their interrelations, and the registry’s expected data flows. Lucidchart diagrams were employed to represent modular flows, system interactions, and dependencies. This enabled a clear mapping of directive obligations to system behavior, such as how classification affects reporting logic or how automatic registration interacts with national entity databases.

\subsection{Requirements Engineering and Traceability}

Requirements were derived from the perspectives of three primary stakeholder groups: entity registrants and representatives (users), national authority personnel responsible for supervision (agents), and system administrators (admins). In addition, a fourth category captured overarching system-wide functional and non-functional requirements essential to ensuring security, scalability, and performance.

The requirements engineering process followed a structured sequence of steps:

\begin{itemize}
    \item \textbf{Domain Definition:} The system was decomposed into distinct operational domains—public, user, agent, and admin. Each domain was independently analyzed to extract role-specific workflows and user journeys, which were subsequently synthesized into an integrated, end-to-end system design.

    \item \textbf{Domain-Specific Requirement Matrices:}
        \begin{itemize}
            \item \textit{Public Resource Requirements:} Defined core functions available to unauthenticated users, including homepage navigation, registration initiation, and profile creation.
            \item \textit{User Requirements:} Captured interactions specific to entity registrants and representatives, including entity submission, access management, and communication with agents.
            \item \textit{Agent Requirements:} Encompassed the functionality needed by competent authority staff to oversee entities. This included statistical dashboards, entity search and filtering, editable entity views, task tracking, and direct messaging capabilities.
            \item \textit{Admin Requirements:} Focused on administrative oversight of automatic registration and notification workflows, covering tasks such as batch processing, error handling, statistical aggregation, and scheduled reporting.
        \end{itemize}

    \item \textbf{System and Security Requirements:} These requirements governed the technical and architectural integrity of the registry. Key areas included single sign-on (SSO), role-based access control (RBAC), audit logging, encryption, validation layers, and resilience measures against data loss or corruption.

    \item \textbf{Algorithm Design:} Prior to implementation, core algorithms—such as classification, size determination, and entity-type transformation—were formally conceptualized. This ensured that the system’s decision logic was transparent, testable, and grounded in the regulatory objectives.

    \item \textbf{Traceability Mapping:} Each requirement was tagged to a specific system module or algorithm to establish traceability between legal obligations, system behavior, and implementation. This enabled both forward (requirements-to-code) and backward (code-to-requirements) tracking, reinforcing compliance and maintainability.
\end{itemize}

These matrices served as the foundation for a structured development backlog, ensuring full legal and technical traceability from the directive’s provisions down to individual code-level features.

\subsection{Algorithm and Workflow Design}

The core intelligence of the registry lies in its algorithmic subsystems. These were categorized into four distinct modules and modeled using BPMN-style flowcharts in Lucidchart. Each model ensured that decision logic, input validation, and output conditions were explicitly defined to support system robustness and regulatory traceability.

\paragraph{User-Driven Registration Logic}

This module supports users in manually entering data. It features:
\begin{itemize}
    \item \textbf{Entity submission workflow:} Data is validated against minimal required fields (name, sector, national ID).
    \item \textbf{Duplication checks:} Leveraging existing national registry APIs to prevent redundant entries.
    \item \textbf{Edit and approval mechanisms:} Submitted entries are categorized as pending and may be modified by agents before approval. This improves administrative efficiency without requiring resubmission.
    \item \textbf{Access control validation:} Upon account creation and submission, users are linked to the entity and must be verified by agents before representative privileges are granted.
    \item \textbf{Notification system:} Email notifications are triggered on key events (submission, approval, access granted).
\end{itemize}

\paragraph{Automated Registration Logic}

This module processes batch or single imports from interoperable national systems or public databases:
\begin{itemize}
\item \textbf{Input parsing engine:} Accepts structured formats (CSV, XML, JSON), validates schema.
\item \textbf{Threshold evaluation:} Cross-references entity data (turnover, headcount, entity activity) against the size-cap rule and Annex I and II criteria.
\item \textbf{Classification scoring:} Applies classification thresholds using rule-based logic.
\item \textbf{Logging and fallback:} Entities failing schema or logic checks are logged for manual review. Failed entries are accessible through administrative dashboards for analysis and feedback.
\end{itemize}

\paragraph{Entity Classification Logic}

Key to compliance with Article 2 and 3, these algorithms were created to determine whether an entity is "essential", "important", or "excluded":
\begin{itemize}
    \item \textbf{Sector matching:} Entity sector code is matched against Annex I or II.
    \item \textbf{Size criteria:} Entity size and criticality are evaluated using thresholds based on turnover, balance sheet, employee count, and Euro conversion benchmarks.
    \item \textbf{Contextual fallback:} In cases where sectors lack a designated entity type under NIS2 (e.g., certain manufacturing sub-sectors), the algorithm prompts fallback responses.
    \item \textbf{Priority sequence:} The classification algorithm classifies entities in a pre-determined sequence to avoid classification failure or conflict. 
\end{itemize}

\paragraph{Notification Logic}

Member states are obligated to notify the European Commission, the Cooperation Group, and ENISA with aggregated reports about registered entities. The registry supports this requirement through a streamlined reporting workflow:

\begin{itemize}
    \item \textbf{Aggregation:} Entity data is automatically aggregated and prepared for export through standardized formats tailored to regulatory reporting.
    \item \textbf{Presentation:} Aggregated data is visualized in an intuitive dashboard with charts and contextual summaries, providing oversight and decision support.
    \item \textbf{Reporting:} Reports can be automatically populated with live data and exported at the click of a button. Each report is customized to meet the requirements of its intended recipient.
\end{itemize}

\paragraph{System Coordination and Integration Logic}

These were created to handle inter-system workflows, including notification, escalation, and coordination with EU-level systems:
\begin{itemize}
    \item \textbf{Trigger-based flows:} From entity data to classification changes propagated throughout the registry.
    \item \textbf{API orchestration layer:} Synchronizes registry data across the system.
\end{itemize}

The orchestration flowchart includes fallback channels and redundancy checks to prevent data mismatches, misclassifications and redundant entries.

\subsection{System Architecture Modeling}

A unified system architecture was developed to encapsulate all algorithmic modules and user interactions. This architectural model emphasizes microservices, event-driven workflows, and modular deployment across national infrastructures and EU.

Key technical components:
\begin{itemize}
\item \textbf{Frontend Portals:} Public, User, Agent, and Admin portals operate on separated authentication layers. UI components reflect requirement matrices with strict role-based access control (RBAC).
\item \textbf{Dashboard Features:} Users can monitor multiple entities, track classification status, submit changes, and view audit logs. Agents have supervisory dashboards with entity filtering, classification override tools, and statistical summaries. Administrators oversee automatic registrations and generate standardized NIS2 reports.
\item \textbf{Algorithm Engine:} Exposes each algorithm as an independent microservice with RESTful interfaces. Services include validation, classification, registration, orchestration, and logging.
\item \textbf{Data Layer:} The system uses MongoDB collections to store and manage structured entity data. It supports versioned records, audit trails, and rollback capabilities through embedded metadata and historical snapshots. The data layer is optimized for fast read operations, particularly for analytics dashboards and reporting views.
\item \textbf{Interoperability Layer:} Built-in adapters support API-level communication with national registries. XML and JSON schemas are enforced for structured exchanges.
\item \textbf{Security Backbone:} The system enforces authentication, role-based authorization, data validation, and structured logging. Deployment is intended to include HTTPS encryption for all communication, and secure storage with regular backups. Planned features also include real-time alerting and access control auditing.
\end{itemize}

\subsection{Benchmarking and Accuracy Evaluation}

To evaluate the effectiveness of the implemented algorithms, a structured benchmarking methodology was applied:

\begin{itemize}
    \item \textbf{Data Sources:} Benchmarking was conducted using real-world data from the Regnskapsregisteret and Enhetsregisteret to evaluate the accuracy of size an
    d activity classifications.
    \item \textbf{Sample Selection:} A random sample of 100 entities was drawn from Kapital’s 2024 list of Norway’s 500 largest companies, ensuring high-quality and complete data coverage.
    \item \textbf{Scoring Framework:} Each entity was scored across three categories—size, activity, and classification. A score of 1 was assigned for correct outcomes and 0 for discrepancies, with a maximum possible score of 3 per entity.
    \item \textbf{Evaluation Output:} The process revealed sector-specific classification challenges, notably within the manufacturing and health domains. These insights informed iterative improvements to prompt formulation and fallback logic.
\end{itemize}

This quantitative assessment enhanced the reliability and transparency of the system’s classification performance.

\section{System Development Framework}

While the system architecture and technology stack are discussed in detail in Chapter 3, this section briefly outlines the overall development framework.

An iterative development model was followed, enabling incremental refinement of features in response to evolving insights gained from regulatory analysis and testing feedback. Requirements were translated into actionable development tasks and tracked using a structured backlog derived from the requirement matrices presented earlier.

\begin{figure}[H]
    \centering    
    \includegraphics[width=1.0\textheight, width=0.75\textwidth]{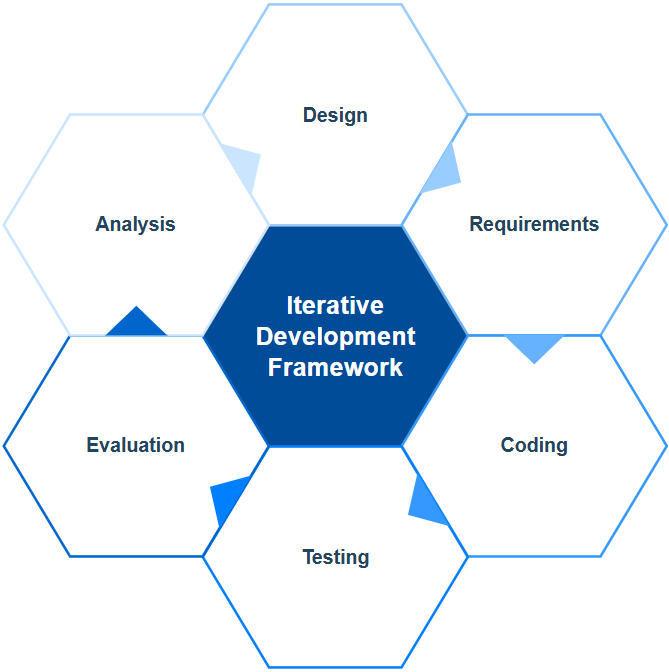}
    \caption{Development Framework}
    \label{fig:Development_Framework}
\end{figure}

Each iteration targeted a combination of functional domains (public, user, agent, admin) and core algorithmic modules (registration, labeling, classification). Continuous integration and testing were applied throughout to ensure correctness, maintainability, and alignment with regulatory and performance objectives. Visual models created during earlier phases were used as implementation blueprints, ensuring consistency between regulatory interpretation and technical execution.

The complete system, along with its algorithms and user interfaces, was built in parallel with benchmarking procedures to ensure accuracy, compliance, and practical viability.

Version control was managed using Git, with the repository hosted on GitHub. This ensured traceability, collaboration, and safe iterative development through branches and pull requests. Key milestones and features were versioned using tags, supporting clear documentation and rollback if needed.

Additionally, the registry is developed by following best practices. We adhere to JavaScript best practices as outlined in the Mozilla Developer Network \cite{JavaScriptBestP}, TypeScript best practices as detailed in their official documentation \cite{TypeScriptBestP}, and Vue.js best practices as recommended by learnvue \cite{VueBestP}.

\section{Summary}

This chapter detailed the structured and technically grounded methodology adopted to develop the national registry with scalability to a pan-European registry in alignment with the NIS2 Directive. By embedding evaluation metrics directly into algorithm development, employing benchmarking against authoritative sources, and implementing clear workflows for registration and supervision, we ensured that the resulting registry is not only legally compliant but also operationally verifiable. The combined use of flowcharts, modular design, and dashboard-driven supervision delivers a transparent, scalable, and measurable system for the classification and oversight of critical entities.

\chapter{Design \& Implementation}
% Insert text here
This chapter is dedicated to synthesizing the NIS2 Directive into a system design and requirements. The latter part of this chapter will combine findings from the NIS2 Directive with the insights gathered from the literature review into a complete registry system. Following is an analysis of the NIS2 Directive.

\section{The NIS2 Directive}

% Introduce the NIS2 Directive and how it is structured.

The NIS2 Directive \cite{NIS2} is composed of the sections recitals, articles, and annexes. Recitals provide the rationale of the legislation and provide the reader with an understanding of the directive's intent. Each recital address a specific aspect of the directive, but numerous recitals might address the same aspect. 

Articles are the operative and legally binding provisions of the document. They define the rules, rights, and procedures that must be implemented for a member state to be compliant. Each article is a unique section and covers specific aspects of the directive, and they are usually structured into paragraphs, sub-paragraphs, and points to provide granularity. 

Annexes provide supplementary information to support the main provisions of the document. Annexes are legally binding and are often supplementary to multiple articles. Article to annex relationships are established through reference. An article will reference an entire annex or a specific part of an annex as a supplement to implement that article. In the NIS2 Directive, the annexes provide a list of essential and important entity types, and the sector and sub-sector in which they belong. Below is a concept matrix to link relevant recitals, articles, and annexes to concepts that appear in the NIS2 directive and are important to this thesis's objectives and problem statement. 

\begin{figure}[H]
    \centering
    \includegraphics[height=0.9\textheight, width=0.6\textwidth]{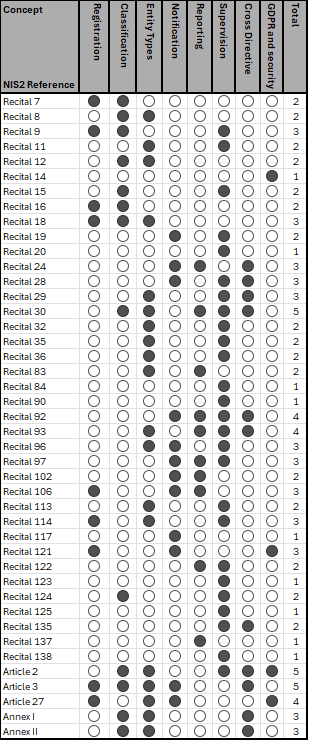}
    \caption{NIS2 Concept Matrix}
    \label{fig:Concept_Matrix}
\end{figure}

The concept matrix consists of eight key concepts derived from their prominence in the NIS2 directive and their relevance to this thesis. These concepts are deliberately chosen to strike a balance between breadth and specificity—broad enough to encompass the essential aspects of NIS2 while remaining precise enough to provide a meaningful structure. This approach prevents excessive fragmentation caused by overly narrow concepts while avoiding the vagueness that can result from concepts that are too general. Later sections will go into details about each concept and its components. Following is a concise description of each concept. 

\paragraph{Registration}

The concept of registration is addressed in recital 7, 9, 16, 18, 106, 114, 121, and in article 3 and 27. Registration encompasses the need for a single point of registration and registry information collection requirements. 

\paragraph{Classification}

The concept of classification is addressed in recital 7, 8, 9, 12, 15, 16, 18, 30, 124, Article 2 and 3, and in Annex I and II. Classification is a rather large and complex concept that defines the rule set for classification. It includes all factors used by the rule sets to determine the correct classification class. Additionally, it defines the classification classes and is closely connected to the concept of registration and entity types. The entity types defines economic and societal activities that are within the scope of the NIS2 Directive, and thereby, required to undergo classification. 

\paragraph{Entity Types}

The concept of entity types is addressed in recital 8, 11, 12, 18, 29, 30, 32, 35, 36, 83, 93, 96, 113, 114, Article 2, 3, 37, and in Annex I and II. Entity types are categorized into sector and sub-sector, and an entity type is a description of their economic and societal activity. The concept of entity types is often interpreted in the context of classification and thus, appears in just under half the relevant sections of the NIS2 directive. 

\paragraph{Notification}

The concept of notification is addressed in recital 19, 24, 28, 92, 96, 97, 102, 106, 117, 121, Article 3 and 27. Notification elaborates on information sharing between the competent authorities, CSIRTs, and Union regulatory bodies. It also defines what information must be shared, with whom, and at what intervals. This concept is closely linked to registration, as it helps determine the required information during the registration process.

\paragraph{Reporting}

The concept of reporting is addressed in recital 24, 30, 83, 92, 93, 97, 102, 106, 122, and 137. It is also addressed in Article 23, which is excluded from the scope of this thesis. Reporting outlines what to report, when to report, and to whom. It also describes the broader reason for reporting and its impact on cross-border cooperation and systemic risks. 

\paragraph{Supervision}

The concept of supervision is addressed in recital 9, 11, 15, 19, 20, 28, 29, 30, 32, 35, 36, 84, 90, 92, 93, 96, 97, 113, 114, 122, 123, 124, 125, 135, 138, and Article 2. This is the largest concept and encompasses all forms of supervision performed by the competent authorities on entities within the scope of the NIS2 Directive. It is linked to classification in that supervision differs based on an entity's classification class. 

\paragraph{Cross Directive}

The concept of cross directive is addressed in recital 24, 28, 29, 30, 92, 93, 135, Article 2, 3, and in Annex I and II. This concept emerge when multiple directives address a certain aspect of the NIS2 directive. For example, entities identified as critical by the CER Directive are deemed essential by the NIS2 Directive \cite{NIS2}. 

\paragraph{GDPR and security}

The concept of GDPR and security is addressed in recital 14 and 121, and Article 2 and 27. This concept is based on the GDPR and security criteria imposed on the implementation of the NIS2 pan-European registry. 
\\
\\
Having established a structured understanding of these key concepts, we will now move on to examining the most crucial parts of these concepts in developing a NIS2 compliant pan-European registry. 

\subsection{Recitals}

% Explain the importance of receitals and their purpose. Then, explain the most important ones. 

The recitals in the NIS2 Directive provides a contextual understanding when interpreting articles. They help interpreting the articles within a context and provide increased granularity. An example of this is recital 28 on the application of Union legal acts \cite{NIS2}. A reference is made to the DORA regulation as a Union Legal act. Even though, article 2 on the subject of scope informs that all entities exempted from certain obligations in NIS2 Directive, DORA is not mentioned to be a Union legal act \cite{NIS2}. The same happens under article 4 on sector-specific Union legal acts, underscoring the importance of including recitals in our research. 

Recitals are not legally binding, but they can help the competent authorities perform their supervisory tasks under nuanced classifications. This sub-section is therefore meant to supplement a key part of our registry, manual assessments. Below are the interpretations of recitals that can aid in this task. 

\subsubsection{Public Administration Entities}

A public administration entity is excluded from the scope of this directive if their primary activities are in the areas of national or public security, defense or law, but not if the primary activity is only marginally related to those areas as stated in Recital 8 of the NIS2 Directive \cite{NIS2}. Additionally, any entities that provides services exclusively to a public administration entity excluded from the scope of NIS2, on that basis, should also be excluded as stated in Recital 9 of the NIS2 Directive \cite{NIS2}. As a final note, Recital 11 states that public administration entities that provide trust services defined to be within the scope of Regulation (EU) No 910/2014\cite{RECITAL11} should fall within the scope of the NIS2 Directive \cite{NIS2}. 

Manual assessments, combined with feedback from the entities involved, are necessary to ensure accurate classification in these nuanced cases. In this regard, the registry has functionality for both. Additionally, entities will be asked if they provide services exclusively to entities within the scope of this subsection. 

\subsubsection{Risk Assessments}

Risk assessments are mentioned in Recital 15, which is an integral part of correct classification \cite{NIS2}. Entities that are classified as important entities should undergo risk-assessments to ensure that entities with a higher risk profile are re-classified as essential. Furthermore, Recital 124 underscores the importance of distinguishing between essential entities based on their risk-profile to prioritize supervisory tasks \cite{NIS2}. 

This peculiarity can be captured by requesting a manual risk assessment for every important entity. Additionally, a comprehensive labeling system will allow the assessor to attach a risk-label to any entity considered to have a higher risk profile. This labeling scheme increases entity understanding and helps appropriate and proportionate supervision of entities. 

\subsubsection{Size Disqualification}

According to Recital 16, entities that have partner enterprises or that are linked enterprises in which their network and information systems in relation to their entity type are considered independent, should be disqualified from their initial size estimation \cite{NIS2}. Realistically, their new size estimation must be assessed with regards to the independence of their network and information systems. Entities with revised size estimates may be excluded or reclassified under the NIS2 Directive.

Unfortunately, this can only be captured through manual assessment or manual registration. In those cases, the entity is responsible for correctly informing under manual registration, or to request a classification change through the user portal. 
\\
\\
In summary, the recitals in the NIS2 Directive provide critical interpretative guidance that informs nuanced classifications within the registry. By addressing exclusions, risk assessments, and size-based disqualifications, the registry can ensure accurate and context-sensitive classification of entities. These insights underscore the importance of combining automated processes with manual assessments and feedback mechanisms to handle complex classifications effectively.

\subsection{Article 2: Scope}

% Explain this article and its relevance to this thesis

Article 2 of the NIS2 Directive defines the scope of the directive to clarify the directive's boundary \cite{NIS2}. Particularly, it defines the sector, sub-sectors, and entity types that are subject to the rules outlined in other articles by specifying them explicitly or by reference to annex I and II. Additionally, article 2 gives reference to the size-cap defined in Article 2 of the Annex in Recommendation 2003/361/EC \cite{NIS2}. The size-cap rule defines the criteria that outlines which size an entity belongs \cite{SizeCapRule}. Furthermore, Article 2 specifies special entity cases, cases of heightened risk, entities governed by multiple directives, and rules member states must follow when including and excluding certain entities from the scope in its entirety or from specific articles in the NIS2 Directive. Following is an elaboration on the cases covered by the scope:
\begin{itemize}
    \item \textbf{Size-Cap Rule:} The size of an entity is categorized into above medium, medium, and below medium size \cite{NIS2}. The criteria for size is based on the entity's annual turnover, annual balance sheet, and it's total number of employees within the Union \cite{SizeCapRule}. Entity size is then applied to the entity type in Annex I or II to determine if that entity is within scope of this directive.
    \begin{itemize}
        \item \textbf{Above Medium Size:} Any entity in Annex I or II that has at least 250 employees and more than EUR 50M in annual turnover or an annual balance sheet of more that EUR 43M \cite{SizeCapRule}. 
        \item \textbf{Medium Size:} Any entity in Annex I or II that has between 50 and 249 employees and an annual turnover between EUR 10M and 50M or an annual balance sheet between EUR 10M and 43M. 
        \item \textbf{Below Medium Size:} Any entity in Annex I or II that has less than 50 employees and an annual turnover below EUR 10M or an annual balance sheet below EUR 10M.
    \end{itemize}
    \item \textbf{Special Entity Cases:} Entity size is ignored for the entity types public electronic communications networks, publicly available electronic communications services, trust service provides, top-level domain (TLD) name registers, domain name system (DNS) service providers, domain name registration services, central government public administration, regional level government public administration, and research organizations \cite{NIS2}. 
    \item \textbf{Cases of Heightened Risk:} Entities that are essential for the maintenance of critical societal or economic activities, entities that provides services that could upon disruption have significant impact on public safety, public security, or public health, entities that provides services that could upon disruption have significant systemic and particularly cross-border impact, and entities that have a special importance at national or regional level for the particular sector or type service they provide \cite{NIS2}.
    \item \textbf{Entities Governed by Multiple Directives:} Entities identified as critical under CER \cite{NIS2}. 
\end{itemize}

Article 2 is closely related to Article 3 on essential and important entities, and these articles provide the basis for our classification framework. 

\subsection{Article 3: Essential and Important Entities}

Article 3 of the NIS2 Directive defines a framework for classifying entities into two groups: essential and important entities \cite{NIS2}. The classification framework is predicated on the scope established in article 2 and the sectors, sub-sectors, and entity types in Annex I and II. The result is a proportional regulatory environment for classifying entities based on their criticality for member state and Union wide cybersecurity. Additionally, the article briefly state entity registration requirements and sets the timeline and reporting obligations imposed on member states. 

\subsubsection{Entity Classification}

The classification of entities is a central concept in this thesis. Under the concept of classification, every entity must be assigned to exactly one of the three mutually exclusive classes in the figure below. Additionally, cases of classification conflict will invoke the following resolution hierarchy: Essential $>$ Important $>$ Excluded.

\begin{figure}[H]
    \centering
    \includegraphics[width=1.0\linewidth]{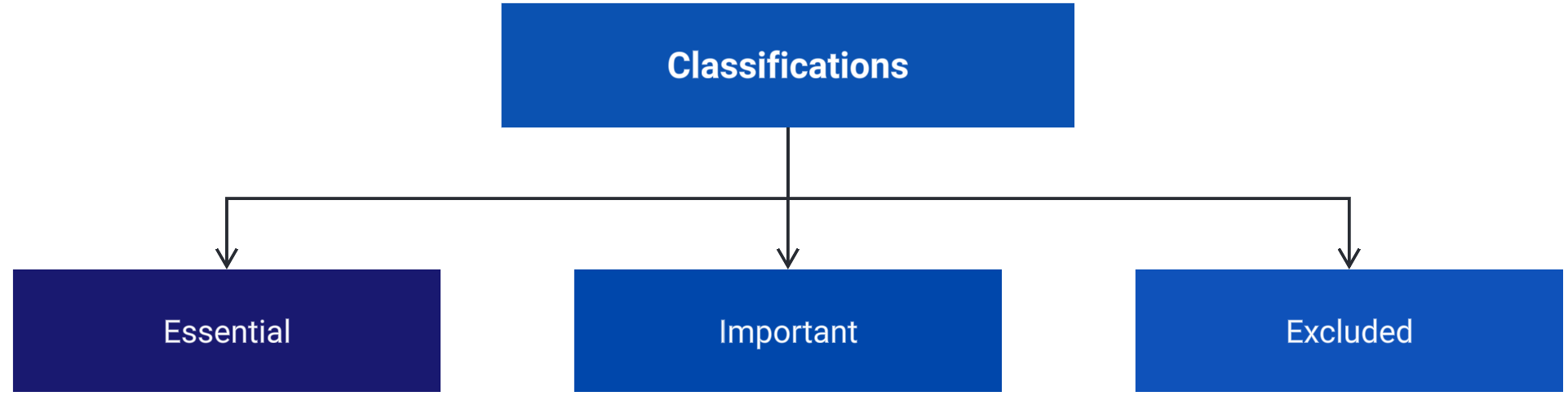}
    \caption{Entity Classification Classes}
    \label{fig:Classification_Classes}
\end{figure}

This ensures that each entity has a clear and unambiguous classification based on its criticality for the societal and economic functioning in the member state in which it operates, and in the Union as a whole. Classification is mutually exclusive to ensure that an entity cannot simultaneously belong to more than one class. For instance, an entity classified as 'essential' cannot also be classified as 'important' or 'excluded'. This rule eliminates classification overlaps and ensures the correct classification to guide appropriate supervision of entities. Additionally, the classification system is exhaustive, meaning that every entity in a member state must belong to one of the three classes. Exhaustiveness means that every entity is considered for inclusion to ensure that all entities in a member state are evaluated for inclusion. A figure is provided below to illustrate the characteristics of classification.

\begin{figure}[H]
    \centering
    \includegraphics[width=0.4\linewidth, height=3.5cm, keepaspectratio]{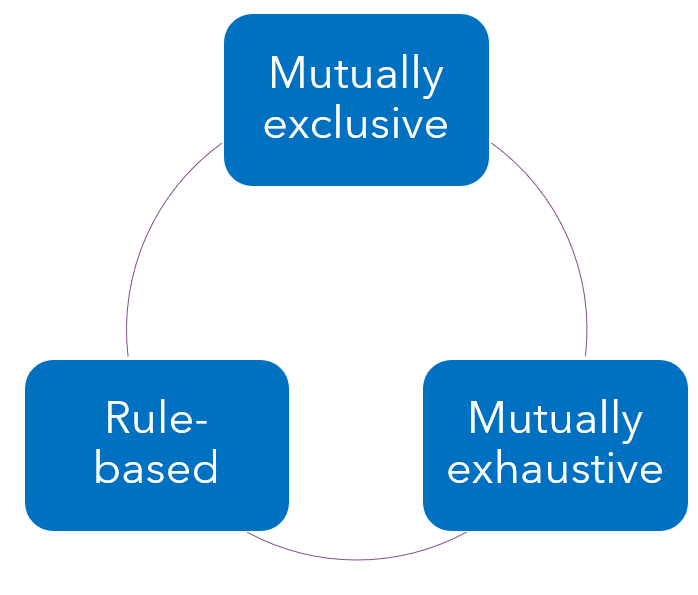}
    \caption{Entity Classification Characteristics}
    \label{fig:Classification_characteristics}
\end
{figure}

Furthermore, the correct classification is determined by evaluating a range of factors and their sub-factors. The factors for classification are stated in the recitals, Article 2 on scope, Article 3 on essential and important entities, and in Annex I and II \cite{NIS2}. The following bubble map presents the factors and their corresponding sub-factors that are assessed to determine the appropriate classification.

\begin{figure}[H]
    \centering
    \includegraphics[width=1.0\linewidth]{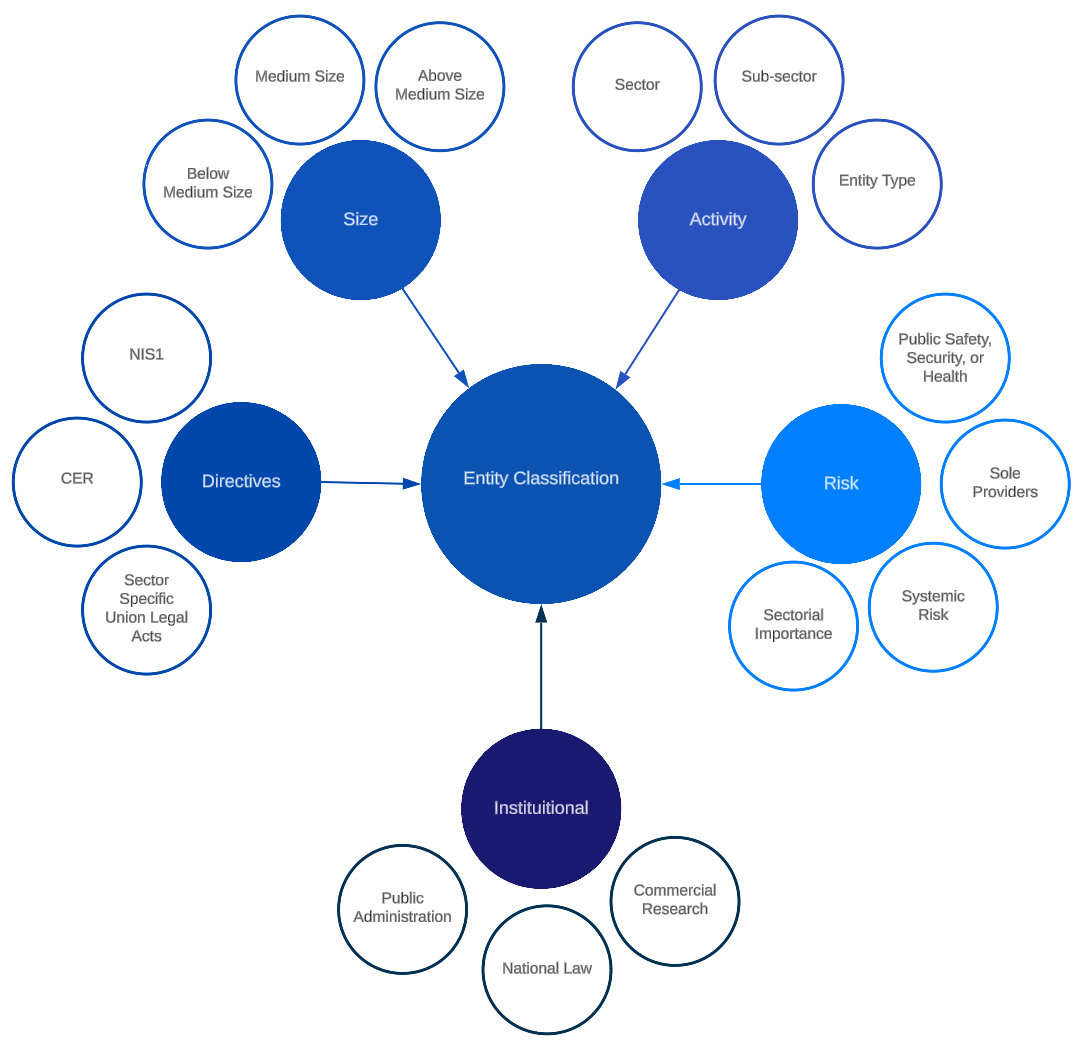}
    \caption{Entity Classification Factors}
    \label{fig:Classification_Factors}
\end{figure}

A rule set is applied to evaluate all or some of these factors to determine an entity's classification class. Each classification class has multiple rule sets designed to assess whether an entity belongs to that classification. 

\subsubsection{Essential Entity Classification} 

Essential entities are defined as those who are essential for the societal and economic functions of member states and the Union. The criteria for essential classification are based on entity type criticality for the functioning of societal and economic activities, the size of an entity, and cases of increased risk \cite{NIS2}. Additionally, entities identified as essential pursuant to the NIS1 Directive and entities identified as critical pursuant to the CER Directive are determined to be essential pursuant to the NIS2 Directive \cite{NIS2}. 

The NIS2 Directive allows member states some flexibility in classifying entities. Cases that enable member states to assess whether an entity is essential are entities identified as essential pursuant to national laws, commercial research entities, and local public administration entities \cite{NIS2}. Furthermore, member states are currently required to assess the risk of entities themselves, allowing for more flexibility to decide when an entity is essential. Below is a list of  rule sets for essential classification \cite{NIS2}:
\begin{itemize}
    \item \textbf{Above Medium-Sized:} Entity types in Annex I that are above medium in size. 
    \item \textbf{Medium-Sized:} Medium sized entities of entity type: provider of public electronic communication networks or services.
    \item \textbf{Risk-Based:} Entities that have heightened risk as stated in the section about Article 2 on Scope. Below are the cases of heightened risk that meet the criteria for essential classification \cite{NIS2}:
    \begin{itemize}
        \item \textbf{Safety and Security:} Disruption of the entities activities can have a significant impact on public safety, security, or health. 
        \item \textbf{Sole Provider:} The entity is a sole provider of a service which is essential for the maintenance of critical societal or economic activities.
        \item \textbf{Systemic Risk:} Disruption of the entities activities can induce systemic risk or cross-border impact.
        \item \textbf{Sectorial Importance:} The entity is of special importance for a particular sector or entity type at a national or regional level. 
    \end{itemize}
    \item \textbf{Public Administration Sector:} Entities in the public administration sector defined to be of the central government entity type. 
    \item \textbf{Entity Types:} Trust Service Providers, TLD name registers, DNS service providers, and DNS registration services.
    \item \textbf{Referenced Directives:} Entities identified as critical pursuant to the CER Directive or essential pursuant to the NIS1 Directive.
    \item \textbf{Flexible Cases:} Entities identified as essential per national laws, and commercial research and local public administration entities chosen by the member state.
\end{itemize}

The preceding list contains the rules that govern essential entity classification. Essential entity classification is the most comprehensive and perhaps the most important component of entity classification. Entities identified to be essential are most critical to the well-being of a member state's economy and societal functioning and they are subject to the highest level of supervision. Following is an illustration of the concept of essential entity classification. 

\begin{figure}[H]
    \centering
    \includegraphics[width=1.0\linewidth]{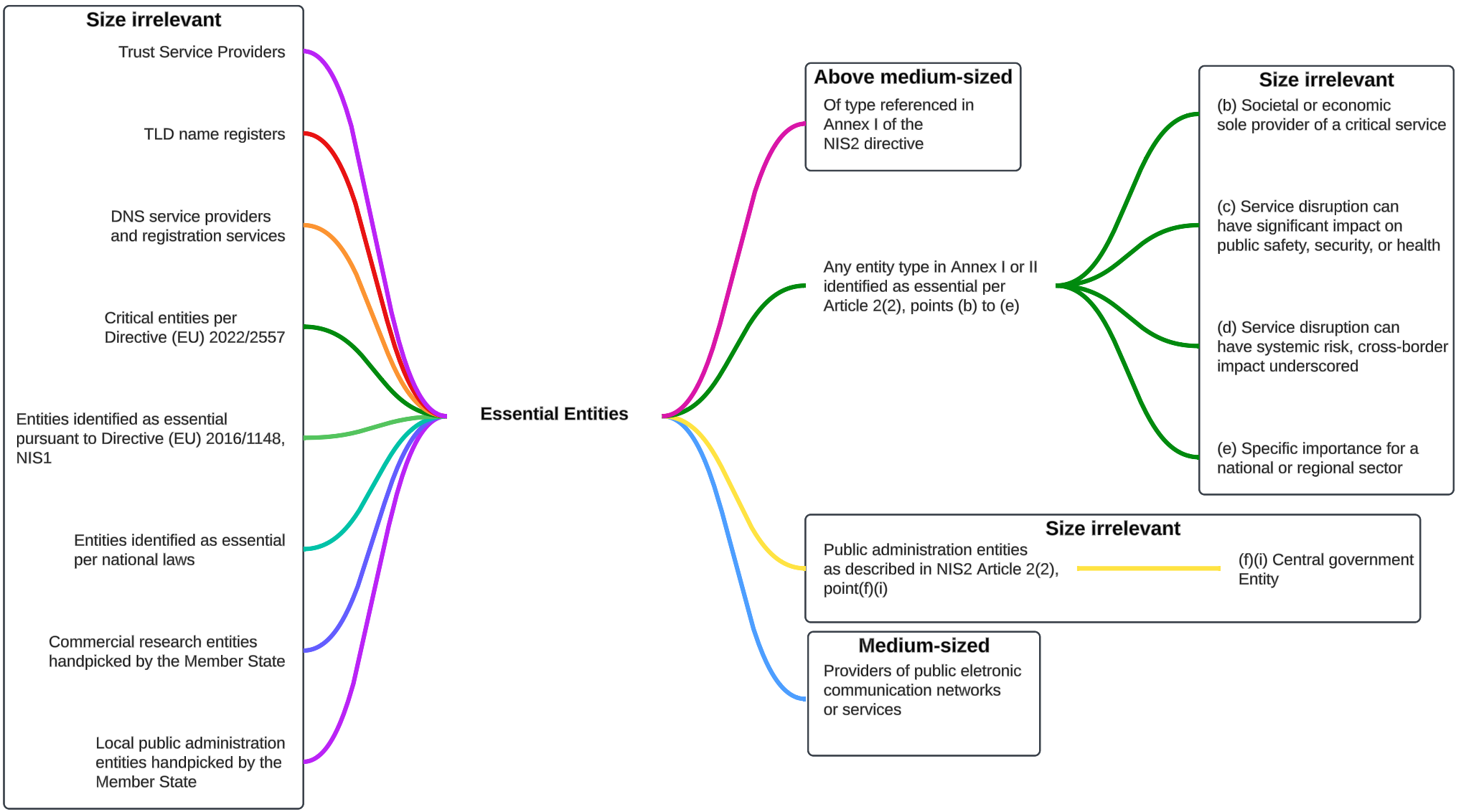}
    \caption{Essential Entity Classification}
    \label{fig:Essential_Classification}
\end{figure}

Entities that do not meet the criteria for essential classification must be screened for important classification. 

\subsubsection{Important Entity Classification:}

Article 3 of the NIS2 Directive \cite{NIS2} primarily defines an important entity as an entity of entity type in annex I or II that do not meet the criteria for essential classification. Classification of important entities consist of 4 cases \cite{NIS2}:
\begin{itemize}
    \item \textbf{Above Medium-Sized:} Entity types in Annex I or II that are above medium in size and do not meet the criteria for essential classification. 
    \item \textbf{Medium-Sized:} Entity types in Annex I or II that are medium in size and do not meet the criteria for essential classification. 
    \item \textbf{Below Medium-Sized:} Providers of public electronic communication networks or services that are below medium in size. 
    \item \textbf{Public Administration Sector:} Any entity in the public administration sector that are determined to be regional pursuant to national law and a risk-assessment. 
\end{itemize}

\begin{figure}[H]
    \centering
    \includegraphics[width=1.0\linewidth]{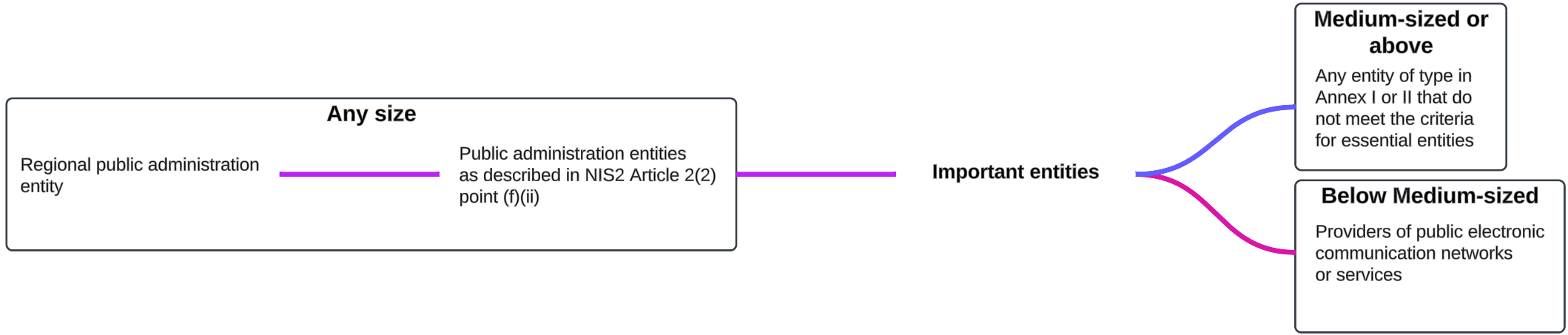}
    \caption{Important Entity Classification}
    \label{fig:Important_Classification}
\end{figure}

In the case an entity does not meet the criteria for either essential or important classification, it is classified as excluded. 

\subsubsection{Excluded Entity Classification}

There are four cases of exclusion besides entities that do not meet the criteria for essential or important classification \cite{NIS2}. The first case is entities exempted from the scope of the CER directive. The second is entities in the public administration sector which primary activity is in national security, public security, defense or law enforcement. The third is entities that provide services exclusively to excluded public administration sector entities. An additional and fourth case of exclusion is for entities that fall outside the scope of the NIS2 Directive. This addition ensures that the classification system is exhaustive and screens all entities for inclusion. Following is an illustration of the concept of excluded entity classification. 

\begin{figure}[H]
    \centering
    \includegraphics[width=1.0\linewidth]{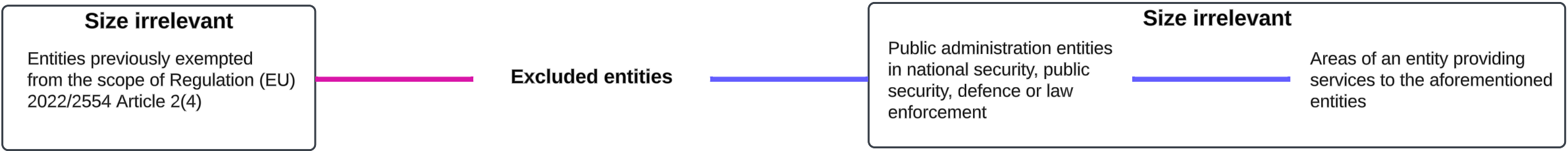}
    \caption{Excluded Entity Classification}
    \label{fig:Excluded_Classification}
\end{figure} 

Observe that the fourth and additional case of exclusion is not a part of the illustration. This is intended because exclusion based on falling outside the scope of the NIS2 Directive is not considered to be part of the classification rule set, it is rather a byproduct of the system's design. 

\subsubsection{Labels}

The factors used to classify entities can be utilized beyond classification. This information can enhance supervision of entities and provide a contextual understanding of an entity. To achieve this, the system will use this information as labels. Labels have exclusivity types, which defines when and whether a label can appear together with another label of the same category. For example, size-based labels have mutual exclusivity among other size-based labels, but are non-exclusive with other labels. Additionally, labels can be assigned automatically or manually, which is referred to as mechanism. Below is a table of the labels used to increase entity understanding. 

\begin{table}[H]
    \centering
    \resizebox{\textwidth}{!}{
    \rotatebox{90}{
    \begin{tabular}{|p{4.0cm}|p{3.0cm}|p{3.0cm}|p{4.8cm}|p{3.5cm}|} % 14.45cm
        \hline
        \textbf{Label} & \textbf{Category} & \textbf{Type} & \textbf{Description} & \textbf{Mechanism} \\ 
        \hline
        Above Medium Sized & Size & Mutually Exclusive & Above medium sized per the size-cap rule & Automatic/Manual \\
        \specialrule{0.4pt}{0pt}{0pt}
        Medium Sized & Size & Mutually Exclusive & Medium sized per the size-cap rule & Automatic/Manual \\
        \specialrule{0.4pt}{0pt}{0pt}
        Below Medium Sized & Size & Mutually Exclusive & Below medium sized per the size-cap rule & Automatic/Manual \\
        \specialrule{0.4pt}{0pt}{0pt}
        High Criticality & Activity & Non-exclusive & Entity type is in annex I & Automatic/Manual \\
        \specialrule{0.4pt}{0pt}{0pt}
        Critical & Activity & Non-exclusive & Entity type is in Annex II & Automatic/Manual \\
        \specialrule{0.4pt}{0pt}{0pt}
        Sector & Activity & Hierarchical Dependency & The primary sector of an entity & Automatic/Manual \\
        \specialrule{0.4pt}{0pt}{0pt}
        Sub-Sector & Activity & Hierarchical Dependency & The sub-sector of an entity & Automatic/Manual \\
        \specialrule{0.4pt}{0pt}{0pt}
        Entity Type & Activity & Hierarchical Dependency & The entity types of an entity as prescribed in Annex I or II & Automatic/Manual \\
        \specialrule{0.4pt}{0pt}{0pt}
        CER & Cross-Directive & Non-exclusive & Entities critical per the CER Directve & Automatic/Manual \\
        \specialrule{0.4pt}{0pt}{0pt}
        NIS1 & Cross-Directive & Non-exclusive & Entities essential per the NIS1 Directive & Automatic/Manual \\
        \specialrule{0.4pt}{0pt}{0pt}
        National Law & Cross-Directive & Non-exclusive & Entities essential per National Law & Automatic/Manual \\
        \specialrule{0.4pt}{0pt}{0pt}
        Research Organization & Cross-Directive & Non-exclusive & Entities performing critical commercial research activity & Automatic/Manual \\
        \specialrule{0.4pt}{0pt}{0pt}
        Security operator & Cross-Directive & Non-exclusive & Entity is involved in national or public security & Automatic/Manual \\
        \specialrule{0.4pt}{0pt}{0pt}
        Sole Provider & Risk & Non-exclusive & Sole Provider of a critical service & Automatic/Manual \\
        \specialrule{0.4pt}{0pt}{0pt}
        Public Safety, Security, or Health & Risk & Non-exclusive & Critical for public safety, security, or health & Automatic/Manual \\
        \specialrule{0.4pt}{0pt}{0pt}
        Systemic Risk & Risk & Non-exclusive & Disruption can cause systemic risks & Automatic/Manual \\
        \specialrule{0.4pt}{0pt}{0pt}
        Sectorial Importance & Risk & Non-exclusive & Special importance to a primary- or sub-sector & Automatic/Manual \\ 
        \specialrule{0.4pt}{0pt}{0pt}
        Representative & None & Non-exclusive & Supervision of the entity is through a representative & Automatic/Manual \\
        \specialrule{0.4pt}{0pt}{0pt}
        Custom Label & Any & Non-exclusive & A custom label defined by an agent & Manual \\
        \hline
    \end{tabular}}}
    \caption{Entity Labels}
    \label{tab:Entity_Labels}
\end{table}

Label types are applicable within the labels label category and defines its relation to other labels of the same category.

\paragraph{Mutually exclusive}

The mutually exclusive label type determines that a label cannot appear with other labels of the same category. Essentially, this means that an entity can only be assigned one label from the size category. 

\paragraph{Hierarchical Dependency}

The hierarchical dependency label type informs that the labels within this category have an hierarchical dependency with each other. Label dependency is implicitly mentioned in table \ref{tab:Entity_Labels} by the ordering of labels. This means that an entity type is a child of a sub-sector and that a sub-sector is the child of a sector. Specific sub-sectors belong to specific sectors and specific entity types belong to specific sub-sectors. 
\\
\\
Furthermore, agents can define custom labels. These labels are intended for cases where a pre-defined label does not exist, and when a pre-defined label would bring contextual value to entity supervision. 

\subsection{Annex I and II}
% Explain the Annexes and their relevance to this thesis
Annex I and II contains a list of sectors, sub-sectors, and the entity types within a sector or sub-sector that are within the scope of the NIS2 Directive \\ \cite{NIS2}. Entity types are definitions of entity activity. Entities that recognize that their activity match an entity type in Annex I or II must consider themselves to be within the scope of the NIS2 directive, unless other rules imply that they are not. Entity types are therefore used to determine whether an entity's activity is within the scope of the NIS2 Directive \\ \cite{NIS2}. 

The sectors, sub-sectors, and entity types in Annex I and II are those considered to be essential and important to the societal and economic well-being of a member state within the Union, and for the Union as a whole \cite{NIS2}. Entity type definitions are predominantly defined in other Union directives and regulations and we have rewritten the entity types by extracting their definitions from those documents. The rationale behind this choice was twofold. First, entity type definitions must be examined for us to understand them. Second, entity type definitions must be extracted and rewritten to enable manual and automatic registration of entity types. Additionally, the manual registration mechanism is supplemented with explanations from these documents to ensure that a registrant understands the definition of an entity type. This allows the registrant to make an informed decision about the type of entity they are registering. 

To facilitate a clearer understanding of the scope of the NIS2 Directive, the following sections provide a detailed overview of the sectors, sub-sectors, and entity types listed in Annex I and II. This breakdown aims to provide registrants and stakeholders with the necessary context to identify whether an entity's activities align with the defined entity types. Annex I contains the sectors, sub-sectors, and entity types that are considered to be of high criticality for the societal and economic well-being of a member state and the Union \cite{NIS2}. Annex II contains the sectors, sub-sectors, and entity types that are considered to be critical for the societal and economic well-being of a member state and the Union \\ \cite{NIS2}. Additionally, entity types that need additional clarification are assigned an Annotation Number, which corresponds to a row in the explanation table.

\subsubsection{Annex I on Sectors of High Criticality}

Annex I contains all the sectors, their respective sub-sectors and entity types that are of high criticality \cite{NIS2}. If an entity is of one or multiple entity types in Annex I, it is an entity of high criticality to the societal and economic well-being of the member state in which it has its activities \\ \cite{NIS2}. Following are the sectors and their respective entity types in Annex I. 

\paragraph{The Energy Sector}

The energy sectors consists of 5 sub-sectors and 18 entity types and is the largest sector in Annex I. The following outlines the clarified entity types for the energy sector.

\begin{table}[H]
    \centering
    \resizebox{\textwidth}{!}{
    \begin{tabular}{|p{2cm}|p{2cm}|p{11cm}|}
        \hline
        \textbf{Sector} & \textbf{Sub-sector} & \textbf{Entity Type} \\ \hline
        Energy          & Electricity         & Sale of Electricity \\
        \specialrule{0.4pt}{0pt}{0pt}
                        &                     & Operation or maintenance of electricity distribution systems \\
                        \specialrule{0.4pt}{0pt}{0pt}
                        &                     & Operation or maintenance of electricity transmission systems \\
                        \specialrule{0.4pt}{0pt}{0pt}
                        &                     & Generation of electricity \\
                        \specialrule{0.4pt}{0pt}{0pt}
                        &                     & Nominated electricity market operator [1] \\
                        \specialrule{0.4pt}{0pt}{0pt}
                        &                     & Aggregation, demand-response, or energy storage services \\
                        \specialrule{0.4pt}{0pt}{0pt}
                        &                     & Operator of electricity recharging point \\
                        \specialrule{0.4pt}{0pt}{0pt}
                        & District Heating 
                          and cooling         & Distribution of thermal energy in steam or hot/cold liquid form \\
                          \specialrule{0.4pt}{0pt}{0pt}
                        & Oil                 & Operators of transmission pipelines \\
                        \specialrule{0.4pt}{0pt}{0pt}
                        &                     & Operators of productions, refinement, treatment, storage, and/or                          transmission facilities \\
                        \specialrule{0.4pt}{0pt}{0pt}
                        &                     & Acquisition, maintenance, or sale of oil stocks \\
                        \specialrule{0.4pt}{0pt}{0pt}
                        & Gas                 & Carry out the function of gas supply \\
                        \specialrule{0.4pt}{0pt}{0pt}
                        &                     & Carry out the function of gas distribution \\
                        \specialrule{0.4pt}{0pt}{0pt}
                        &                     & Carry out the function of gas transmission \\
                        \specialrule{0.4pt}{0pt}{0pt}
                        &                     & Carry out the function of gas storage \\
                        \specialrule{0.4pt}{0pt}{0pt}
                        &                     & Carry out the function of liquefaction of natural gas, including                          import, offload, and re-gasification of LNG \\
                        \specialrule{0.4pt}{0pt}{0pt}
                        &                     & Production, transmission, distribution, supply, purchase, and/or                          storage of natural gas including LNG \\
                        \specialrule{0.4pt}{0pt}{0pt}
                        &                     & Operator of natural gas refinement or treatment facilities \\
                        \specialrule{0.4pt}{0pt}{0pt}
                        & Hydrogen            & Operator of hydrogen production, storage, and transmission \\
                        \specialrule{0.4pt}{0pt}{0pt}
        \hline
    \end{tabular}}
    \caption{The Energy Sector}
    \label{tab:Energy_Sector}
\end{table}

\begin{table}[H]
    \centering
    \resizebox{\textwidth}{!}{
    \begin{tabular}{|p{0.5cm}|p{14.95cm}|}
        \hline
        \textbf{ID} & \textbf{Entity Type} \\ \hline
        1         & A market operator designated by the competent authority to carry out tasks related to                     single day-ahead or single intraday coupling \\ 
        \hline
    \end{tabular}}
    \caption{Entity type explanations for the Energy Sector}
    \label{tab:Energy_Sector_Explanations}
\end{table}

\paragraph{The Transport Sector}

The transport sector is the second largest sector in Annex I and it consists of 4 sub-sectors and 15 entity types. 

\begin{table}[H]
    \centering
    \resizebox{\textwidth}{!}{
    \begin{tabular}{|p{2cm}|p{2cm}|p{11cm}|}
        \hline
        \textbf{Sector} & \textbf{Sub-sector} & \textbf{Entity Type} \\ \hline
        Transport       & Air                 & Air carrier holding a valid operating license \\
        \specialrule{0.4pt}{0pt}{0pt}
                        &                     & Administration and management of airports, airport network                                infrastructure, or operator controls \\
                        \specialrule{0.4pt}{0pt}{0pt}
                        &                     & Airport and core airports, including ancillary installations [2]                          \\
                        \specialrule{0.4pt}{0pt}{0pt}
                        &                     & Air Traffic Control (ATC) services \\
                        \specialrule{0.4pt}{0pt}{0pt}
                        & Rail                & Establishment, maintenance, or management of railway                                      infrastructure \\
                        \specialrule{0.4pt}{0pt}{0pt}
                        &                     & Traffic management, Control-Command, or signaling \\
                        \specialrule{0.4pt}{0pt}{0pt}
                        &                     & Service or transport of goods or passengers which ensure traction                         or provide traction alone \\
                        \specialrule{0.4pt}{0pt}{0pt}
                        &                      & Operator or supplier of railway service facility \\
                        \specialrule{0.4pt}{0pt}{0pt}
                        & Water               & Inland, sea or coastal passenger or freight water transport \\
                        \specialrule{0.4pt}{0pt}{0pt}
                        &                     & Manager of port(s) \\
                        \specialrule{0.4pt}{0pt}{0pt}
                        &                     & Manager of port facilities \\
                        \specialrule{0.4pt}{0pt}{0pt}
                        &                     & Operator of port works or equipment \\
                        \specialrule{0.4pt}{0pt}{0pt}
                        &                     & Operator of vessel traffic services (VTS) \\
                        \specialrule{0.4pt}{0pt}{0pt}
                        & Road                & Road authority responsible for planning, control, or management                           of roads, excluding non-essential public entity activity \\
                        \specialrule{0.4pt}{0pt}{0pt}
                        &                     & Operator of intelligent transport systems applied to road                                 infrastructure, vehicles, users, traffic management, mobility                             management, or interfaces \\
                        \specialrule{0.4pt}{0pt}{0pt}
        \hline
    \end{tabular}}
    \caption{The Transport Sector}
    \label{tab:Transport_Sector}
\end{table}

\begin{table}[H]
    \centering
    \resizebox{\textwidth}{!}{
    \begin{tabular}{|p{0.5cm}|p{14.95cm}|}
        \hline
        \textbf{ID} & \textbf{Entity Type} \\ \hline
        2         & Core airports are listed in Annex II of Regulation (EU) No 1315/2013 \cite{A1TA3}\\ 
        \hline
    \end{tabular}}
    \caption{Entity type explanations for the Transport Sector}
    \label{tab:Transport_Sector_Explanations}
\end{table}

\paragraph{The Banking Sector}

The banking sector contains zero sub-sectors and one entity type as seen below.

\begin{table}[H]
    \centering
    \resizebox{\textwidth}{!}{
    \begin{tabular}{|p{2cm}|p{2cm}|p{11cm}|}
        \hline
        \textbf{Sector} & \textbf{Sub-sector} & \textbf{Entity Type} \\ \hline
        Banking          &                  & Credit Institution \\
        \hline
    \end{tabular}}
    \caption{The Banking Sector}
    \label{tab:Banking_Sector}
\end{table}

\paragraph{The Financial Market Infrastructure Sector}

The financial sector contains zero sub-sectors and two entity types as seen below. 

\begin{table}[H]
    \centering
    \resizebox{\textwidth}{!}{
    \begin{tabular}{|p{2cm}|p{2cm}|p{11cm}|}
        \hline
        \textbf{Sector} & \textbf{Sub-sector} & \textbf{Entity Type} \\ \hline
        Financial Market Infrastructures          &                  & Operator of MTF or OTF trading venue \\
        \specialrule{0.4pt}{0pt}{0pt}
                        &                  & Central Counter Party that acts as the buyer to every seller and                          seller to every buyer \\
        \hline
    \end{tabular}}
    \caption{The Financial Market Infrastructure Sector}
    \label{tab:Financial_Sector}
\end{table}

\paragraph{The Health Sector}

The health sector has zero sub-sectors and five entity types. However, some of the entity types in the health sector requires additional explanations to be interpretable. 

\begin{table}[H]
    \centering
    \resizebox{\textwidth}{!}{
    \begin{tabular}{|p{2cm}|p{2cm}|p{11cm}|}
        \hline
        \textbf{Sector} & \textbf{Sub-sector} & \textbf{Entity Type} \\ \hline
        Health          &                     & Healthcare provider \\
        \specialrule{0.4pt}{0pt}{0pt}
                        &                     & Laboratory [3] \\
                        \specialrule{0.4pt}{0pt}{0pt}
                        &                     & Research or development of medicinal products \\
                        \specialrule{0.4pt}{0pt}{0pt}
                        &                     & Manufacturer of basic pharmaceutical products or pharmaceutical                        preparations \\
                        \specialrule{0.4pt}{0pt}{0pt}
                        &                     & Manufacturing of medical devices considered to be critical during                       a public health emergency [4]\\
                        \specialrule{0.4pt}{0pt}{0pt}
        \hline
    \end{tabular}}
    \caption{The Health Sector}
    \label{tab:Health_Sector}
\end{table}

\begin{table}[H]
    \centering
    \resizebox{\textwidth}{!}{
    \begin{tabular}{|p{0.5cm}|p{14.95cm}|}
        \hline
        \textbf{ID} & \textbf{Entity Type} \\ \hline
        3         & An EU reference laboratory as stated in               Regulation (EU) 2022/2371 \cite{A1H2} \\
        \specialrule{0.4pt}{0pt}{0pt}
        4         & Medical Device Shortages Steering Group publish these on a                    dedicated webpage \cite{A1H5} \\
        \hline
    \end{tabular}}
    \caption{Entity type explanations for the Health Sector}
    \label{tab:Health_Sector_Explanations}
\end{table}

\paragraph{The Drinking Water Sector} 

The drinking water sector has one entity type. 

\begin{table}[H]
    \centering
    \resizebox{\textwidth}{!}{
    \begin{tabular}{|p{2cm}|p{2cm}|p{11cm}|}
        \hline
        \textbf{Sector} & \textbf{Sub-sector} & \textbf{Entity Type} \\ \hline
        Drinking Water Sector          &                  & Supplier or distributor of water intended for human consumption unless it is a non-essential part of general activity \\
        \hline
    \end{tabular}}
    \caption{The Drinking Water Sector}
    \label{tab:Drinking_Water_Sector}
\end{table}

\paragraph{The Waste Water Sector}

The waste water sector has one entity type.

\begin{table}[H]
    \centering
    \resizebox{\textwidth}{!}{
    \begin{tabular}{|p{2cm}|p{2cm}|p{11cm}|}
        \hline
        \textbf{Sector} & \textbf{Sub-sector} & \textbf{Entity Type} \\ \hline
        Waste Water Sector          &                  & Collection, disposal, or treatment of urban, domestic, or industrial waste unless it is a non-essential part of general activity \\
        \hline
    \end{tabular}}
    \caption{The Waste Water Sector}
    \label{tab:Waste_Water_Sector}
\end{table}

\paragraph{The Digital Infrastructure Sector}

The digital infrastructure sector has zero sub-sectors and nine entity types. Evidently, multiple entity types in this sector are governed by special rule sets as seen under the subsections about essential and important entity classification. 

\begin{table}[H]
    \centering
    \resizebox{\textwidth}{!}{
    \begin{tabular}{|p{2cm}|p{2cm}|p{11cm}|}
        \hline
        \textbf{Sector} & \textbf{Sub-sector} & \textbf{Entity Type} \\ \hline
        Digital Infrastructure          &                  & Internet Exchange Point Provider \\
        \specialrule{0.4pt}{0pt}{0pt}
                                &                  & DNS service provider or DNS registration service, root name servers excluded \\
                                \specialrule{0.4pt}{0pt}{0pt}
                                &                  & TLD name register \\
                                \specialrule{0.4pt}{0pt}{0pt}
                                &                  & Cloud computing service provider \\
                                \specialrule{0.4pt}{0pt}{0pt}
                                &                  & Data centre service provider \\
                                \specialrule{0.4pt}{0pt}{0pt}
                                &                  & Content delivery network provider \\
                                \specialrule{0.4pt}{0pt}{0pt}
                                &                  & Trust service provider \\
                                \specialrule{0.4pt}{0pt}{0pt}
                                &                  & Provider of public electronic communications networks \\
                                \specialrule{0.4pt}{0pt}{0pt}
                                &                  & Provider of publicly available electronic communications services \\
        \hline
    \end{tabular}}
    \caption{The Digital Infrastructure Sector}
    \label{tab:Digital_Infrastructure_Sector}
\end{table}

\paragraph{The Information and Communications Technology (ICT) Service Management Sector}

The ICT Service management sector has two entity types.

\begin{table}[H]
    \centering
    \resizebox{\textwidth}{!}{
    \begin{tabular}{|p{2cm}|p{2cm}|p{11cm}|}
        \hline
        \textbf{Sector} & \textbf{Sub-sector} & \textbf{Entity Type} \\ \hline
        ICT Service management (B2B)          &                  & Managed service provider \\
        \specialrule{0.4pt}{0pt}{0pt}
                                      &                  & Managed security service provider [5]\\
        \hline
    \end{tabular}}
    \caption{The ICT Service Management Sector}
    \label{tab:ICT_Service_Management_Sector}
\end{table}

\begin{table}[H]
    \centering
    \resizebox{\textwidth}{!}{
    \begin{tabular}{|p{0.5cm}|p{14.95cm}|}
        \hline
        \textbf{ID} & \textbf{Entity Type} \\ \hline
        5         &  Entities that provide these services as part of national or public security, defense or law might be eligible to exemption from certain obligations in the NIS2 directive \cite{NIS2}. \\
        \hline
    \end{tabular}}
    \caption{Entity type explanations for the ICT Service management Sector}
    \label{tab:ICT_Service_Management_Sector_Explanations}
\end{table}

\paragraph{The Public Administration Sector}

The public administration sector has three entity types and each of them has its own rule set as seen in the subsections about essential and important classification. 

\begin{table}[H]
    \centering
    \resizebox{\textwidth}{!}{
    \begin{tabular}{|p{2cm}|p{2cm}|p{11cm}|}
        \hline
        \textbf{Sector} & \textbf{Sub-sector} & \textbf{Entity Type} \\ \hline
        Public Administration         &                  & Central government public administration [6] \\
        \specialrule{0.4pt}{0pt}{0pt}
                               &                  & Regional level public administration [6] \\
                               \specialrule{0.4pt}{0pt}{0pt}
                               &                  & Local level public administration [6] \\
        \hline
    \end{tabular}}
    \caption{The Public Administration Sector}
    \label{tab:Public_Administration_Sector}
\end{table}

\begin{table}[H]
    \centering
    \resizebox{\textwidth}{!}{
    \begin{tabular}{|p{0.5cm}|p{14.95cm}|}
        \hline
        \textbf{ID} & \textbf{Entity Type} \\ \hline
        6           & Entities with primary activities in national or public security, defense or law enforcement              are excluded from NIS2 supervision                 \cite{NIS2} \\
        \hline
    \end{tabular}}
    \caption{Entity type explanations for the Public Administration Sector}
    \label{tab:Public_Administration_Sector_Explanations}
\end{table}

\paragraph{The Space Sector}

The space sector has one entity type. 

\begin{table}[H]
    \centering
    \resizebox{\textwidth}{!}{
    \begin{tabular}{|p{2cm}|p{2cm}|p{11cm}|}
        \hline
        \textbf{Sector} & \textbf{Sub-sector} & \textbf{Entity Type} \\ \hline
        Public Administration          &                  & Operator of ground-based infrastructure supporting provision of space-based service, excluding public electronic communications network providers \\
        \hline
    \end{tabular}}
    \caption{The Space Sector}
    \label{tab:Space_Sector}
\end{table}

\subsubsection{Annex II on Critical Sectors}

Annex II contains all the sectors, their respective sub-sectors and entity types that are critical \cite{NIS2}. If an entity is of one or multiple entity types in Annex II, its activities are critical for the societal and economic well-being of the member state in which it has its activities \\ \cite{NIS2}. Following are the sectors and their respective entity types in Annex II. 

\paragraph{The Postal and Courier Services Sector}

The postal and courier services sector has one entity type.

\begin{table}[H]
    \centering
    \resizebox{\textwidth}{!}{
    \begin{tabular}{|p{2cm}|p{2cm}|p{11cm}|}
        \hline
        \textbf{Sector} & \textbf{Sub-sector} & \textbf{Entity Type} \\ \hline
        Postal and courier services          &                  & Clearance, sorting, transport, or delivery of postal items \\
        \hline
    \end{tabular}}
    \caption{The Postal and Courier Services Sector}
    \label{tab:Postal_Courier_Services_Sector}
\end{table}

\paragraph{The Waste Management Sector}

The waste management sector contains one entity type. 

\begin{table}[H]
    \centering
    \resizebox{\textwidth}{!}{
    \begin{tabular}{|p{2cm}|p{2cm}|p{11cm}|}
        \hline
        \textbf{Sector} & \textbf{Sub-sector} & \textbf{Entity Type} \\ \hline
        Waste management          &                  & Collection, transport, recovery, or disposal of waste, or the supervision of either tasks \\
        \hline
    \end{tabular}}
    \caption{The Waste Management Sector}
    \label{tab:Waste_Management_Sector}
\end{table}

\paragraph{The Manufacture, Production, and Distribution of Chemicals Sector}

This sector has one entity type which requires further explanation upon registration. 

\begin{table}[H]
    \centering
    \resizebox{\textwidth}{!}{
    \begin{tabular}{|p{2cm}|p{2cm}|p{11cm}|}
        \hline
        \textbf{Sector} & \textbf{Sub-sector} & \textbf{Entity Type} \\ \hline
        The Manufacture, Production, and Distribution of Chemicals          &                  & Manufacturer or distributor of substances or articles [7] \\
        \hline
    \end{tabular}}
    \caption{The Manufacture, Production, and Distribution of Chemicals Sector}
    \label{tab:Chemicals_Sector}
\end{table}

\begin{table}[H]
    \centering
    \resizebox{\textwidth}{!}{
    \begin{tabular}{|p{0.5cm}|p{14.95cm}|}
        \hline
        \textbf{ID} & \textbf{Entity Type} \\ \hline
        7           & An article is a produced object with a shape that determines its function to a greater degree than its chemical composition \cite{A2MPDC1} \\
        \hline
    \end{tabular}}
    \caption{Entity type explanations for the Manufacture, Product, and Distribution of Chemicals Sector}'
    \label{tab:Chemicals_Sector_Explanations}
\end{table}

\paragraph{The Production, Processing, and Distribution of Food Sector}

This sector contains one entity type. 

\begin{table}[H]
    \centering
    \resizebox{\textwidth}{!}{
    \begin{tabular}{|p{2cm}|p{2cm}|p{11cm}|}
        \hline
        \textbf{Sector} & \textbf{Sub-sector} & \textbf{Entity Type} \\ \hline
        Production, processing and distribution of food          &                  & Any activity related to the production, processing, or distribution of food \\
        \hline
    \end{tabular}}
    \caption{The Production, Processing, and Distribution of Food Sector}
    \label{tab:Food_Sector}
\end{table}

\paragraph{The Manufacturing Sector}

The manufacturing sector is a large sector with six sub-sectors. Each sub-sector has one entity type. However, each entity type is further defined in other articles like NACE Rev.2 \cite{A2M2}. 

\begin{table}[H]
    \centering
    \resizebox{\textwidth}{!}{
    \begin{tabular}{|p{2cm}|p{2cm}|p{11cm}|}
        \hline
        \textbf{Sector} & \textbf{Sub-sector} & \textbf{Entity Type} \\ \hline
        Manufac- turing          & Manufac- ture of medical devices or in vitro diagnostic medical devices                & Manufacture of medical devices or accessory for medical devices [8] \\
        \specialrule{0.4pt}{0pt}{0pt}
                       & Manufact- ure of computer, electronic, or optical devices                 & Any economic activity in [9]   \\
                       \specialrule{0.4pt}{0pt}{0pt}
                       & Manufact- ure of electronic equipment                 & Any economic activity in [10] \\
                       \specialrule{0.4pt}{0pt}{0pt}
                       & Manufact- ure of machinery and equipment                 & Any economic activity in [11] \\
                       \specialrule{0.4pt}{0pt}{0pt}
                       & Manufact- ure of motor vehicles, trailers, and semi-trailers                 & Any economic activity in [12]\\
                       \specialrule{0.4pt}{0pt}{0pt}
                       & Manufact- ure of transport equipment                 & Any economic activity in [13] \\
        \hline
    \end{tabular}}
    \caption{The Manufacturing Sector}
    \label{tab:Manufacturing_Sector}
\end{table}

\begin{table}[H]
    \centering
    \resizebox{\textwidth}{!}{
    \begin{tabular}{|p{0.5cm}|p{14.95cm}|}
        \hline
        \textbf{ID} & \textbf{Entity Type} \\ \hline
        8           & Regulation (EU) 2017/745 Article 2 point 1,2 \cite{A2M1} provides the definitions of medical devices and accessories. \\
        \specialrule{0.4pt}{0pt}{0pt}
        9          & Computer, electronic, optical product. Electronic components, boards, loaded boards. Peripheral computer, and communication equipment. Consumer electronics, watches, and clocks. Measuring, testing, optical, and navigation instruments. Irradiation, electro-medical, electro-therapeutic, and photographic equipment. Optical media.   \\
        \specialrule{0.4pt}{0pt}{0pt}
        10          & Electrical equipment, electric motors, generators,  transformers, electricity distribution, and control apparatus. Batteries, accumulators, wiring, wiring devices, electronic lighting equipment, domestic appliances, electri domestic appliances, non-electric domestic appliances, and other electrical equipment. \\
        \specialrule{0.4pt}{0pt}{0pt}
        11          & Machine-yard equipment, general purpose machinery, fluid power equipment, pumps, compressors, taps, valves, bearing gears, gearning, driving elements, ovens, furnaces, furnace burners, lifting equipment, handling equipment, office machinery and equipment, power-driven hand tools, non-domestic cooling and ventilation equipment, agricultural and forestry machines, metal forming machine-yard tools, metal forming machinery, machine tools, special-purpose machinery, metallurgy machinery, mining machinery, quarrying and construction machinery. Food, beverage, and tobacco machinery. Textile, apparel, and leather machinery. Machinery for paper and paperboard production. Plastic and rubber machinery. Non aircraft, vehicle, cycle: engines, turbines. \\
        \specialrule{0.4pt}{0pt}{0pt}
        12          & Motor vehicles, trailer, and semi-trailers. Motor vehicle bodies, parts, accessories, electronic equipment, and other parts.\\
        \specialrule{0.4pt}{0pt}{0pt}
        13          & Transport equipment, ship and boat building, ship and floating structure, pleasure and sporting boats, railway locomotives and rolling stock. Air, spacecraft and related machinery. Military-fighting vehicles, transport equipment, motorcycles, bicycles and invalid carriages, other transport equipment. \\
        \hline
    \end{tabular}}
    \caption{Entity type explanations for the Manufacturing Sector}
    \label{tab:Manufacturing_Sector_Explanations}
\end{table}

\paragraph{The Digital Providers Sectors}

The digital providers sector is a small sector with three entities. However, it is a special sector as most of the entities are established in other countries \cite{NIS2}. In light of this, recital 114 defines how to determine the main establishment of a cross-border entity \\ \cite{NIS2}. There are three cases that must be considered. These are presented in a hierarchy of prioritization below. 

\begin{itemize}
    \item \textbf{Option 1:} The main establishment is determined to be in the member state where decisions related to cybersecurity risk-management measures are predominantly taken.
    \item \textbf{Option 2:} The main establishment is determined to be in the member state where cybersecurity operations are carried out. 
    \item \textbf{Option 3:} The main establishment is determined to be in the member state where the entity has the highest number of employees. 
\end{itemize}

Additionally, these entities can be have a representative in the member states in which they operate. This aspect is important for registration and member state reporting as seen in a later sector. Following are the entity types in the digital providers sector.

\begin{table}[H]
    \centering
    \resizebox{\textwidth}{!}{
    \begin{tabular}{|p{2cm}|p{2cm}|p{11cm}|}
        \hline
        \textbf{Sector} & \textbf{Sub-sector} & \textbf{Entity Type} \\ \hline
        Digital providers          &                  & Provider of online marketplace \\
        \specialrule{0.4pt}{0pt}{0pt}
                           &                  & Provider of online search engine \\
                           \specialrule{0.4pt}{0pt}{0pt}
                           &                  & Provider of social networking services platforms \\
        \hline
    \end{tabular}}
    \caption{The Digital Providers Section}
    \label{tab:Digital_Providers_Sector}
\end{table}

\paragraph{The Research Sector}

The research sector has one entity. However, the research and particularly educational institutions that are engaged in commercial research, are considered an integral part of value chains \cite{NIS2}. Recital 36 notes the importance of entities involved in commercial research \\ \cite{NIS2}. However, these has to be identified manually or by manual registration. Following is a tabular view of the research sector. 

\begin{table}[H]
    \centering
    \resizebox{\textwidth}{!}{
    \begin{tabular}{|p{2cm}|p{2cm}|p{11cm}|}
        \hline
        \textbf{Sector} & \textbf{Sub-sector} & \textbf{Entity Type} \\ \hline
        Research          &                  & Research organization \\
        \hline
    \end{tabular}}
    \caption{The Research Sector}
    \label{tab:Research_Sector}
\end{table}

% Explain this article and its relevance to this thesis

\subsection{Article 27: Registry of Entities}
% Explain this article and its relevance to this thesis
Article 27 of the NIS2 Directive governs the registry of all entities, and in particular the registration of entities in the digital providers, digital infrastructure and ICT service management sector \cite{NIS2}. However, entity registration can only be understood analyzing information from the recitals, article 2, 3, and 27. 

Furthermore, article 3 outlines procedural and notification requirements. By 17th of April 2025, Member States are required to compile and regularly update a comprehensive list of essential and important entities \cite{NIS2}. This list serves as the foundation for implementing sector-specific cybersecurity measures and must be reviewed at least biennially. Entities must submit critical information, including contact details, sectoral classification, and geographical scope of operations, to competent authorities, ensuring dynamic risk management \\ \cite{NIS2}.

\subsubsection{Registration of Entities}

The act of registering and the registry of entities should be a single system as stated in recital 106 \cite{NIS2}. This system must process personal data in accordance with the GDPR regulation \\ \cite{NIS2}. Additionally, the list of all entities should be established by the 17th of April 2025 and updated biannually. The list of entities that must be reported to ENISA should be established by the 17th of January 2025. To comply with general NIS2 registration requirement, registration must collect the information presented in the table below either manually, automatically, or a combination of both. 

\begin{table}[H]
    \centering
    \resizebox{\textwidth}{!}{
    \begin{tabular}{|p{8.8cm}|p{6.5cm}|}
        \hline
        \textbf{Registration Information Requirement} & \textbf{Change Notification Deadline} \\ 
        \hline
        Name of the entity & 2 weeks \\
        \specialrule{0.4pt}{0pt}{0pt}
        Address of the main establishment or address of the representative, if applicable & 3 months \\
        \specialrule{0.4pt}{0pt}{0pt}
        Addresses of the other establishments in the Union & 3 months \\
        \specialrule{0.4pt}{0pt}{0pt}
        Email addresses & 3 months \\
        \specialrule{0.4pt}{0pt}{0pt}
        Telephone numbers & 3 months \\
        \specialrule{0.4pt}{0pt}{0pt}
        IP ranges & 3 months \\
        \specialrule{0.4pt}{0pt}{0pt}
        Sector, sub-sector, and entity type & 3 months \\
        \specialrule{0.4pt}{0pt}{0pt}
        Other member states they provide services & 3 months \\
        \hline
    \end{tabular}}
    \caption{Registry Information Requirements}
    \label{tab:Registry_Information_Requirements}
\end{table}

However, initial classification only requires the following information to be collected.

\begin{table}[H]
    \centering
    \resizebox{\textwidth}{!}{
    \begin{tabular}{|p{15.7cm}|}
        \hline
        \textbf{Classification Information Requirements} \\ 
        \hline
        The primary sector \\
        \specialrule{0.4pt}{0pt}{0pt}
        The sub-sector, if it exists \\
        \specialrule{0.4pt}{0pt}{0pt}
        The entity type \\
        \specialrule{0.4pt}{0pt}{0pt}
        Number of employees in the Union \\
        \specialrule{0.4pt}{0pt}{0pt}
        Latest annual turnover in the Union \\
        \specialrule{0.4pt}{0pt}{0pt}
        Latest annual balance sheet in the Union \\
        \hline
    \end{tabular}}
    \caption{Classification Information Requirements}
    \label{tab:Classification_Information_Requirements}
\end{table}

In addition to the information requirements presented above, entities that have been assessed to be in any of the heightened risk categories identified in the section on Essential Entity Classification, must submit additional information through the registration portal \cite{NIS2}. The additional information required from these entities are presented below. 

\begin{table}[H]
    \centering
    \resizebox{\textwidth}{!}{
    \begin{tabular}{|p{15.7cm}|}
        \hline
        \textbf{Risk Information Requirements} \\ 
        \hline
        Statement on heightened risk (yes/no) \\ 
        \specialrule{0.4pt}{0pt}{0pt}
        Statement on public safety, security, and health importance of entity \\
        \specialrule{0.4pt}{0pt}{0pt}
        Statement on sole provider of a service \\
        \specialrule{0.4pt}{0pt}{0pt}
        Statement on the potential of systemic risk, particularly cross-border impact \\
        \specialrule{0.4pt}{0pt}{0pt}
        Statement on sectorial importance \\
        \specialrule{0.4pt}{0pt}{0pt}
        Types of services provided \\
        \specialrule{0.4pt}{0pt}{0pt}
        The provision of services provided \\
        \hline
    \end{tabular}}
    \caption{Risk Information Requirements}
    \label{tab:Risk_Information_Requirements}
\end{table}

Furthermore, the Norwegian ecosystem requires us to collect some additional information to determine the validity of an entity, cross-reference it across the registry and to automatically register it. To achieve this we use the Norwegian national identifier for legal entities, organization number.

\subsubsection{Member State Notification Requirements}

\paragraph{Notifying ENISA}

The ENISA is obligated to maintain a registry that contains information about the entity types below \cite{NIS2}.

\begin{table}[H]
    \centering
    \resizebox{\textwidth}{!}{
    \begin{tabular}{|p{15.7cm}|}
        \hline
        \textbf{Enisa Entity Type Notification} \\ 
        \hline
        DNS Service Providers \\ 
        \specialrule{0.4pt}{0pt}{0pt}
        TLD Name Reigsters \\
        \specialrule{0.4pt}{0pt}{0pt}
        Domain Name Registration Services \\
        \specialrule{0.4pt}{0pt}{0pt}
        Cloud Computing Service Providers \\
        \specialrule{0.4pt}{0pt}{0pt}
        Data Center Service Providers \\
        \specialrule{0.4pt}{0pt}{0pt}
        Content Delivery Network Providers \\
        \specialrule{0.4pt}{0pt}{0pt}
        Managed Service Providers \\
        \specialrule{0.4pt}{0pt}{0pt}
        Managed Security Services Providers \\
        \specialrule{0.4pt}{0pt}{0pt}
        Providers of Online Market Places \\
        \specialrule{0.4pt}{0pt}{0pt}
        Providers of Online Search Engines \\
        \specialrule{0.4pt}{0pt}{0pt}
        Providers of Social Networking Services Platforms \\
        \hline
    \end{tabular}}
    \caption{Enisa Entity Type Notification}
    \label{tab:Enisa_Entity_Type_Notification}
\end{table}

The rationale behind this is stated in recital 96 and 117 of the NIS2 Directive \cite{NIS2}. ENISA intends to have an overview of those entities to account for their role in network communications and the increased attack surface some of the entity types provide \cite{NIS2}. Member states are obligated to immediately forward all the information stated in the table 'Registry Information Requirements', except IP ranges, to ENISA at the point of registration. 

\paragraph{Notifying the European Commission and the Cooperation Group}

Certain information must be aggregated to produce a biannual statistics notification for the European Commission and the Cooperation Group. The notification must include the total number of essential and important entities \cite{NIS2}. Additionally, the notification must contain statistics on the total number of essential  and important entities aggregated by sector, sub-sector, and entity type \\ \cite{NIS2}. This notification must execute on the 17th of April 2025 and biannually thereafter.

\paragraph{Notifying the European Commission}

The European Commission requires member states to produce an additional biannual report about entities with increased risk. The report must include the information below \cite{NIS2}. 
\begin{itemize}
    \item The total number of essential and important entities in increased risk categories
    \item The sector and sub-sector in which they belong
    \item The type of service each entity provides and the provision of those services
    \item Aggregated presentation of the total number of essential entities per risk category
\end{itemize}

This notification must execute on the 17th of April 2025 and biannually thereafter.
\\
\\
The concept of registration and notification are visualized below. Note that it solely based on requirements extracted from the NIS2 Directive \\ \cite{NIS2}.

\begin{figure}[H]
    \centering
    \includegraphics[width=1.0\linewidth]{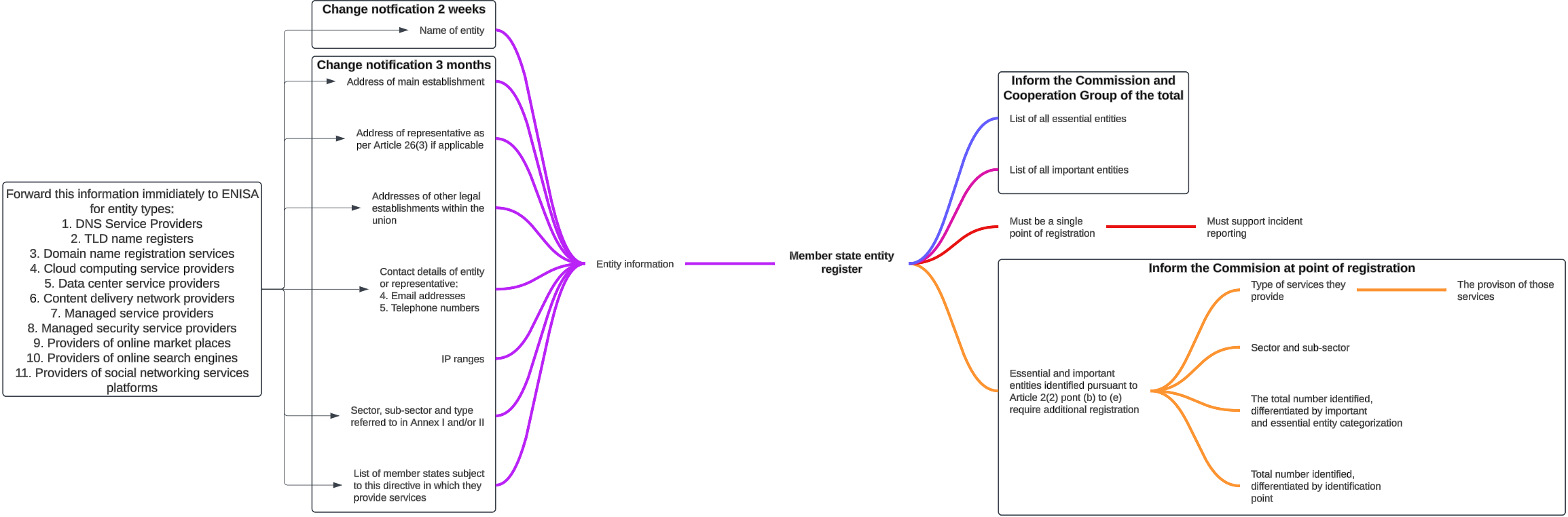}
    \caption{The Concept of NIS2 Registration and Notification}
    \label{fig:Registration_Concept}
\end{figure}

The current outline of registration and notification accounts for entity registration requirements, change notification timelines, the nature of a single registration portal, and the notification requirements imposed on member states. 

\section{The Supervision of Entities}

Supervision encompasses the governance activities a competent authority exerts on an entity within the scope of the NIS2 Directive. This concept is relevant to the registry’s implementation, as the registry must be designed to facilitate and support supervisory functions. To achieve this, the registry must provide adequate information to both entities and agents to ensure correct classification and prioritization of entities. Correct classification and prioritization of entities ensure appropriate and proportionate supervision of entities. To achieve this the below elements must be part of the registry. 
\begin{itemize}
    \item \textbf{Entity Information:} Information about entities must support informed supervision. For this purpose, we recommend a comprehensive labeling system that can provide adequate and actionable information about an entity.
    \item \textbf{Entity Information Management:} The registry must have features to manage entity information. Entities must be able to request changes to their labels and agents must be able to request additional information from entities to ensure that entities are accurately represented. Additionally, agents from the competent authority of the member state concerned must be able to manually change information when it is needed.
    \item \textbf{Information Sharing:} Features that enable sharing specific and aggregated information about entities must be part of the registry to ensure timely and effective supervision of entities. Such features must meet national needs and be compliant with Union guidelines on information sharing. 
\end{itemize}

The concept of supervision will not be fully explored and is limited to the scope of this thesis. The rationale for exploring this concept is to ensure that the registry is implemented to suppert the supervision of entities. 

\section{The Cross Directive Nature of NIS2}

The NIS2 Directive is dependent on numerous directives that support its implementation \cite{NIS2}. This concept is limited to exploring its direct implications for the governance of entities, particularly, classification and supervision. Sector-specific Union legal acts and the directives CER and NIS1 influence the classification class given to entities and how they are supervised \\ \cite{NIS2}. Entities previously classified as critical by the CER Directive or essential by the NIS1 Directive are classified as essential pursuant to the NIS2 Directive \cite{NIS2}. Additionally, sector-specific Union legal acts that overlap with the supervision and reporting requirements imposed by the NIS2 Directive must be accounted for when performing supervisory tasks \cite{NIS2}. The rationale behind this is to reduce the administrative burden imposed on entities and to avoid duplicate requirements \cite{NIS2}. For example, the aviation and finance sector has sector-specific Union legal acts with supervisory requirements that overlap with the NIS2 directive \cite{NIS2}. To account for this, a comprehensive labeling system was introduced in this chapter. Additionally, overlapping directives need features for real-time information sharing to fully perform their supervisory tasks, efficiently. 

Furthermore, the NIS2 Directive integrates seamlessly with the CER Directive \\ \cite{CER}, which designates certain entities as critical. These entities automatically fall under the essential category, avoiding regulatory duplication and ensuring coherence across the European Union’s cybersecurity landscape. Furthermore, overlaps with sectoral regulations such as the Digital Operational Resilience Act (DORA) highlight the need for careful cross-regulatory alignment to prevent conflicting compliance obligations. 

\section{The Registry, GDPR, and Security}

In consideration of the scope of this thesis, the handling of personally identifiable and other sensitive information must comply with the GDPR directive. Additionally, the registry must use state-of-the-art cybersecurity controls to ensure the confidentiality, integrity, availability, and authenticity of the system. 

\section{Stakeholders}
% Why stakeholders? Explain.
Stakeholders are parties that have an interest in, are influenced by, or will use the registry, or are otherwise affected by its existence. We are not going into depth about different stakeholders, but knowing who they are and their role is crucial for the development of a NIS2 pan-European registry. A simple tabular presentation of identified stakeholders and a description of their roles is presented below. 

\begin{table}[H]
    \centering
    \resizebox{\textwidth}{!}{
    \renewcommand{\arraystretch}{1.4}
    \begin{tabular}{|l|p{10cm}|}
        \hline
        \textbf{Stakeholder Type} & \textbf{Role in the Registry} \\ 
        \hline
        Norwegian entities & All entities in Norway will be evaluated for inclusion and can be subject to supervision. They must have access to the portal, have the ability to make an informed registration, and be able to register themselves. \\ 
        \specialrule{0.4pt}{0pt}{0pt}
        Users & Users are entities that have been classified and are subject to supervision. Users must have access to the user portal to review their entity and request changes. \\ 
        \specialrule{0.4pt}{0pt}{0pt}
        The competent authority & The Norwegian competent authority is Nasjonal Sikkerhetsmyndighet (NSM). Authorized employees in NSM must have access to the agent portal to review, assess, and change entity information. Additionally, NSM requires integrated notification capabilities. \\ 
        \specialrule{0.4pt}{0pt}{0pt}
        Norwegian CSIRTs & Norwegian National Cyber Security Centre (NCSC) is the main CSIRT in Norway and it is a part of NSM. They are not directly relevant to the functioning of the registry. However, NCSC is an integral part of entity incident reporting and might need access to the registry.  \\ 
        \specialrule{0.4pt}{0pt}{0pt}
        Union CSIRTs & An integral part of the NIS2 Directive requires CSIRTs across the EU to cooperate on cross-border incidents. The list of designated CSIRTs in the European Union can be found at https://tools.enisa.europa.eu/topics/incident-response/csirt-inventory/certs-by-country-interactive-map. While their direct connection to the registry is limited, they may require access to information in the registry via NCSC, which serves as the national coordination point for cybersecurity in Norway. \\ 
        \specialrule{0.4pt}{0pt}{0pt}
        ENISA & As stated on their webpage, ENISA is the Union's agency dedicated to achieving a high common level of cybersecurity across Europe. The registry must have the capabilities to meet the notification obligations imposed by ENISA. \\ 
        \specialrule{0.4pt}{0pt}{0pt}
        The Commission & The European Commission is the executive body of the European Union. The registry must have the capability to meet the notification obligations imposed by the European Commission. \\ 
        \specialrule{0.4pt}{0pt}{0pt}
        The Cooperation Group & The NIS Cooperation Group exists to facilitate cross-border cooperation and information exchange. The registry must have the capability to meet the notification obligations imposed by the NIS Cooperation Group. \\ 
        \hline
    \end{tabular}}
    \caption{Stakeholder Mapping for the Registry}
    \label{tab:Stakeholder_Map}
\end{table}

\section{Proposed System}

The literature review and the NIS2 Directive provides a comprehensive framework for designing a NIS2-compliant pan-European registry of essential and important entities. A high-level workflow of the system is illustrated below. The illustration covers the system's workflow and its essential components.  

\begin{figure}[H]
    \centering
    \includegraphics[width=1.0\linewidth]{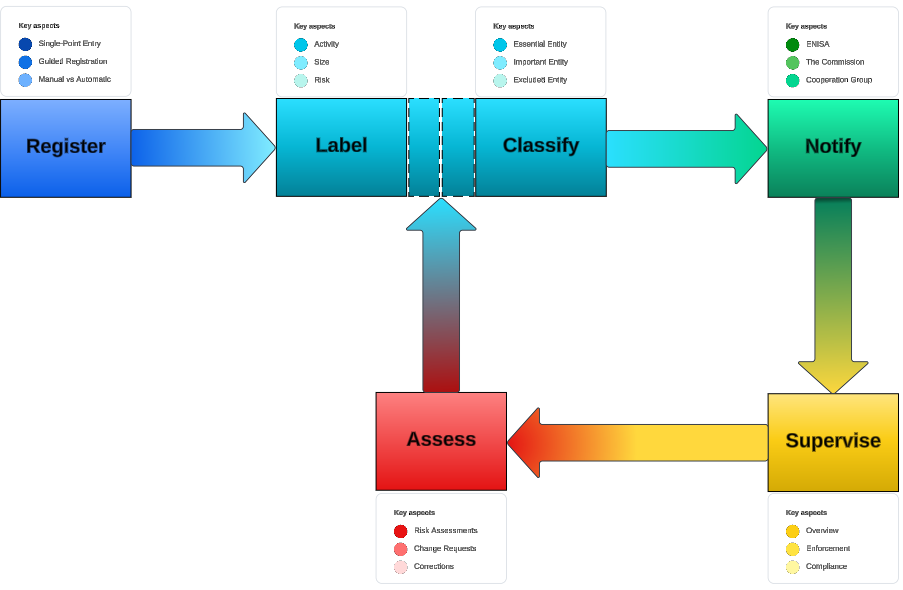}
    \caption{High-level Registry Workflow}
    \label{fig:Registry_Workflow}
\end{figure}

The starting node in the registry is automatic or manual registration, or a combination of both. Information gathered through registration will be used to label and classify the registered entity. After initial classification, entity information in table \ref{tab:Registration_Requirements}, except IP ranges, will be aggregated and forwarded to ENISA for every entity of entity type in table \ref{tab:Enisa_Entity_Type_Notification}, or after a satisfactory number of entities has been processed. After classifying all entities or a satisfactory number of entities, a statistics report will be produced and forwarded to the European Commission and the Cooperation Group. Additionally, a risk report will be produced and forwarded to the European Commission for all entities that are attributed to a higher risk. 

Furthermore, any entity classified as essential or important will be subject to supervision by that member state's competent authority. For the purpose of establishing this registry, our implementation of supervision is scoped to develop the functionality that enables supervisory tasks. Supervision and assessment are joined for that purpose and the requirements for achieving this is stated in table \ref{tab:Supervision_Requirements}. Should an entity's assessment conclude with changes that affect its labeling or classification status, it must be re-labeled and re-classified. To that end, the workflow restarts at either node. 

\subsection{NIS2 System Requirements}

In this section we will elaborate on NIS2 system requirements. Requirements are categorized per concept and provides a concrete overview of the requirements imposed on the registry. 
\\
\\
The requirements for establishing a registry pursuant to what has been introduced so far are presented in five tables below. 

\begin{table}[H]
    \centering
    \resizebox{\textwidth}{!}{
    \rotatebox{90}{
    \begin{tabular}{|p{1cm}|p{4cm}|p{7.7cm}|p{2.5cm}|p{5.5cm}|}
    \hline
    \textbf{ID} & \textbf{Name} & \textbf{Description} & \textbf{NIS2 Reference} & \textbf{NIS2 Rationale} \\
    \hline
        R-1 & Single-point Registration & Registration must happen in a single system. & Recital 106 & A single-entry point is a technical means to reduce the administrative burden imposed on entities. \\
        \specialrule{0.4pt}{0pt}{0pt}
        R-2 & Manual Registration & Registration must support manual registration. & Recital 106 & Entities should have a technical means to register themselves. \\  
        \specialrule{0.4pt}{0pt}{0pt}
        R-3 & Automatic Registration & Registration must support automatic registration. & Recital 106 & Automated registration is a technical means to reduce the administrative burden imposed on entities. \\
        \specialrule{0.4pt}{0pt}{0pt}
        R-4 & Guidance & The registration process must be guided and supplemented with adequate information so that the act or registration is informed. The information provided as explanations in the chapters about Annex I and II provides a starting point. & NIS2 & This is to ensure data accuracy. \\
        \specialrule{0.4pt}{0pt}{0pt}
        R-5 & Data Collection & Data collection must satisfy table  \ref{tab:Registry_Information_Requirements}, table \ref{tab:Classification_Information_Requirements}, and entities must provide their organizational number. & Recital 18, and Article 2, 3, and 27 & The complete registration data required about entities are provided in these articles. \\ 
        \specialrule{0.4pt}{0pt}{0pt}
        R-6 & Input Restriction & An Input field must be unique and restricted to accepting a single data point referenced in registration requirement N-5. This is true for all input fields. & NIS2 & This is to ensure data accuracy. \\ 
        \specialrule{0.4pt}{0pt}{0pt}
        R-7 & Risk Data & Registration must allow registrants to provide the statement on risk in table. \ref{tab:Risk_Information_Requirements} & Recital 15, 124, and Article 2 & Risk assessments can change classification outcomes and help prioritize supervisory tasks. \\
        \specialrule{0.4pt}{0pt}{0pt}
        R-8 & GDPR & Data processing must comply with the GDPR Directive. & Recital 14, 121, and Article 2 & The processing of sensitive information is governed by the GDPR Directive. \\
        \specialrule{0.4pt}{0pt}{0pt}
        R-9 & Security Controls & The manual registration mechanism is an entry point and must be protected by adequate security controls. & Recital 117 and 121 & Data, data access and the accuracy and functioning of the registry must be protected. \\
    \hline
    \end{tabular}}}
    \caption{Registration Requirements}
    \label{tab:Registration_Requirements}
\end{table}

\begin{table}[H]
    \centering
    \resizebox{\textwidth}{!}{
    \rotatebox{90}{
    \begin{tabular}{|p{1cm}|p{4cm}|p{7.7cm}|p{2.5cm}|p{5.5cm}|}
    \hline
    \textbf{ID} & \textbf{Name} & \textbf{Description} & \textbf{NIS2 Reference} & \textbf{NIS2 Rationale} \\
    \hline
        C-1 & Classes & Entities must be classified and distinguished based on the classes in figure \ref{fig:Classification_Classes}. & Recital 15 and Article 2, and 3 & Classification should inform if they are within the scope of the NIS2 Directive and their criticality for societal and economic functioning. \\
        \specialrule{0.4pt}{0pt}{0pt}
        C-2 & Factors & Classification must be based on the factors in figure \ref{fig:Classification_Factors}. & Recital 7, Article 2 and 3 & Classification factors should be uniform. \\ 
        \specialrule{0.4pt}{0pt}{0pt}
        C-3 & Rule set & Entities must be classified based on the rule sets in figure \ref{fig:Essential_Classification}, \ref{fig:Important_Classification}, and \ref{fig:Excluded_Classification}. & Recital 7, and Article 2 and 3 & Classification criteria should be uniform. \\
        \specialrule{0.4pt}{0pt}{0pt}
        C-4 & Labeling & The labels referred to in table \ref{tab:Entity_Labels} and additional information gathered upon registration must be used as labels to contextualize classification. & Recital 15, 30, 122, 123, 124, 125, 135, and Article 2 & Entity information beyond classification is required to prioritize supervision and determine supervisory tasks. \\
        \specialrule{0.4pt}{0pt}{0pt}
        C-5 & Partitioning & Classification must be mutually exclusive and exhaustive. & Recital 15, and Article 2 or 3 & To distinguish between entities criticality for societal and economic functioning. \\
        \specialrule{0.4pt}{0pt}{0pt}
        C-6 & Automatic Classification & The classification rule-sets referenced in requirement C-3 must be automated to the extent it is possible and does not adversely impact nuanced classifications. & Recital 14, and 106 & Automation should be implemented to reduce the administrative burden on entities and competent authorities. \\
        \specialrule{0.4pt}{0pt}{0pt}
        C-7 & Intervention & Manual intervention must be possible and automatically assigned in cases of classification uncertainty to avoid classification failure. & Article 2, 3, and 27 & Certain aspects of these articles, particularly risk assessments and legal matters cannot be automated.
        \\
    \hline
    \end{tabular}}}
    \caption{Classification Requirements}
    \label{tab:Classification_Requirements}
\end{table}

\begin{table}[H]
    \centering
    \resizebox{\textwidth}{!}{
    \rotatebox{90}{
    \begin{tabular}{|p{1cm}|p{4cm}|p{10cm}|p{2.5cm}|p{5.5cm}|}
    \hline
    \textbf{ID} & \textbf{Name} & \textbf{Description} & \textbf{NIS2 Reference} & \textbf{NIS2 Rationale} \\
    \hline
        N-1 & General Statistics & The total number of essential and important entities must be aggregated and forwarded to The European Commission and Cooperation Group once classification is completed and every second year thereafter. & Recital 18 and Article 3(5), and 14 & To enable Union-wide cybersecurity governance and coordination, and to produce biannual reports. \\
        \specialrule{0.4pt}{0pt}{0pt}
        N-2 & ENISA & All information except IP Ranges in table \ref{tab:Registry_Information_Requirements} must be forwarded immediately and automatically to ENISA upon registration of an entity of entity type in table \ref{tab:Enisa_Entity_Type_Notification}. & Recital 117, and Article 27 & ENISA is required to maintain registry of key service providers and entities at Union level. \\ 
        \specialrule{0.4pt}{0pt}{0pt}
        N-3 & ENISA Update & In the case any information about an entity in table \ref{tab:Registry_Information_Requirements} is updated, the registry must automatically forward an updated notification of type in requirement N-2 to ENISA. & Recital 117, and Article 27 & ENISA's registry must be actionable and therefore up-to-date. \\
        \specialrule{0.4pt}{0pt}{0pt}
        N-4 & The Commission & The total number of essential entities with a risk label, their sector, sub-sector, the type of service they provide, and the provision of those services must be forwarded immediately and automatically to The European Commission once they have been identified and every second year thereafter. Additionally, the total number of essential entities per risk category must be aggregated and forwarded as well. & Recital 18, Article 3(5), 3(6) & To enable Union-wide cybersecurity governance and coordination, and to produce biannual reports. \\ 
        \specialrule{0.4pt}{0pt}{0pt}
        N-5 & Reports & The notifications discussed in row N-1, N-2, N-3, and N-4 must be presentable in standardized report templates. & Recital 106, Article 3(5), 14, and 27 & Reports should be standardized to simplify and streamline the reporting process and to reduce administrative burden. \\ 
        \specialrule{0.4pt}{0pt}{0pt}
        N-6 & Manual Notification & It must be possible to produce reports manually that can contain all or specific information about an entity or set of entities. & Recital 106, Article 3(5), 14, and 27 & Reporting flexibility is required to meet ad hoc reporting needs.  \\
        \specialrule{0.4pt}{0pt}{0pt}
        N-7 & Configuration & It must be possible to configure notification dates, recipients, and the information provided in reports. & Recital 106, Article 3(5), 14, and 27 & Reporting flexibility is required to meet ad hoc reporting needs. \\
    \hline
    \end{tabular}}}
    \caption{Notification Requirements}
    \label{tab:Notification_Requirements}
\end{table}

\begin{table}[H]
    \centering
    \resizebox{\textwidth}{!}{
    \rotatebox{90}{
    \begin{tabular}{|p{1cm}|p{4cm}|p{10cm}|p{3.5cm}|p{8cm}|}
    \hline
    \textbf{ID} & \textbf{Name} & \textbf{Description} & \textbf{NIS2 Reference} & \textbf{NIS2 Rationale} \\
    \hline
        S-1 & Request Information & Agents must be able to request additional information from entity registrants. & Recital 106, and Article 27, 32, and 33 & Single entry point to reduce administrative burden and to be able to request information if required. \\
        \specialrule{0.4pt}{0pt}{0pt}
        S-2 & Request Change & Entity registrants must be able to request changes to their classification and labels. & Recital 106, and Article 27(3) and 3(3) & Single entry point to reduce administrative burden and to maintain accurate entity information. \\
        \specialrule{0.4pt}{0pt}{0pt}
        S-3 & Change Notification & Registrants must have adequate functionality to update entity information within the time frames stated in table \ref{tab:Registry_Information_Requirements}. & Recital 18, and Article 27(3) and 3(4) & Keep information up-to-date and accurate and to enable entities to notify of any change to their information. \\
        \specialrule{0.4pt}{0pt}{0pt}
        S-4 & Editable & Agents must be able to edit information in the registry. & Recital 125, and Article 27 & The competent authorities are responsible for ensuring accurate and complete registry information. \\
        \specialrule{0.4pt}{0pt}{0pt}
        S-5 & Sectoral Risk Assessments & Agents must have adequate functionality to apply risk assessments to sectors. & Recital 124, 125, and Article 31(2) & Supervision should follow a risk-based approach and it should be possible to apply a statement on risk to sectors. \\
        \specialrule{0.4pt}{0pt}{0pt}
        S-6 & Non-repudiation & Changes to entity information must be recorded to ensure non-repudiation. & Recital 117 and 125 & The competent authorities are responsible for ensuring accurate and complete registry information. \\ 
        \specialrule{0.4pt}{0pt}{0pt}
        S-7 & Assignment & The registry and privileged agents must be able to assign agents to entities or tasks. & Recital 14, 117 and 121 & The registry must comply with the GDPR Directive and competent authorities are responsible for access control. \\
        \specialrule{0.4pt}{0pt}{0pt}
        S-8 & Information Sharing & Agents must be able to produce aggregated reports about entities for sharing. & Recital 18, 117, Article 3(5), 3(6), 14, 27, and Article 29 & To enable notifications and ad hoc reporting. \\
        \specialrule{0.4pt}{0pt}{0pt}
        S-9 & Uploading & Agents must have the functionality to support an entity's context through uploading documents. & Recital 124, 125, Article 27, 31, 32, and 33. & Additional information about entities from audits, reports, risk assessments, and other relevant documents can increase entity understanding and thereby enhance supervision, particularly the prioritization of supervisory tasks. \\ 
        \specialrule{0.4pt}{0pt}{0pt}
        S-10 & Presentation & Classifications, labels, and additional information about an entity must be available to agents and to a lesser degree registrants to provide contextual information. & Recital 15, 30, 122, 123, 124, 125, 135, and Article 2 and 3. & Entity information is required to prioritize supervision and determine supervisory tasks. \\
        \specialrule{0.4pt}{0pt}{0pt}
        S-11 & Authentication & Registry access must be authenticated. & Recital 117 and 121 & Access to the registry must be restricted to authorized users. \\
        \specialrule{0.4pt}{0pt}{0pt}
        S-12 & Authorization & The degree of registry access must be enforced through authorization. & Recital 117 and 121 & Access to registry information and functionality ensured to those who are authorized. \\
    \hline
    \end{tabular}}}
    \caption{Supervision Requirements}
    \label{tab:Supervision_Requirements}
\end{table}

\subsection{Literature Review Requirements}

Multiple insights gathered from the literature review are in agreement with requirements synthesized from the NIS2 Directive. This section will not regurgitate those requirements, but will provide a tabular overview of those that are not overlapping. 

\begin{table}[H]
    \centering
    \resizebox{\textwidth}{!}{
    \rotatebox{90}{
    \begin{tabular}{|p{1cm}|p{4cm}|p{7.7cm}|p{2.5cm}|p{4.5cm}|}
    \hline
    \textbf{ID} & \textbf{Name} & \textbf{Description} & \textbf{Literature Review Reference} & \textbf{Literature Review Rationale} \\
    \hline
        LR-1 & Artificial Intelligence & The registry must be powered by artificial intelligence to enable automatic registration. & Bianchi et al. (2022) & NCP's integration with AI enhanced anomaly detection has demonstrated the benefits of integrating AI into the registry. \\
        \specialrule{0.4pt}{0pt}{0pt}
        LR-2 & Dynamic Classifications & Classifications, labels, and other information must provide adequate information to enable dynamic and nuanced classifications. & Chen et al. (2023) & Underscores the limitation of rule-based classification, and advocates for dynamic classification to address the dynamic nature of cyber threats. \\
        \specialrule{0.4pt}{0pt}{0pt}
        LR-3 & Interoperability & Standardized data formats are critical for interoperability & Oltramari et al., 2021 & Highlight the importance of standardized data formats to enable interoperability between automation and expert judgment. \\
        \specialrule{0.4pt}{0pt}{0pt}
        LR-4 & Scalability & The registry must be designed to handle an increasing number of entities while maintaining efficiency and compliance. & Van Dijk et al. (2021) & Outlines scalability as key principle for compliance registries. \\
        \specialrule{0.4pt}{0pt}{0pt}
        LR-5 & Labeling System & The labeling system must be developed to support predictive analysis, particularly AI-powered assessments. & Kshetri (2023), NCP, Ferreura et al., 2023, and Chen et al. (2023) & Predictive analysis can enable dynamic risk assessments, thereby enhancing the registry responsiveness to on-going threats. \\
    \hline
    \end{tabular}}}
    \caption{Literature Review Requirements}
    \label{tab:Literature_Review_Requirements}
\end{table}

Five requirements can be extracted from the literature review that provides more insights into the design of a NIS2-compliant pan-European registry of essential and important entities beyond those established through an analysis of the NIS2 Directive. 
\\
\\
The list of both NIS2 and literature review requirements provide a solid overview of the requirements of a NIS2-compliant pan-European registry of essential and important entities. Additionally, the requirements strike a sufficient balance between breadth and specificity to guide the development of the registry. 

\subsection{System Design}

The purpose of the system is to register and classify entities to enable supervision by competent authorities. A key aspect of this purpose is to align it with the requirements outlined so far in this thesis. To support development, a set of design principles for the registry are listed below.

\begin{itemize}
    \item \textbf{Automation: } Any part of the system that can be automated, must be automated, particularly, classification and registration. Additionally, automation must not adversely impact registration and classification accuracy. 
    \item \textbf{Intervention: } Cases of failure and uncertainty must be logged and redirected for manual assessment. This is particularly important for nuanced classification and cases that can not easily be determined by a rule-set. 
    \item \textbf{Uniformity: } Simple classification cases must be uniformly classified based on the same factors and rule-sets. This ensures that no other variables are used for classification and that classification decisions are based on the same rule-set. This excludes manual classifications, as these require dynamic assessments. 
    \item \textbf{Contextualization: } The factors, variables, and other data used for classification must be translated into labels to provide effective and contextualized supervision. 
    \item \textbf{AI-Enhanced Functionality: } Advances in artificial intelligence provides an opportunity to use AI or design the registry to support AI integrations.  
    \item \textbf{Security: } The registry must be developed to ensure the confidentiality, integrity, availability, and authenticity of the system through secure development life cycles. 
\end{itemize}

The system serves as a national registry for essential and important entities, supporting compliance with the NIS2 directive. It facilitates automated and manual entity registration and classification. Additionally, risk assessments and supervision by competent authorities are supported through manual assessments. To enable effective supervision, entities are classified and labeled as specified earlier. Furthermore, the system is designed with automation, uniformity, and security in mind, ensuring efficient and accurate entity classification while allowing for human intervention in complex cases. 

\subsection{Architecture}

\begin{figure}[H]
    \centering
    \begin{minipage}[b]{0.48\linewidth}
        \centering
        \includegraphics[height=7cm]{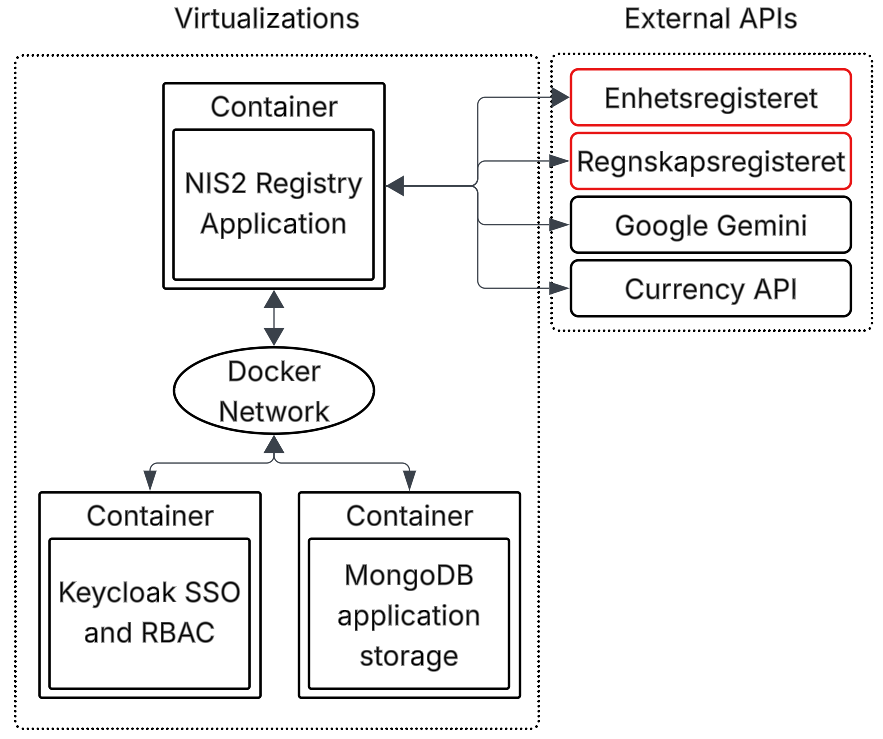}
        \caption{High-level Application Architecture}
        \label{fig:Application_Architecture}
    \end{minipage}
    \hfill
    \begin{minipage}[b]{0.48\linewidth}
        \centering
        \includegraphics[height=7cm]{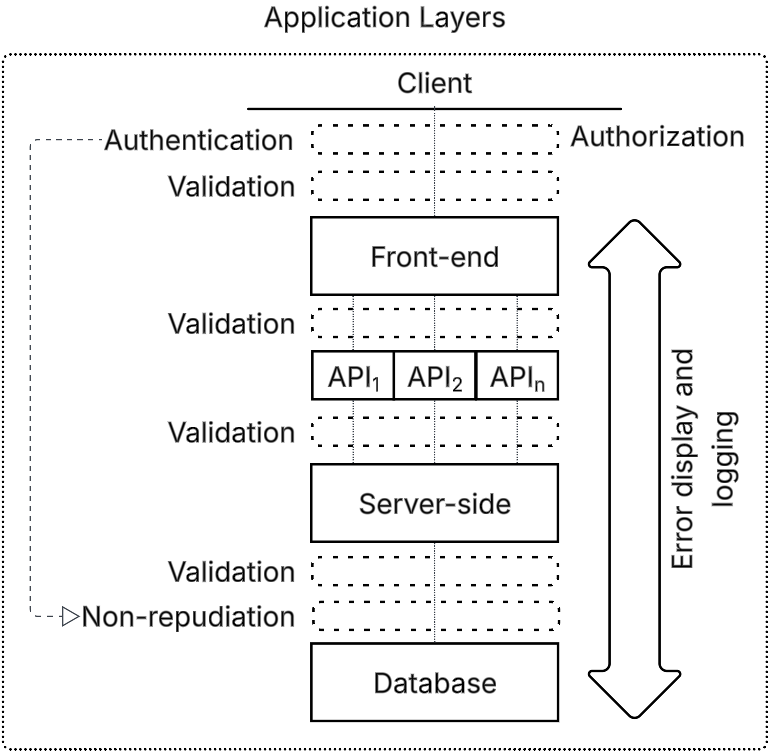}
        \caption{Layered Application Architecture}
        \label{fig:Application_Layers}
    \end{minipage}
\end{figure}

The application runs in a Dockerized environment, consisting of three isolated containers that are interconnected through a shared Docker network. Figure \ref{fig:Application_Architecture} provides a high-level overview of this container-based architecture, while Figure \ref{fig:Application_Layers} illustrates the logical layering and data flow within the application.

In addition to internal components, the application integrates with several external APIs, including public registries (e.g., Enhetsregisteret and Regnskapsregisteret), a language model (Google Gemini), and a currency exchange service. These integrations are handled exclusively by the NIS2 Registry Application container, ensuring that external communication remains encapsulated and that access control can be centrally enforced. Public registries are highlighted in red to indicate their specificity to the Norwegian ecosystem. For deployments in other member states, these are the only components that require modification to achieve full interoperability.

The public registries are used to retrieve information about organizations, which supports automated registration, labeling, and classification of entities. However, the sector, sub-sector, and entity type classifications retrieved from Enhetsregisteret are not directly compatible with the NIS2 specification \cite{NIS2}. To address this, prompt engineering is applied using the Gemini 2.0 Flash language model to transform the data into NIS2-compliant descriptions of sectors, sub-sectors, and entity types \cite{NIS2}. Furthermore, accountancy data obtained from Regnskapsregisteret may be reported in varying currencies, primarily USD and NOK. To ensure accurate size estimation in compliance with NIS2's euro-based thresholds, the CurrencyAPI is used to convert reported values into their euro equivalents.

The architecture also enforces layered validation and access control, extending from the client interface down to data storage. At its foundation, the application ensures data integrity, supports validation mechanisms, and enables non-repudiation. It plays a critical role in securely managing application data, access logs, and domain-specific information. These operations are tightly integrated with the server-side logic and protected through policy-based access control, detailed auditing, and error logging. Additionally, the system's internal APIs are modularly designed to handle specific interactions between the front-end and the business logic layer, promoting maintainability and a clear separation of concerns. 

To implement the application architecture described above, a range of modern technologies and external services have been employed. The technology stack has been carefully selected to ensure modularity, scalability, maintainability, and security across all layers of the application. Table~\ref{tab:technology_stack} provides an overview of the primary technologies, frameworks, and external dependencies used throughout the system.

\begin{table}[H]
    \centering
    \resizebox{\textwidth}{!}{
    \renewcommand{\arraystretch}{1.4}
    \begin{tabular}{|p{4cm}|p{10cm}|}
        \hline
        \textbf{Technology Area} & \textbf{Technologies and Dependencies} \\ 
        \hline
        \multicolumn{2}{|c|}{\textbf{Application Core}} \\
        \hline
        Application Framework & Nuxt 3 (Full-stack development: front-end, internal APIs, back-end) with Vue 3 and TypeScript \\ 
        \specialrule{0.4pt}{0pt}{0pt}
        Styling & CSS, Tailwind CSS \\
        \specialrule{0.4pt}{0pt}{0pt}
        Database & MongoDB (Document-based NoSQL storage) with Mongoose (ODM for MongoDB) \\
        \specialrule{0.4pt}{0pt}{0pt}
        Containerization & Docker, Docker Compose \\
        \hline
        \multicolumn{2}{|c|}{\textbf{Security and Identity Management}} \\
        \hline
        Identity Management & Keycloak (SSO, RBAC) \\
        \specialrule{0.4pt}{0pt}{0pt}
        Validation & Zod and custom TypeScript implementations \\
        \specialrule{0.4pt}{0pt}{0pt}
        Logging & Custom TypeScript implementations \\
        \hline
        \multicolumn{2}{|c|}{\textbf{External Integrations and Services}} \\
        \hline
        Public Data Sources & Enhetsregisteret API, Regnskapsregisteret API \\
        \specialrule{0.4pt}{0pt}{0pt}
        Data Transformation & Google Gemini 2.0 Flash (prompt-engineered classification mapping) \\
        \specialrule{0.4pt}{0pt}{0pt}
        Currency Conversion & CurrencyAPI \\
        \specialrule{0.4pt}{0pt}{0pt}
        Mailing Service & nuxt-mail \\
        \hline
        \multicolumn{2}{|c|}{\textbf{Front-end and User Interaction}} \\
        \hline
        Form Handling & FormKit \\
        \specialrule{0.4pt}{0pt}{0pt}
        Statistics and Charts & Highcharts \\
        \specialrule{0.4pt}{0pt}{0pt}
        PDF Generation & pdfmake \\
        \hline
    \end{tabular}}
    \caption{Overview of Technologies and Dependencies Used}
    \label{tab:technology_stack}
\end{table}

\subsubsection{Front-end}

The front-end of the application is built with Nuxt 3, utilizing Vue 3’s Composition API for component development and Nuxt’s file-based routing system for navigation. Each visitable page is composed of multiple reusable Vue components and is protected through Keycloak-based single sign-on (SSO) and role-based access control (RBAC). An overview of the application's visitable pages is illustrated in the sitemap below.

\begin{figure}[H]
    \centering
    \includegraphics[width=1.0\linewidth]{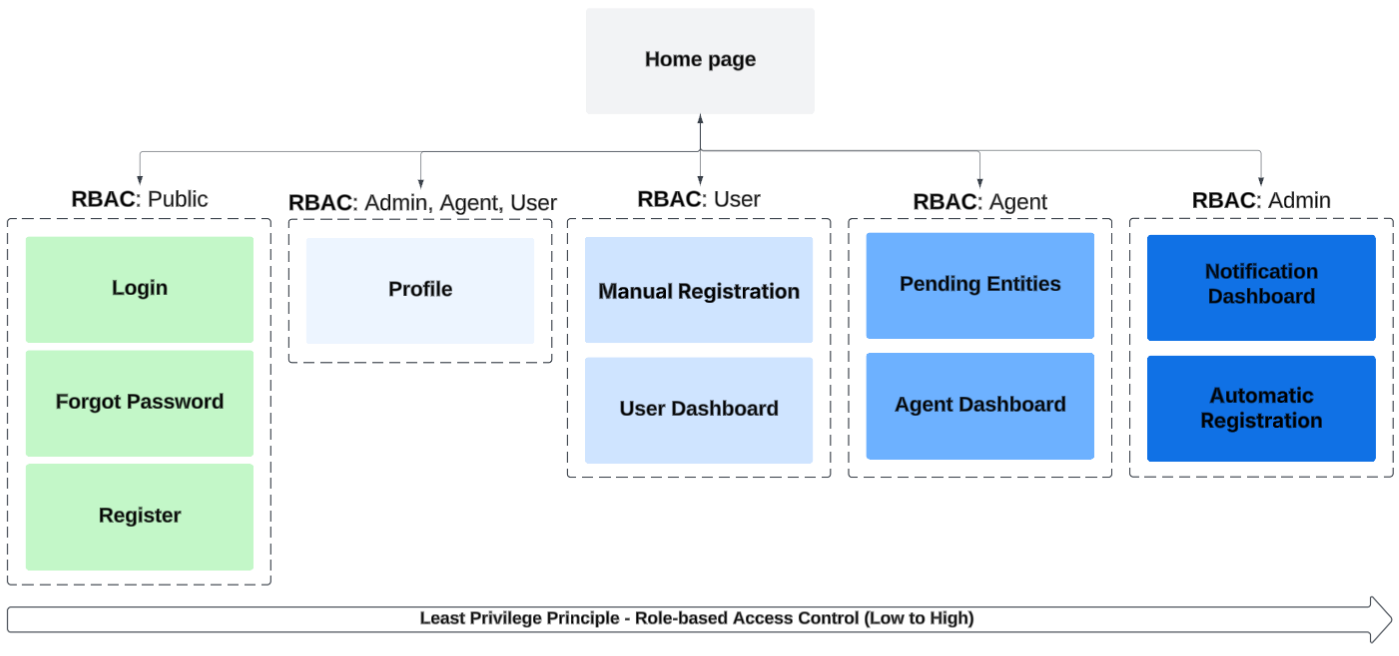}
    \caption{Sitemap}
    \label{fig:sitemap}
\end{figure}

\subsubsection{Back-end}

The back-end consists of internal APIs connected to modular business logic, organized by separation of concerns. It includes utility functions, database operations, type declarations, and schema definitions. While a detailed breakdown of all server-side functionality is beyond the scope of this section, the core responsibilities include data aggregation, entity labeling and classification, security, and interaction with the underlying database. These operations are primarily implemented using Nuxt 3’s server route architecture, ensuring a unified full-stack structure.

\paragraph{Data aggregation}

Data is aggregated at multiple points within the application to support various core functionalities. The most significant aggregation use cases are outlined below:

\begin{itemize}
    \item \textbf{Automatic Registration:} Data fetched from public registries is validated and transformed to conform with internal data structures and NIS2 requirements. In particular, data from external sources is processed using prompt engineering techniques with Gemini Flash 2.0 to map sectors, sub-sectors, and entity types to the definitions specified in NIS2.
    
    \item \textbf{Automatic Notification:} Following entity registration, processed data is stored and statistical insights are aggregated to meet the notification requirements outlined earlier in this chapter. These statistics are displayed on the Notification Dashboard, and a standardized report can be generated at the press of a button.
\end{itemize}

\paragraph{Entity Labeling and Classification}

Entity data, along with aggregated labels, is stored in the database to provide contextual understanding, facilitate statistical analysis, and enable automated classification. This process is fully automated to ensure consistency across the registry. It guarantees correct data capture, aggregation, labeling, classification, and notification, thus completing the registry workflow described in Figure~\ref{fig:Registry_Workflow}.

Labels are generated as shown in Table~\ref{tab:Entity_Labels}. Entity classification is based on the factors depicted in Figure~\ref{fig:Classification_Factors} and results in the classifications presented in Figure~\ref{fig:Classification_Classes}, according to the rules outlined in Figures~\ref{fig:Essential_Classification},~\ref{fig:Important_Classification}, and~\ref{fig:Excluded_Classification}.

Additionally, the priority sequence illustrated below ensures classification accuracy during both the initial classification and the classification with complete entity information.

\begin{figure}[H]
    \centering    
    \includegraphics[width=1.0\textheight, width=1.0\textwidth]{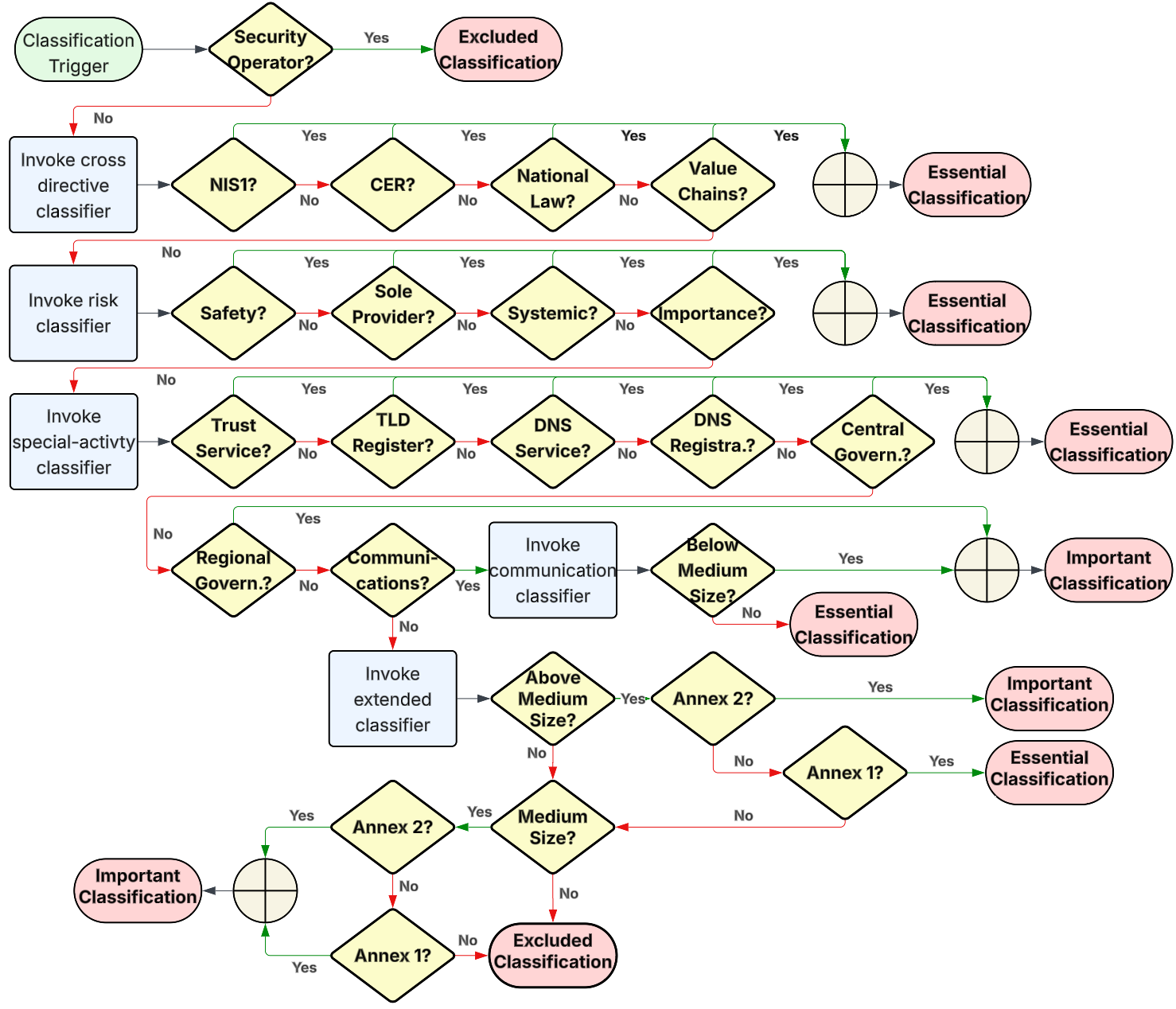}
    \caption{Classification Priority Sequence}
    \label{fig:Classification_Priority_Sequence}
\end{figure}

\paragraph{Security}

Security is implemented throughout the application and not confined solely to the back-end. However, a significant portion of security enforcement is facilitated through back-end operations. 

\begin{itemize}
    \item \textbf{Identity Management:} New registrants are automatically assigned the user role, allowing them to register entities. Access to specific entities must be approved by an agent. Higher-privilege roles (agent and admin) can only be assigned through the Keycloak admin panel to ensure centralized control. All authentication and role management are handled by Keycloak and enforced programmatically throughout the application.
    \item \textbf{Validation:} Application integrity is safeguarded through comprehensive client-side and server-side validation. Client-side validation provides immediate feedback to users, preventing incorrect data from propagating deeper into the application. Server-side validation ensures robust protection against injection attacks and guarantees that only validated data is processed.
    \item \textbf{Non-repudiation:} Non-repudiation mechanisms are implemented by logging user activities with associated user identifiers. This guarantees that actions can be traced back to individual users, ensuring accountability and responsibility tracking.
    \item \textbf{Logging:} Comprehensive logging is implemented across all layers of the application. Logging often supports non-repudiation, facilitates auditing, enables post-event assessments, and assists in debugging.
    \item \textbf{Encryption:} At this stage of development, the application does not implement encryption at rest or encryption in transit. This decision is based on the fact that the system is still under active development and has not yet been deployed to a production environment. During this phase, access is restricted to a limited set of developers, and the system operates within a secured, isolated environment.

    As part of the planned deployment strategy, encryption in transit will be ensured through the use of HTTPS, and encryption at rest will be applied to all persistent data storage. These measures will be incorporated to align with security best practices and to meet relevant regulatory and compliance requirements prior to production deployment.
\end{itemize}

\paragraph{Interaction with the database}

Database interactions are managed using Mongoose, with distinct NoSQL schemas defined for each document type. Each document supports tailored CRUD operations (create, read, update, delete) aligned with its specific structure. Utility modules encapsulate database logic per document, and most operations are exposed through dedicated API endpoints. Where functionality overlaps, shared endpoints are used to minimize redundancy.

All database interactions are logged to support error handling and non-repudiation, and all inbound and outbound data is validated using Zod to ensure structural integrity and type safety.

\subsubsection{Data Storage}

Authentication, authorization, and user-related information is exclusively managed by Keycloak and stored in its relational database. All other application-specific data is stored in MongoDB, structured into collections. Each collection serves a dedicated purpose, as outlined in the table below.

\begin{table}[H]
    \centering
    \resizebox{\textwidth}{!}{
    \renewcommand{\arraystretch}{1.4}
    \begin{tabular}{|p{5.5cm}|p{8.5cm}|}
        \hline
        \textbf{Collection Name} & \textbf{Purpose} \\ 
        \specialrule{0.4pt}{0pt}{0pt}
        \texttt{agentlogs} & Records all changes made to entity documents by users with the agent role. \\ 
        \specialrule{0.4pt}{0pt}{0pt}
        \texttt{apperrorlogs} & Logs application-level errors for diagnostics and monitoring. \\
        \specialrule{0.4pt}{0pt}{0pt}
        \texttt{entities} & Central collection that stores all structured information about registered entities. \\
        \specialrule{0.4pt}{0pt}{0pt}
        \texttt{errorlogs} & Captures errors specifically related to automatic registration and entity processing. \\
        \specialrule{0.4pt}{0pt}{0pt}
        \texttt{informationrequests} & Tracks information requests made against entities, including answers provided. \\
        \specialrule{0.4pt}{0pt}{0pt}
        \texttt{organizationaccessrequests} & Maintains user-to-entity access requests and their approval status. \\
        \specialrule{0.4pt}{0pt}{0pt}
        \texttt{orgnumlists} & Holds organizational number lists governed by directives other than NIS2. \\
        \specialrule{0.4pt}{0pt}{0pt}
        \texttt{pendingentities} & Contains manually registered entities that are pending review. \\
        \specialrule{0.4pt}{0pt}{0pt}
        \texttt{userrequests} & Logs change requests submitted by users regarding entities they have access to. \\
        \hline
    \end{tabular}}
    \caption{MongoDB Collections and Their Responsibilities}
    \label{tab:mongodb_collections}
\end{table}

\section{Summary}

With the theoretical background, analysis, application requirements, and architecture defined, the following chapters present the results of the implementation.

\chapter{Results}

This chapter presents the results of implementing the registry outlined in Chapter 4. It begins by examining requirements traceability matrices to evaluate how well the implementation fulfills the defined functional and non-functional requirements. Subsequent sections highlight the core system features, performance benchmarks, and the outcomes of automated classification and registration.

\section{Requirement Traceability}

To evaluate how well the implementation fulfills the system’s goals, a traceability overview summarizes the degree of requirement satisfaction across five key categories: registration, classification, notification, supervision, and literature-derived objectives. This provides a clear snapshot of overall system alignment with the defined goals. 

Across a total of 40 requirements, 36 were successfully met, 3 were not fulfilled, and 1 was not assessed. The unmet requirements primarily relate to auxiliary or non-essential features—such as file uploads and entity-specific report generation—rather than core compliance functionality. These omissions were the result of deliberate prioritization during development, constrained by the project’s limited timeframe.

\begin{figure}[H]
    \centering
    \includegraphics[width=1.0\linewidth]{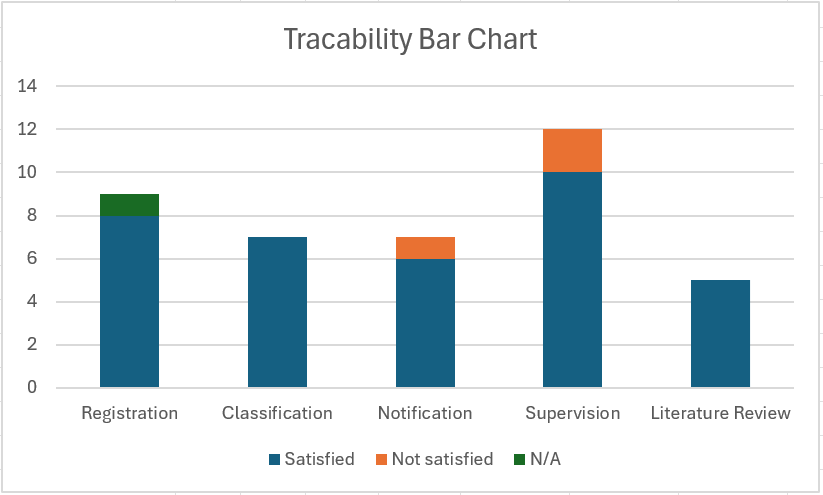}
    \caption{Traceability Bar Chart}
    \label{fig:Traceability_Bar_Chart}
\end{figure}

Each requirement is assessed based on two core dimensions:

\begin{itemize}
    \item \textbf{Implementation:} Whether the functionality exists in the deployed system.
    \item \textbf{Alignment:} Whether tthe implemented functionality fulfills the intent and scope of the requirements as described in \ref{tab:Registration_Requirements}, \ref{tab:Classification_Requirements}, \ref{tab:Notification_Requirements}, \ref{tab:Supervision_Requirements}, and \ref{tab:Literature_Review_Requirements}. 
\end{itemize}

A requirement is marked as \textbf{'Satisfied'} if it is both implemented and correctly aligned with its description. Requirements marked \textbf{'Not Satisfied'} are either missing or only partially implemented in a way that does not meet the intended specification. \textbf{'N/A'} is used for requirements deemed outside of the project's current scope. 

\subsection{Requirement Traceability in-depth}

The registration requirements were designed to address the research question by establishing mechanisms necessary to support and facilitate the registration process. 

As summarized in Table~\ref{tab:Registration_Requirements_Traceability_Matrix}, 8 out of 9 requirements (89\%) were successfully satisfied. The one requirement not assessed pertains to GDPR compliance, which was deemed not applicable since the application does not process personal or sensitive data.

\begin{table}[H]
    \centering
    \resizebox{\textwidth}{!}{
    \begin{tabular}{|p{2cm}|p{11cm}|p{2cm}|}
        \hline
        \textbf{Req. ID} & \textbf{Comment} & \textbf{Result} \\ \hline
        R-1 & All registration occurs through the application. & Satisfied \\
        \specialrule{0.4pt}{0pt}{0pt}
        R-2 & Users can manually register entities. & Satisfied \\
        \specialrule{0.4pt}{0pt}{0pt}
        R-3 & Administrators can perform automatic registration of entities. & Satisfied \\
        \specialrule{0.4pt}{0pt}{0pt}
        R-4 & Annex explanations are integrated into the manual registration interface. & Satisfied \\
        \specialrule{0.4pt}{0pt}{0pt}
        R-5 & Automatic registration satisfies the data collection criteria in Table~\ref{tab:Classification_Information_Requirements} and partially in Table~\ref{tab:Registry_Information_Requirements}. Manual registration satisfies both. & Satisfied \\
        \specialrule{0.4pt}{0pt}{0pt}
        R-6 & Client-side validation is applied to all input fields. & Satisfied \\
        \specialrule{0.4pt}{0pt}{0pt}
        R-7 & Manual registration supports data required in Table~\ref{tab:Risk_Information_Requirements}. & Satisfied \\
        \specialrule{0.4pt}{0pt}{0pt}
        R-8 & The database does not handle sensitive personal information. & N/A \\
        \specialrule{0.4pt}{0pt}{0pt}
        R-9 & All entry points are secured through validation, authentication, and authorization. & Satisfied \\ 
        \hline
    \end{tabular}}
    \caption{Registration Requirements Traceability Matrix}
    \label{tab:Registration_Requirements_Traceability_Matrix}
\end{table}

In addition to registration, classification serves as a core component of the registry and directly responds to the research question by defining how classification criteria should be structured and operationalized in support of a harmonized, NIS2-compliant registry. The implemented classification logic is grounded in well-defined rule sets aligned with NIS2 \cite{NIS2}, enabling automated, consistent, and context-aware outcomes. The table below presents a traceability assessment of the classification-related requirements and their implementation status.

\begin{table}[H]
    \centering
    \resizebox{\textwidth}{!}{
    \begin{tabular}{|p{2cm}|p{11cm}|p{2cm}|}
        \hline
        \textbf{Req. ID} & \textbf{Comment} & \textbf{Result} \\ \hline
        C-1 & Classification computes 1 out of the 3 classes shown in Figure~\ref{fig:Classification_Classes}. & Satisfied \\
        \specialrule{0.4pt}{0pt}{0pt}
        C-2 & Classification is based on the factors illustrated in Figure~\ref{fig:Classification_Factors}. & Satisfied \\
        \specialrule{0.4pt}{0pt}{0pt}
        C-3 & Entities are classified based on the rule sets in Figures~\ref{fig:Essential_Classification}, \ref{fig:Important_Classification}, and \ref{fig:Excluded_Classification}. & Satisfied \\
        \specialrule{0.4pt}{0pt}{0pt}
        C-4 & Labels listed in Table~\ref{tab:Entity_Labels} and additional metadata are used to contextualize entities. & Satisfied \\
        \specialrule{0.4pt}{0pt}{0pt}
        C-5 & Classification outcomes are mutually exclusive and exhaustive for all processed entities. & Satisfied \\
        \specialrule{0.4pt}{0pt}{0pt}
        C-6 & All classification rule sets are automatically computed. & Satisfied \\
        \specialrule{0.4pt}{0pt}{0pt}
        C-7 & Agents can edit entity data to influence classification results. & Satisfied \\
        \hline
    \end{tabular}}
    \caption{Classification Requirements Traceability Matrix}
    \label{tab:Classification_Requirements_Traceability_Matrix}
\end{table}

The ability to automatically aggregate and disseminate information is key to enabling timely notifications and proportionate supervision—two central goals of the NIS2 directive and this research. The requirements listed below focus on how the application supports aggregation of entity-level and statistical data, as well as its ability to generate and distribute notification reports. This functionality directly supports the part of the research question concerning data aggregation and its role in enabling efficient, context-aware oversight while minimizing administrative overhead.

\begin{table}[H]
    \centering
    \resizebox{\textwidth}{!}{
    \begin{tabular}{|p{2cm}|p{11cm}|p{2cm}|}
        \hline
        \textbf{Req. ID} & \textbf{Comment} & \textbf{Result} \\ \hline
        N-1 & General statistics are automatically aggregated and visually displayed. & Satisfied \\
        \specialrule{0.4pt}{0pt}{0pt}
        N-2 & ENISA-compliant statistics are automatically aggregated and visually displayed. & Satisfied \\
        \specialrule{0.4pt}{0pt}{0pt}
        N-3 & Temporal changes in ENISA statistics are not yet tracked or versioned. & Not satisfied \\
        \specialrule{0.4pt}{0pt}{0pt}
        N-4 & Risk-related data is aggregated and presented in a dedicated statistics view. & Satisfied \\
        \specialrule{0.4pt}{0pt}{0pt}
        N-5 & General, ENISA, and risk-based reports can be automatically generated in PDF format. & Satisfied \\
        \specialrule{0.4pt}{0pt}{0pt}
        N-6 & Predefined report templates can be generated automatically. & Satisfied \\
        \specialrule{0.4pt}{0pt}{0pt}
        N-7 & Generated reports can be downloaded and sent via email. & Satisfied \\
        \hline
    \end{tabular}}
    \caption{Notification Requirements Traceability Matrix}
    \label{tab:Notification_Traceability_Matrix}
\end{table}

Additionally, the application must enable effective supervision, a core concern of the research question. Specifically, the system must support mechanisms that allow agents to evaluate and interact with entities in a timely, proportionate, and context-aware manner. This includes features for initiating information and change requests, performing risk assessments, maintaining detailed audit trails, and accessing comprehensive entity profiles. These capabilities not only reduce the administrative burden on entities by streamlining interaction but also ensure that supervisory tasks are informed and traceable. The traceability matrix below maps each supervision requirement to the implementation, illustrating how the system supports supervisory workflows in line with NIS2 objectives.

\begin{table}[H]
    \centering
    \resizebox{\textwidth}{!}{
    \begin{tabular}{|p{2cm}|p{11cm}|p{2cm}|}
        \hline
        \textbf{Req. ID} & \textbf{Comment} & \textbf{Result} \\ \hline
        S-1 & The Agent Dashboard supports sending information requests to specific entities. & Satisfied \\
        \specialrule{0.4pt}{0pt}{0pt}
        S-2 & The User Dashboard supports submission of change requests. & Satisfied \\
        \specialrule{0.4pt}{0pt}{0pt}
        S-3 & The Agent Dashboard supports reviewing and approving change requests. & Satisfied \\
        \specialrule{0.4pt}{0pt}{0pt}
        S-4 & The Agent Dashboard supports editing entity data. & Satisfied \\
        \specialrule{0.4pt}{0pt}{0pt}
        S-5 & Risk assessments are currently constrained to the entity level only. & Not satisfied \\
        \specialrule{0.4pt}{0pt}{0pt}
        S-6 & All changes to entity data are logged with the modifier's user ID. & Satisfied \\
        \specialrule{0.4pt}{0pt}{0pt}
        S-7 & The Agent Dashboard includes a task log linking tasks to their corresponding entities. & Satisfied \\
        \specialrule{0.4pt}{0pt}{0pt}
        S-8 & Reports about specific entities can be generated in the Agent Dashboard & Satisfied \\
        \specialrule{0.4pt}{0pt}{0pt}
        S-9 & The registry does not currently support file uploads. & Not satisfied \\ 
        \specialrule{0.4pt}{0pt}{0pt}
        S-10 & All contextual entity information is clearly displayed. & Satisfied \\
        \specialrule{0.4pt}{0pt}{0pt}
        S-11 & Authentication is implemented using Keycloak. & Satisfied \\
        \specialrule{0.4pt}{0pt}{0pt}
        S-12 & Role-based authorization is implemented using Keycloak. & Satisfied \\
        \hline
    \end{tabular}}
    \caption{Supervision and System Requirements Traceability Matrix}
    \label{tab:Supervision_Requirements_Traceability_Matrix}
\end{table}

Lastly, the following requirements are derived from the literature review and address the research question’s emphasis on automation, supervision, and scalability. Each has been successfully implemented, as shown in the traceability matrix below.

\begin{table}[H]
    \centering
    \resizebox{\textwidth}{!}{
    \begin{tabular}{|p{2cm}|p{11cm}|p{2cm}|}
        \hline
        \textbf{Req. ID} & \textbf{Comment} & \textbf{Result} \\ \hline
        LR-1 & Artificial Intelligence is used to streamline automatic classification & Satisfied \\
        \specialrule{0.4pt}{0pt}{0pt}
        LR-2 & Dynamic classification is made possible through the Agent Dashboard. & Satisfied \\
        \specialrule{0.4pt}{0pt}{0pt}
        LR-3 & Reports are generated based on a standard format to enable timely notifications. & Satisfied \\
        \specialrule{0.4pt}{0pt}{0pt}
        LR-4 & The registry is designed based on Nuxt 3 and Vue 3 principles, making it scalable. & Satisfied \\
        \specialrule{0.4pt}{0pt}{0pt}
        LR-5 & Predictive analysis can be performed on contextualized and aggregated information. & Satisfied \\
        \hline
    \end{tabular}}
    \caption{Literature Review Requirements Traceability Matrix}
    \label{tab:LR_Traceability_Matrix}
\end{table}

Overall, the traceability matrices demonstrate strong alignment with the research question, indicating that the developed application effectively addresses the intended objectives.

\section{Entity Registration and Classification System}

The registry features a complete pipeline for both automatic and manual entity registration. All core functions—labeling, classification, and aggregation—are fully automated. 

Administrators can automatically register any entity listed in Enhetsregisteret and Regnskapsregisteret, provided sufficient public information is available. However, while these registries contain critical data, they do not offer comprehensive coverage for all classification dimensions. As such, only an initial classification can be computed automatically. Two classification factors—risk and directive relevance—must be obtained through manual data input or through direct communication with the entity.

\begin{table}[H]
    \centering
    \renewcommand{\arraystretch}{1.3}
    \begin{tabular}{|c|c|}
        \hline
        \textbf{Factor} & \textbf{Data Availability} \\
        \hline
        Size & Public \\
        Activity & Public \\
        Risk & Not Public \\
        Directive & Not Public \\
        \hline
    \end{tabular}
    \caption{Data Availability per Classification Factor}
    \label{tab:classification_factors_matrix}
\end{table}

As shown in Figures~\ref{fig:Essential_Classification}, \ref{fig:Important_Classification}, and \ref{fig:Excluded_Classification}, the majority of classification logic is based on an entity’s size and activity. However, certain rule sets also depend on risk assessment and cross-directive applicability. To ensure these cases are captured accurately, the registry supports both automatic and manual registration modes. This dual approach increases classification efficiency while preserving accuracy, especially for nuanced or borderline cases. Furthermore, classification rules are codified and automated to guarantee consistency across all outcomes. In addition, institutional factors—such as affiliation with research organizations, national law applicability, or security-related operations—can be determined automatically in simple cases. However, these factors often require further assessment due to the nuanced interpretation of the entity types they govern. This highlights the importance of maintaining manual oversight in the classification process to account for legal and contextual subtleties that fall outside the scope of available public data. 

Manual registration primarily depends on the registrant’s understanding of their own entity. However, the integration of client-side validation and in-form guidance significantly enhances the reliability and consistency of the submitted data, thereby improving the overall accuracy of classification. The performance of the automatic pipeline has also been evaluated through benchmarking, with the results presented below.

\subsection{Automatic Processing Benchmarking}

To evaluate the effectiveness of the application's automated pipeline for registering, labeling, and classifying entities, we assess its performance based on the following three factors:

\begin{itemize}
    \item \textbf{Classification accuracy:} Within the constraints of the information available in Enhetsregisteret and Regnskapsregisteret, we evaluate whether the classification algorithm produces the correct classification.
    \item \textbf{Size calculation accuracy:} Based on data from the Regnskapsregisteret, we assess whether the size calculation algorithm yields accurate results in light of current euro conversion rates.
    \item \textbf{Gemini activity transformation accuracy:} Using data from Enhetsregisteret, we evaluate Gemini's ability to correctly transform business activity data into the appropriate sectors, sub-sectors, and entity types.
\end{itemize}

While Enhetsregisteret and Regnskapsregisteret do not provide all the information necessary for a complete entity classification, they contain sufficient data to enable an initial classification—excluding the risk assessment and cross-directive components.

To perform this benchmark, we select a representative sample of entities that have complete and up-to-date data available in the public registers. Specifically, we selected 100 random companies from Kapital’s list of the 500 largest companies \cite{Top500}.

The benchmark is conducted by automatically registering, labeling, and classifying each entity, and then comparing the result to a manually verified reference outcome. The results are recorded in a scoring table, where each benchmarking variable is assigned a score of either 1 or 0, depending on whether it matches the reference.

A score of 1 indicates that the pipeline produced the same result as the manual assessment, while a score of 0 indicates a discrepancy. An entity can receive a maximum score of 3—one point per benchmark variable—indicating that the pipeline fully matched the reference assessment.

The results are presented below as percentages, while a detailed breakdown is provided in the Appendix.

\begin{figure}[H]
    \centering
    \includegraphics[width=1.0\linewidth]{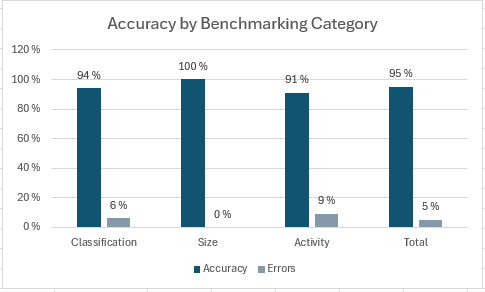}
    \caption{Accuracy results for each automatic pipelining category}
    \label{fig:Benchmarking_Bar_Chart}
\end{figure}

The current application is capable of automatically registering entities, calculating their size, determining their activity, and classifying them with an overall accuracy of 95\%. As classification relies on both size and activity assessments, its accuracy is inherently influenced by the performance of these underlying components.

Among the three evaluation categories, activity determination showed the lowest accuracy at 91\%. In a randomly selected sample of 100 entities from Kapital’s list of Norway’s 500 largest companies \cite{Top500}, the Gemini model encountered particular challenges with entities in the Manufacturing sector (6 out of 9 misclassifications) and, to a lesser extent, the Health sector. In one instance, it failed to return a specific entity type entirely.

The primary issue stems from Gemini correctly identifying an entity as belonging to the manufacturing sector, while that sector lacks a suitable entity type under the NIS2 specification. This mismatch has been addressed by updating the prompt to explicitly handle such cases, although the effectiveness of this solution has not yet been evaluated.

\subsection{Manual Registration Pipeline}

All manually registered entities must undergo a screening process conducted by any agent to ensure the accuracy and completeness of the submitted data. These registrations are initially categorized as pending entities. During this stage, all submitted information is available for review, and agents are granted the ability to either edit or reject entries. This editability feature allows for the correction of minor inaccuracies without requiring complete resubmission, thereby improving administrative efficiency.

To initiate a registration, users are required to create an account. Upon account creation, they are automatically assigned the user role, which grants them permission to submit entity registrations. Under the current authentication and authorization schema, each registrant automatically submits an access request to the entity they registered. This mechanism ensures that an agent can verify whether the user is legitimately associated with the entity before assigning representative privileges.

The application also includes built-in email functionality intended to notify users of key registration events—such as entity approval or rejection, and access grants. Once a registration is approved and the user’s access request is granted, the entity becomes available on the user’s dashboard. It also becomes visible and editable within the agent dashboard, where additional metadata is aggregated for analytical insights and notification generation. 

\begin{figure}[H]
    \centering
    \includegraphics[width=1.0\linewidth, height=0.5\textheight]{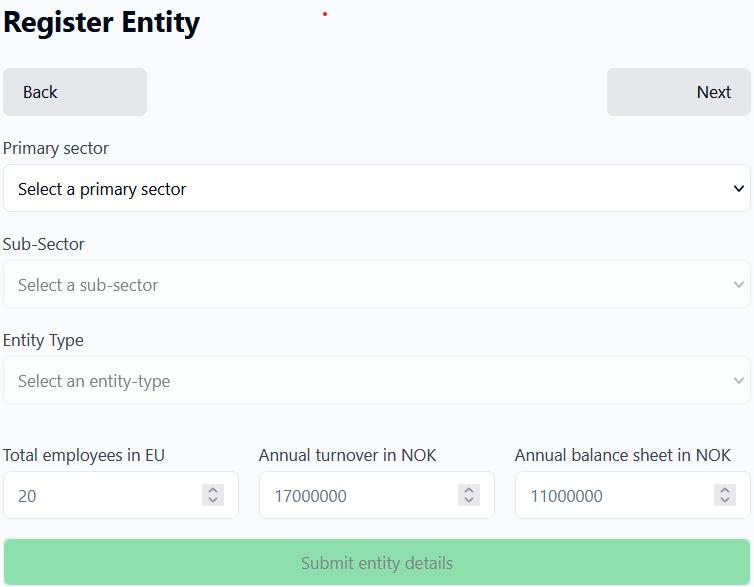}
    \caption{A part of the manual registration form}
    \label{fig:Manual_Registration_Example}
\end{figure}

\section{Supervision, Notification, and Dashboards}

Supervision, notification, assessments, and agent to user interactions are made possible through dashboards. The registry has a user dashboard, an agent dashboard, and two dashboards for the administrator. 

\subsection{User Dashboard}

The user dashboard is designed to allow a single user to manage and access multiple entities. Once access is granted, the user can view general information about each entity, including its classification and computed size. The dashboard also enables users to request changes to entity data, track the status of submitted requests, and respond to information requests initiated by supervising agents. Additionally, a built-in change log provides visibility into all modifications associated with the entity.

This functionality supports core objectives of the NIS2 directive by offering a centralized, transparent interface that simplifies entity interactions, facilitates supervisory communication, and reduces administrative overhead through streamlined access and auditability.

\begin{figure}[H]
    \centering
    \includegraphics[width=1.0\linewidth]{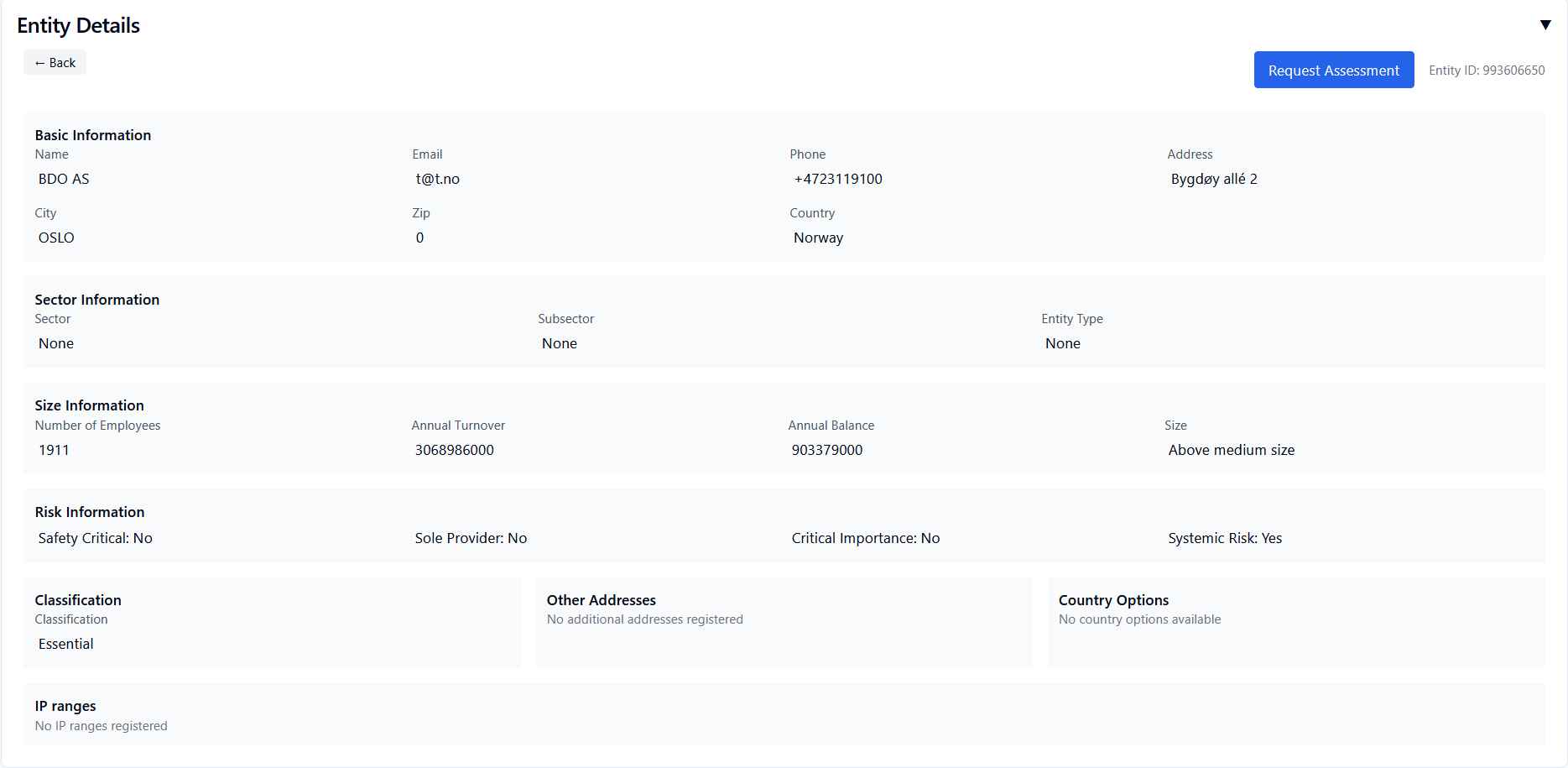}
    \caption{User Dashboard - Entity Display Example}
    \label{fig:User_Entity_Display}
\end{figure}

\begin{figure}[H]
    \centering
    \includegraphics[width=1.0\linewidth]{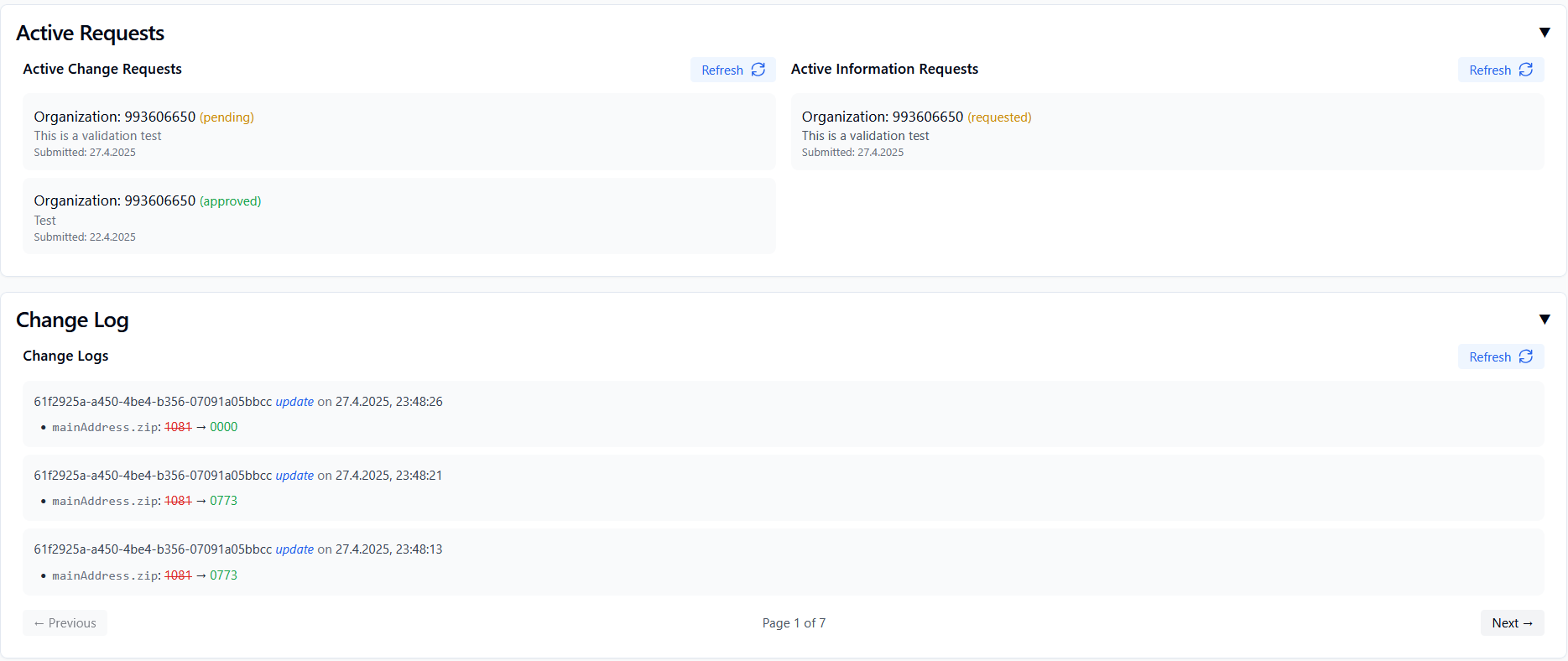}
    \caption{User Dashboard - Requests and Change Log Example}
    \label{fig:User_Entity_Display_2}
\end{figure}

\subsection{Agent Dashboard}

The agent dashboard is designed for use by competent authorities to supervise registered entities. It includes a statistical panel for Smart Insights, a task log to manage entity registrations, access requests, and change approvals. Agents can search, filter, and select individual entities to review their details. Upon selection, agents can re-compute entity labels and classifications, initiate targeted information requests, and directly edit editable entity data. Each data point clearly indicates its source—automated, manual, or edited—and all modifications are logged to support traceability and accountability.

This interface supports key objectives of the NIS2 directive by enabling proportionate, data-driven, and context-aware supervision. It provides authorities with the tools necessary to monitor entities efficiently, respond to evolving risks, and maintain a transparent audit trail for all supervisory actions.

\begin{figure}[H]
    \centering
    \includegraphics[width=1.0\linewidth]{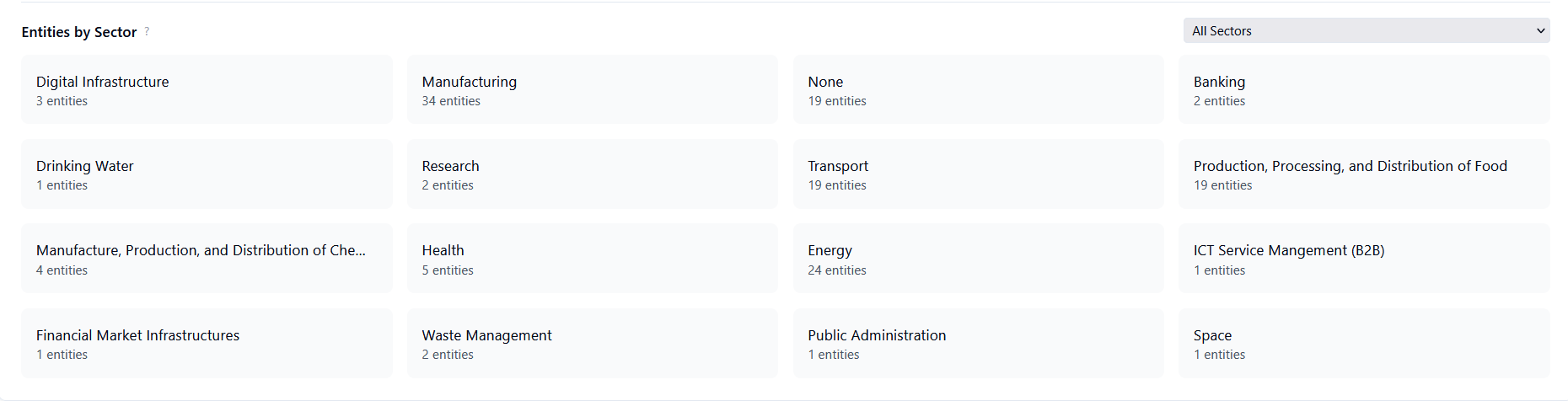}
    \caption{Agent Dashboard - Sector Statistics}
    \label{fig:Agent_Statistics_Display}
\end{figure}

\begin{figure}[H]
    \centering
    \includegraphics[width=1.0\linewidth]{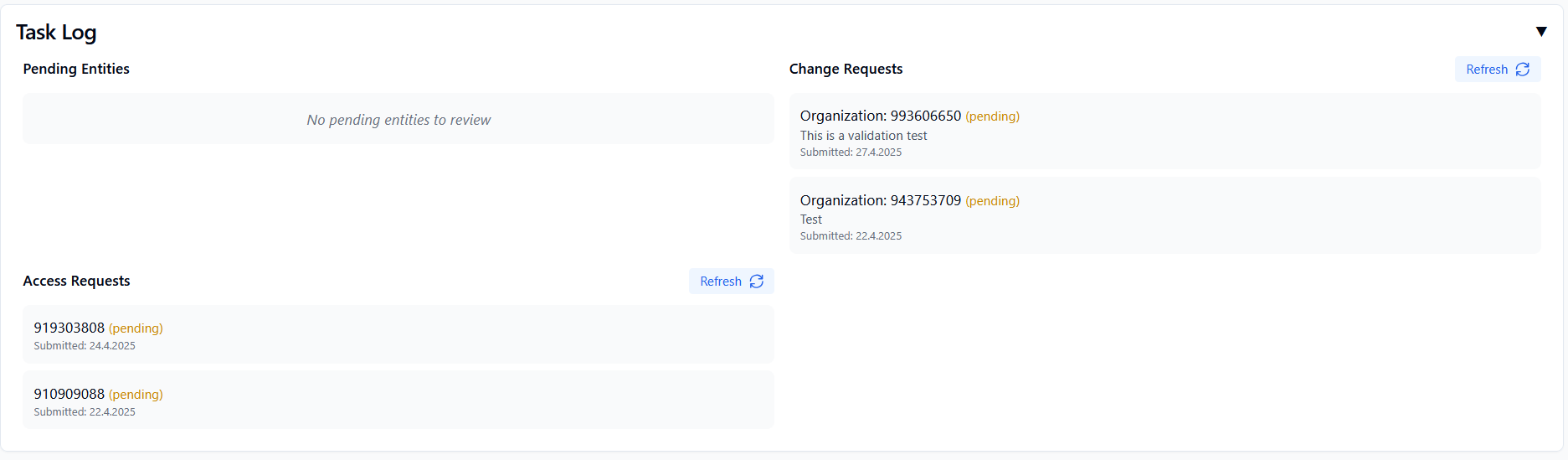}
    \caption{Agent Dashboard - Task Log}
    \label{fig:Agent_TaskLog_Display}
\end{figure}

\begin{figure}[H]
    \centering
    \includegraphics[width=1.0\linewidth]{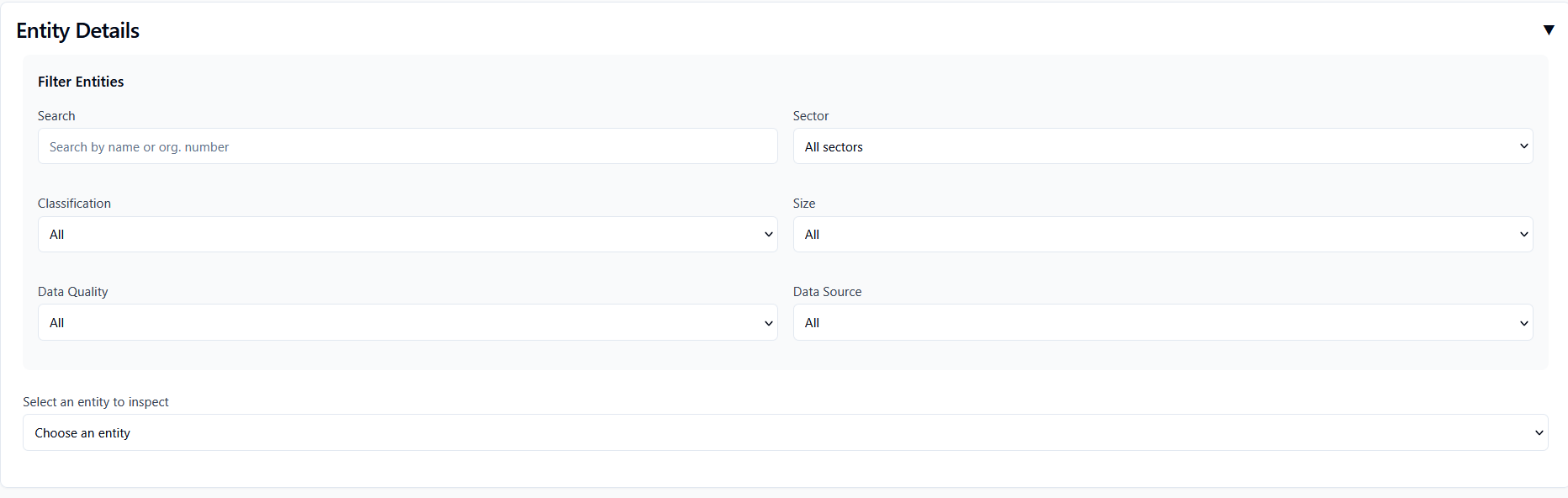}
    \caption{Agent Dashboard - Entity Filter}
    \label{fig:Agent_Filter_Display}
\end{figure}

\begin{figure}[H]
    \centering
    \includegraphics[width=1.0\linewidth]{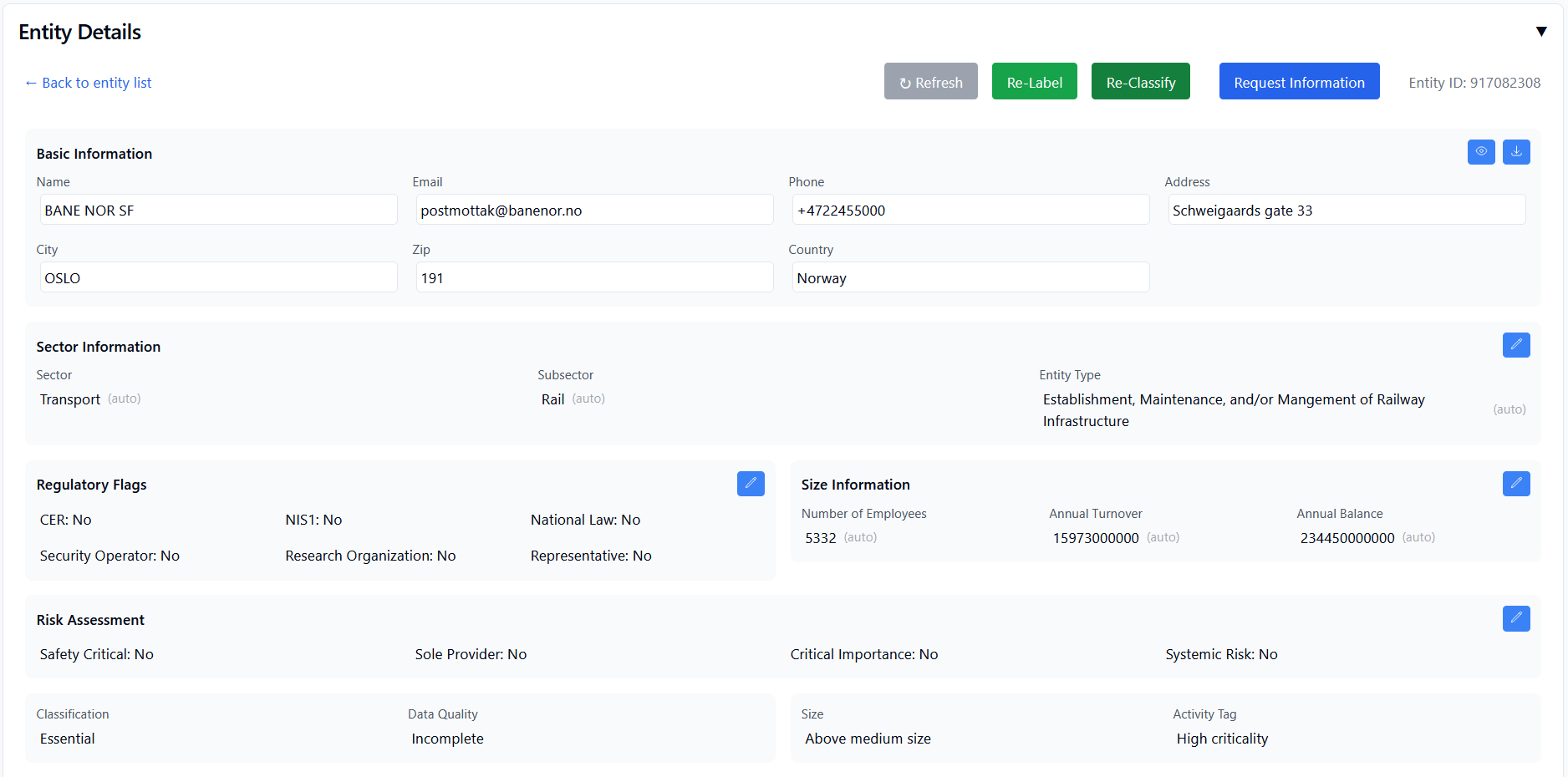}
    \caption{Agent Dashboard - Entity Details}
    \label{fig:Agent_Entity_Details_Display}
\end{figure}

\subsection{Administrator Dashboards}

The administrator dashboards provide system-level control over core processes such as automatic registration and notification management, enabling oversight of data ingestion and alert dissemination. 

\subsubsection{Automatic registration}

Automatic registration is primarily used to register a batch of entities or a single entity by leveraging publicly available data. This feature includes a statistics panel tailored to registration-related metrics and a detailed view of any failed registration attempts. By automating the ingestion of public data from authoritative sources, this functionality supports the research question’s emphasis on reducing administrative burden while maintaining classification accuracy. It also aligns with the NIS2 directive’s objective of achieving harmonized and efficient supervision through streamlined and scalable entity management.

\begin{figure}[H]
    \centering
    \includegraphics[width=1.0\linewidth]{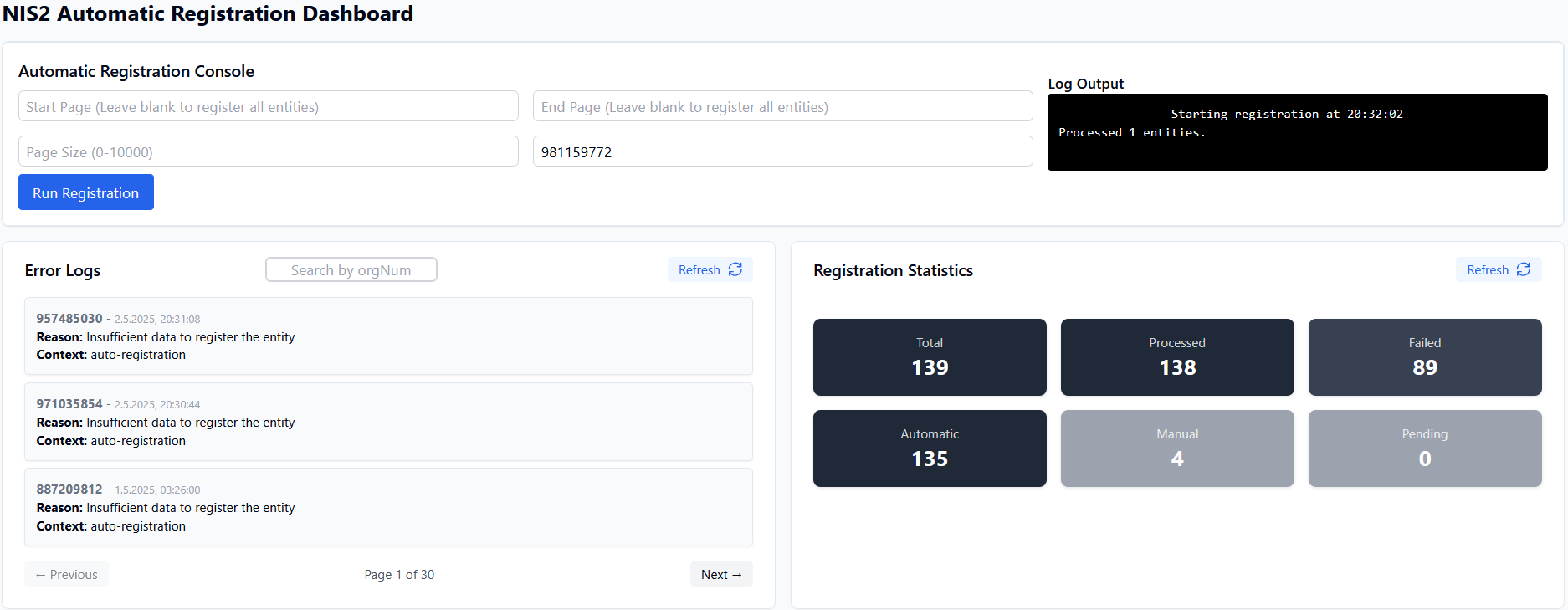}
    \caption{Admin Dashboard - Automatic Registration}
    \label{fig:Admin_Automatic_Registration_Display}
\end{figure}

\subsubsection{Notification Dashboard}

The notification dashboard equips administrators with a centralized view of automatically aggregated statistics aligned with NIS2 reporting requirements, as illustrated in Figure~\ref{fig:Registration_Concept}. Both the statistics and the contents of the generated reports reflect the current state of the registry, enabling timely and informed reporting—whether triggered by directive deadlines or operational insight. Standardized reports can be produced on demand in PDF format, ensuring consistency and efficiency in supervisory communication. 

\begin{figure}[H]
    \centering
    \includegraphics[width=1.0\linewidth]{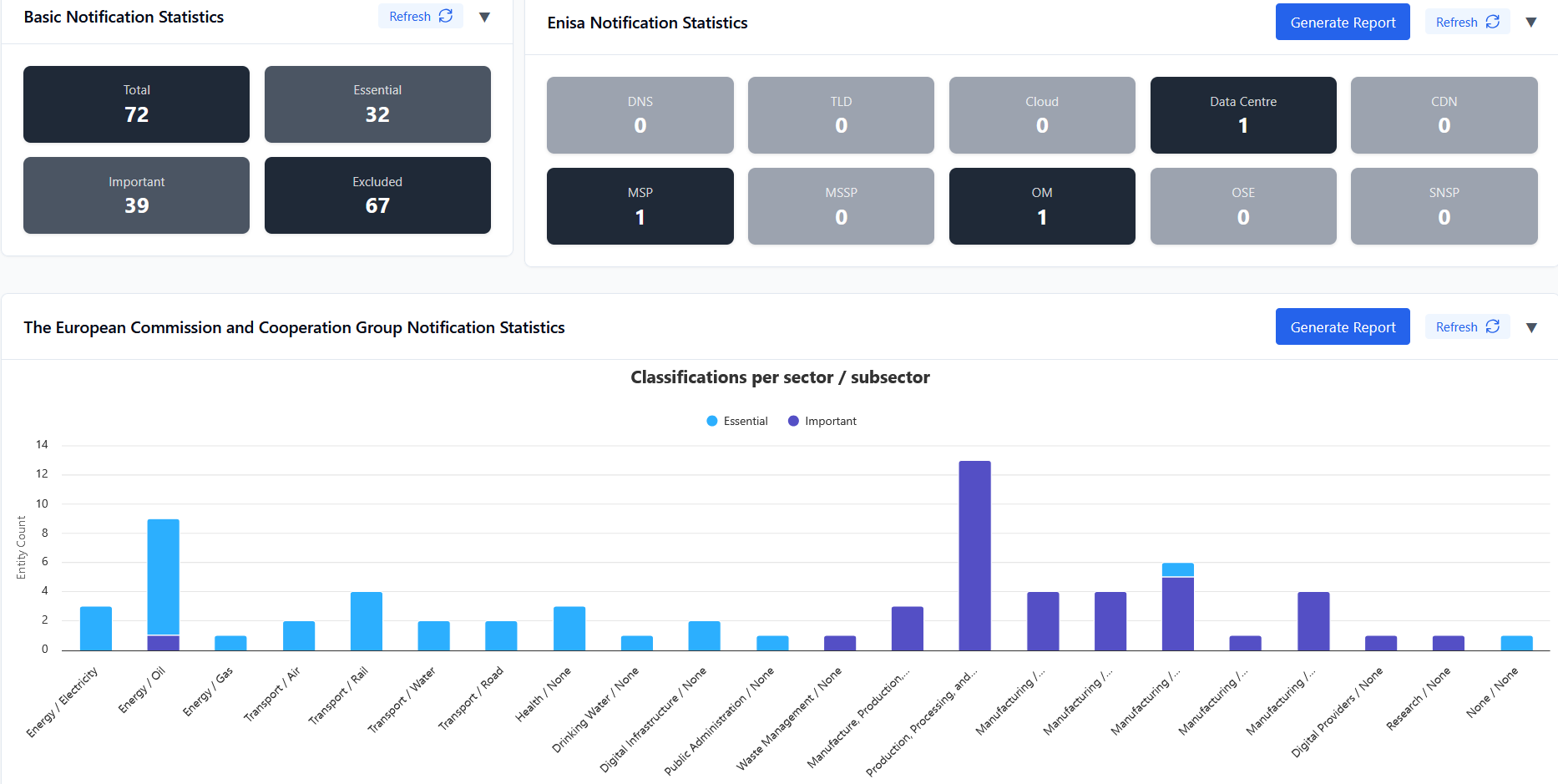}
    \caption{Admin Dashboard - Notification Dashboard}
    \label{fig:Admin_Notification_Dashboard}
\end{figure}

\begin{figure}[H]
    \centering
    \includegraphics[width=1.0\linewidth]{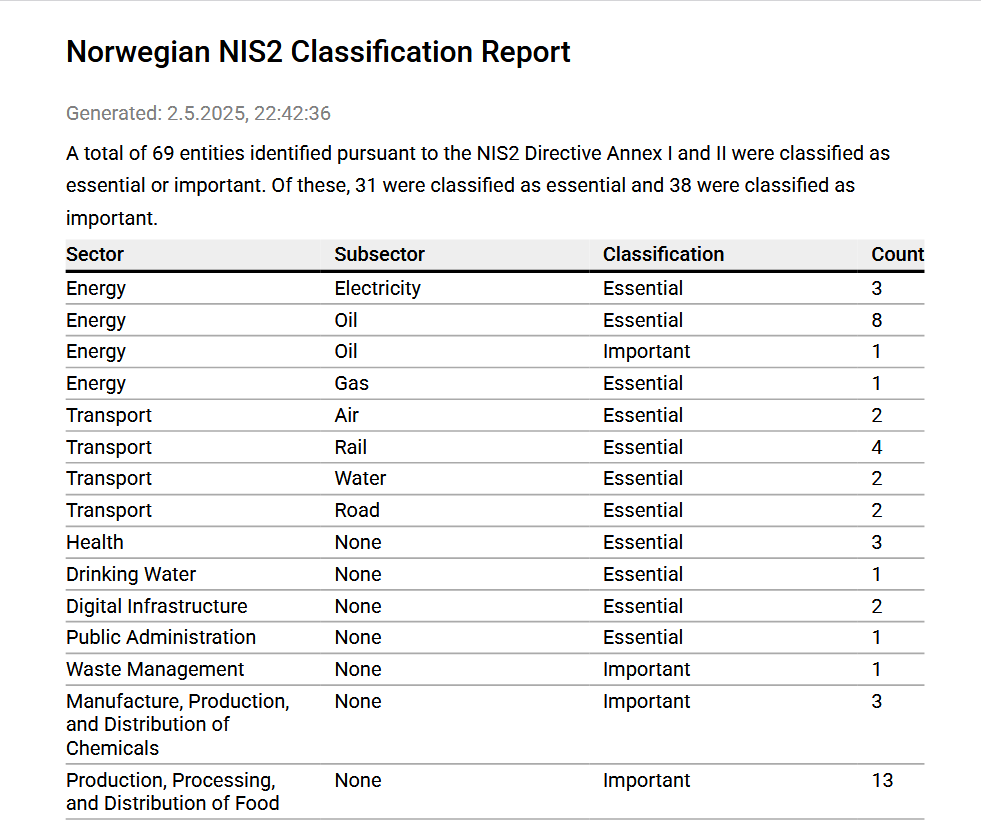}
    \caption{Admin Dashboard - Generated PDF Example}
    \label{fig:Generated_PDF_Example}
\end{figure}

\section{Summary of Results}

In summary, the implemented registry fulfills the majority of its specified requirements and performs effectively across all core system components. With 90\% of the requirements fully satisfied and the remaining gaps limited to auxiliary features, the application demonstrates strong alignment with both the technical goals and regulatory demands of the NIS2 directive. Performance benchmarks confirm that automatic registration, along with the underlying size estimation and classification algorithms, achieves high accuracy. While some limitations remain—particularly regarding risk factors and entity-specific reports—the system establishes a viable, scalable foundation for harmonized, NIS2-compliant registration and supervision. These results validate the system's design and confirm its suitability for the directive's envisioned role in national cybersecurity oversight.

\chapter{Discussion}

This chapter critically evaluates the outcomes of the NIS2-compliant registry system developed through a Design Science Research (DSR) methodology. It interprets the results in light of the legal and technical requirements derived from the NIS2 Directive, assesses the design decisions taken during implementation, and considers broader implications for both national and European cybersecurity governance. Furthermore, it reflects on the methodological effectiveness, the system’s limitations, and potential directions for future development.

\section{Summary of Objectives and Achievements}

The primary goal of this thesis was to design and evaluate a modular registry system to support competent authorities in meeting their supervisory obligations under the NIS2 Directive. The system particularly addresses Article 27 (registry of entities), Article 2 (scope), and Article 3 (entity definitions). This required the translation of complex legal provisions into enforceable technical workflows, with a strong emphasis on automation, traceability, and usability across diverse supervisory roles.

The implemented system demonstrates high functional coverage, fulfilling the vast majority of identified requirements. Its architecture accommodates both manual and automated entity registration, classification of entities according to Articles 2 and 3, integration of entity types as defined in Annexes I and II, and supervision through structured dashboards. These achievements support the central research hypothesis: that a legally grounded, dashboard-based registry system can effectively operationalize the regulatory requirements of the NIS2 Directive within a national supervisory framework.

Additionally, the NIS2 Directive places particular emphasis on establishing a single-point registry and minimizing administrative burden for both entities and supervisory authorities \cite{NIS2}. These priorities are embedded within the core objectives of this research and have been effectively addressed in the implemented system. The solution functions as a true single-point registry by integrating the entire supervisory pipeline—from initial entity registration to notification—within a unified application. All core functionalities, including registration, classification, notification, and supervision, are consolidated into a single interface.

Administrative burden is significantly reduced through automation of critical tasks. This includes automated registration workflows, contextual labeling of entities, rule-based classification aligned with NIS2 definitions, and pre-configured notification mechanisms. These automated components not only streamline supervisory processes but also enhance traceability and consistency in regulatory compliance.

\section{Methodological Reflection}
The development of this registry necessitated a thorough analysis of regulatory texts to extract and formalize concrete system requirements. Given the complexity and ambiguity of legal language, an iterative development process emerged as a natural and necessary methodology. Establishing the requirements, workflows, and design architecture early in the artifact's development ensured alignment with the NIS2 Directive from the outset.

In particular, the classification rules defined in Articles 2 and 3 of the Directive had to be precisely translated into deterministic algorithms. This translation required a careful balance between legal fidelity and technical feasibility. In contrast, more abstract elements—such as the reduction of administrative burden—required interpretation and integration throughout the system design, primarily through automation and user-centric workflows.

The Design Science Research (DSR) methodology proved effective in structuring our problem-solving approach. It provided a disciplined framework for iterating between legal interpretation, system modeling, and implementation. Nonetheless, the process also revealed key challenges. Legal frameworks do not iterate in the same way software does. Laws are often static and context-dependent, whereas system development is dynamic and driven by testable outcomes. Bridging this gap required multiple rounds of refinement to transform legal rules, exemptions, and intentions into consistent and efficient algorithms.

The DSR methodology also helped uncover context-dependent classification challenges that could not be fully integrated into the automated pipeline. For instance, the registry cannot independently determine whether an entity falls under the scope of other directives (such as CER or NIS1), nor can it classify entities based on national law without external input. These dependencies on external legal determinations necessitate manual identification in the initial phase of registration.

To address this, the system incorporates a hybrid approach using labeling and reactive dashboards. Agents can annotate entities as being governed by other directives or national provisions. Once such annotations are confirmed, the system leverages this input to algorithmically adjust classification outcomes. This approach supports the progressive integration of contextual legal knowledge into the automation framework, without compromising compliance or accuracy. 

A similar solution was applied to the handling of risk assessments, which led to the introduction of a mechanism termed the \textit{priority sequence}. This sequence enforces an algorithmic order of operations that prioritizes the most conclusive classification rules before evaluating more complex classification outcomes. By doing so, the system minimizes the likelihood of classification failure or logical conflicts while optimizing the efficiency of the overall classification process. This strategy resembles the “path of least resistance” in that the system prioritizes the simplest, most deterministic classification rules first, thereby resolving the majority of cases efficiently while deferring more complex evaluations only when necessary. However, unlike heuristic shortcuts, the priority sequence is rule-based and designed to ensure compliance and logical consistency.

\section{Interpretation of the Results}
This research has explored the challenges of translating complex regulatory texts into functional, compliant software systems. Specifically, it addressed core issues related to automation, harmonization, and context-dependent supervision within the framework of the NIS2 Directive.

A key insight is that automation in regulatory systems depends on the availability of deterministic classification criteria. The implemented registry demonstrates that automation can significantly reduce administrative burden when applied to structured and rule-based classification scenarios. However, in cases requiring contextual understanding—such as risk assessments or cross-directive classification—manual intervention remains necessary. This combination of automated workflows and human oversight enabled the development of a semi-automated system that is both efficient and capable of handling legal and contextual nuance. The result is a registry that balances scalability with regulatory accuracy, forming a bridge between structured logic and discretionary judgment.

Furthermore, the NIS2 Directive places strong emphasis on harmonization across member states \cite{NIS2}. In this context, harmonization refers not only to aligned classification outcomes but also to consistent processes and criteria used to reach those outcomes. While harmonization is legally mandated, technical design can reinforce it. This registry achieves that reinforcement by employing deterministic algorithms that apply identical classification logic irrespective of user or jurisdiction. By basing decisions on consistent inputs—such as sector, size, and activity codes—the system aligns closely with the Directive’s goal of a harmonized, single-point registry applicable EU-wide.

These findings suggest that legal harmonization efforts can be materially supported by technical determinism. The registry provides a replicable framework that can be adapted across member states while maintaining a unified classification standard. As such, it contributes not only to national implementation efforts but also to broader EU-level objectives in regulatory consistency and cybersecurity governance.

\section{Limitations}
While the development of this registry represents a successful implementation of the NIS2 Directive’s core requirements, several limitations remain. These primarily concern the boundaries of automation, the dependency on external data sources, and the necessary role of manual oversight in classification and supervision tasks.

\subsection{Manual Oversight}
A key objective of both the NIS2 Directive and this thesis is to reduce the administrative burden on entities and competent authorities \newline \cite{NIS2}. This was addressed through the automation of routine tasks and the contextualization of classification workflows to improve usability and decision-making. Although the system achieves these goals in many areas, certain aspects could not be fully automated or contextualized to a satisfactory degree within the scope and timeframe of this thesis.

Supervision, by nature, remains a largely manual process. While the registry supports this with automated data flows and structured dashboards, tasks involving context-specific judgment—particularly those related to cross-directive governance—require manual intervention. For example, entities that fall under the scope of other regulatory frameworks such as the CER Directive, NIS1, or specific national laws must be identified manually. This limitation is partly due to the lack of accessible or standardized registries for those directives, making automated cross-checking infeasible.

Although the registry enables entities to self-declare such status during registration, this information must still be verified by competent authorities to prevent misclassification based on inaccurate or misleading data. As a result, classification outcomes involving overlapping legal scopes rely on a combination of user-supplied data, manual annotation, and oversight. This limits full automation but reflects a necessary trade-off to preserve legal accuracy and compliance integrity.

\subsection{Dynamic Risk Modeling}
Assessing risk is a complex and context-dependent task. The NIS2 Directive outlines four specific risk factors that may influence the classification of an entity (see Figure~\ref{fig:Essential_Classification}). While these factors have been incorporated into the classification logic, the current system does not support dynamic or fully automated risk inference. Instead, the presence of applicable risk factors must be determined manually during registration or supervision.

This limitation primarily stems from the unavailability of standardized, machine-readable risk-related information that could support automated evaluation. As such, competent authorities must assess and verify whether any of the specified risk criteria apply to a given entity. Once identified, these risks can be annotated in the system and subsequently used to refine or override the default classification outcome. However, until such risk is explicitly recorded and tagged, the registry treats the entity as if no exceptional risk applies. This highlights the continued necessity of manual oversight in risk-sensitive classifications.

\subsection{Data Availability and External Dependency}
A key limitation of the registry lies in its reliance on publicly available data. At present, we are not aware of any privileged datasets accessible exclusively to EU institutions or national competent authorities in Norway or other member states. As such, the system operates entirely on public data, and its current implementation is specifically tailored to the Norwegian data ecosystem.

The registry retrieves organizational data from the Enhetsregisteret \newline \cite{Enhetsregisteret} and financial information from the Regnskapsregisteret \cite{Regnskapsregisteret}. While these sources generally provide up-to-date information for most relevant organizations, the completeness and quality of the data may vary. To address this, the current implementation incorporates fallback mechanisms and supports manual registration when necessary. For the purposes of this work, data is considered complete if it satisfies the requirements outlined in Table \ref{tab:Classification_Information_Requirements}. Entries lacking any data about an entity return an empty string. Additionally, in certain cases, addresses may be returned either as a single string or as an array of strings; this variation is handled through appropriate data formatting.

Automatic registration depends on two critical external elements: public company data and Google Gemini Flash 2.0 \cite{Gemini}, which is used to transform unstructured business activity descriptions from Enhetsregisteret into structured formats aligned with NIS2 classification requirements. In cases where data is missing, incomplete, or ambiguous, users may opt for manual registration. Both automatic and manual workflows benefit from shared labeling, contextualization, classification, and aggregation functionalities.

Another dependency involves the computation of the size-cap rule \newline \cite{SizeCapRule}, which requires accurate and current Euro conversion rates. Since companies may report in NOK or other currencies, exchange rate data is required to ensure consistent classification thresholds. Although the choice of exchange rate provider is not critical, CurrencyAPI was selected for its simplicity and reliability \cite{CurrencyApi}.

Overall, while the system has been designed to mitigate some of these dependencies through modular fallback mechanisms, its scalability and cross-border applicability remain contingent on the availability and standardization of relevant public data across member states.

\subsection{Interoperability}
The registry exhibits a high degree of interoperability at the legal and functional levels, as its internal classification logic is fully aligned with the NIS2 Directive. The rule-based mechanisms, entity categorization, and supervision workflows are directive-compliant and therefore applicable across all EU member states.

However, practical interoperability—particularly in relation to automatic registration—requires country-specific adaptation. Each member state maintains its own public databases, data schemas, and formats for organizational and financial information. As a result, the data ingestion layer must be customized to accommodate the structure and availability of national data sources.

The current implementation is tailored for the Norwegian ecosystem and relies on national public databases. However, the registry's architecture, English-language user interface, and core functionality are designed to be modular and reusable. This design facilitates adoption by other member states with minimal adaptation, provided appropriate integrations with national data sources are established.

For member states intending to implement automatic registration, the following table provides an overview of key components and their respective roles within the system.

\begin{table}[H]
    \centering
    \resizebox{\textwidth}{!}{
    \renewcommand{\arraystretch}{1.4}
    \begin{tabular}{|p{5cm}|p{9cm}|}
        \hline
        \textbf{File Name} & \textbf{Description} \\ 
        \hline
        \multicolumn{2}{|c|}{\textbf{Front-end Components}} \\
        \hline
        AutomaticRegistration.vue & Vue.js component responsible for rendering the user interface for automatic entity registration. \\
        \hline
        \multicolumn{2}{|c|}{\textbf{Business Logic}} \\
        \hline
        pipelineEntityProcessing.ts & Defines the primary processing flow for automatic entity registration, supporting both single and batch operations. \\ 
        \specialrule{0.4pt}{0pt}{0pt}
        processEntities.ts & Formats and validates automatically registered entities, integrating both programmatic checks and Gemini Flash 2.0 formatting. \\
        \hline
        \multicolumn{2}{|c|}{\textbf{Types}} \\
        \hline
        FullEntityData.ts & TypeScript interface defining the structure of a fully registered and processed entity. \\
        \specialrule{0.4pt}{0pt}{0pt}
        GeneralEntityData.ts & TypeScript interface for general organizational data (e.g., name, phone number) retrieved from Enhetsregisteret. \\
        \specialrule{0.4pt}{0pt}{0pt}
        EntityEconomyData.ts & TypeScript interface for financial information retrieved from Regnskapsregisteret. \\
        \specialrule{0.4pt}{0pt}{0pt}
        EntityActivityData.ts & TypeScript interface for organizational activity data retrieved from Enhetsregisteret. \\
        \hline
        \multicolumn{2}{|c|}{\textbf{Internal APIs}} \\
        \hline
        fetchEntityData.ts & Internal API for retrieving organizational data from Enhetsregisteret. \\
        \specialrule{0.4pt}{0pt}{0pt}
        fetchEntityEconomy.ts & Internal API for retrieving financial data for a single entity from Regnskapsregisteret. \\
        \specialrule{0.4pt}{0pt}{0pt}
        fetchOrgNums.ts & Internal API for retrieving one or more organizational numbers from Enhetsregisteret. \\
        \hline
        \multicolumn{2}{|c|}{\textbf{Validation Schema}} \\
        \hline
        fullValidatedEntity.ts & Contains validation schemas used to verify automatically registered entities. \\
        \hline
    \end{tabular}}
    \caption{Modules and Interfaces for Automatic Entity Registration}
    \label{tab:automatic_registration_stack}
\end{table}

\section{Future Work}

This section proposes future research and development opportunities grounded in the literature, analysis, implementation, and practical insights from this study.

\subsection{The Size-Cap Rule}
The current size-cap rule referenced in the NIS2 Directive \newline \cite{SizeCapRule} warrants reconsideration. As implemented (policy and technical), the rule relies on static thresholds for the number of employees, annual turnover, and annual balance sheet total. The logic governing these thresholds is based on a combination of “AND” and “OR” conditions, which can lead to ambiguous or inconsistent size determinations.

For instance, an entity may meet the turnover or balance thresholds associated with an “above medium” classification, while simultaneously falling within the employee range of a “medium-sized” entity. This creates uncertainty in classification, particularly for organizations that fall near the boundaries between two categories. As a result, some entities may remain unclassified or misclassified, especially when their operational scale does not align neatly with the directive's predefined thresholds.

Future iterations of the registry could incorporate more flexible or probabilistic models for size estimation, potentially informed by sectoral norms or risk exposure, to better reflect organizational scale and impact. Alternatively, the rule itself may benefit from revision at the policy level to account for edge cases and borderline scenarios. 

The current calculations for size can be seen in Chapter 4, but are summarized here in euro millions to highlight the limitations discussed above.

\begin{itemize}
    \item \textbf{Above:} \((\text{employees} > 250) \land ((\text{annual turnover} > 50) \lor (\text{annual balance} > 43))\).
    \item \textbf{Medium:} \((250 > \text{employees} \geq 50) \land ((50 \geq \text{annual turnover} \geq 10) \lor (43 \geq \text{annual balance} \geq 10))\).
    \item \text{\textbf{Below:}} \((\text{employees} < 50) \land ((\text{annual turnover} < 10) \lor (\text{annual balance} < 10))\)
\end{itemize}

If you inspect the size-cap rule closely, it becomes evident that entities falling within different thresholds on either side of the logical \texttt{AND} operator may remain unclassified. For example, an entity with a small number of employees but a high turnover or balance sheet total may not clearly satisfy any single category. This creates an ambiguous zone where the classification becomes indeterminate—particularly for organizations that straddle category thresholds in different dimensions.

\subsubsection{Corporate Structure}
Additionally, the NIS2 Directive emphasizes that the size and criticality of an entity should also consider the independence of its communications and information systems, regardless of its association with other organizations or its position within a corporate group \cite{NIS2}. This presents a challenge when classifying entities that are part of larger corporate structures or that operate as subsidiaries. A single legal entity may have independent IT infrastructure and operational autonomy, while another may rely entirely on centralized systems managed by the parent company.

Current size estimations based solely on aggregated financial or employment metrics may fail to capture this distinction, leading to over- or under-classification in terms of cybersecurity relevance. Future classification models could benefit from incorporating structural and technical independence as part of the assessment—providing a more accurate reflection of an entity’s actual exposure, control, and systemic risk.

\subsection{Value and Supply Chains}
The current registry partially addresses value chains by allowing competent authorities to label research organizations whose outputs are of particular strategic importance to one or more registered entities. This is operationalized through the custom label Research Organization, which must be applied manually during registration or supervision. However, the registry does not currently assess whether an entity is part of a critical supply chain.

Determining how supply chain dependencies should be incorporated into the registry requires further research and design consideration. One possible approach would be to ask each registrant to provide organizational identifiers for all entities within their known supply chain. Alternatively, each entity could list the organizations on which it depends for essential products or services. However, both methods present challenges: the process is likely to be administratively burdensome, and registrants may lack visibility into the deeper or indirect dependencies that constitute critical supply chain relationships.

Given these limitations, it may be more effective to address supply chain exposure through dynamic risk assessments. These could incorporate data from sectoral analyses, national threat intelligence, or third-party monitoring to identify and flag critical interdependencies. Integrating such mechanisms into the registry could enhance its ability to reflect systemic risks and support more comprehensive supervisory oversight.

\subsection{Integrated Reporting}
Incorporating incident and compliance reporting into the registration pipeline was originally envisioned as part of this thesis. However, due to time constraints and the inherent complexity of designing secure and auditable reporting workflows, this feature was ultimately deemed out of scope for the current implementation.

Nevertheless, the integration of reporting functionalities into a single-point registry represents a significant opportunity to further align with the intent of the NIS2 Directive \cite{NIS2}. Centralizing reporting alongside registration and classification would streamline supervisory processes and reduce redundancy for both entities and competent authorities.

\begin{figure}[H]
    \centering
    \includegraphics[width=1.0\linewidth]{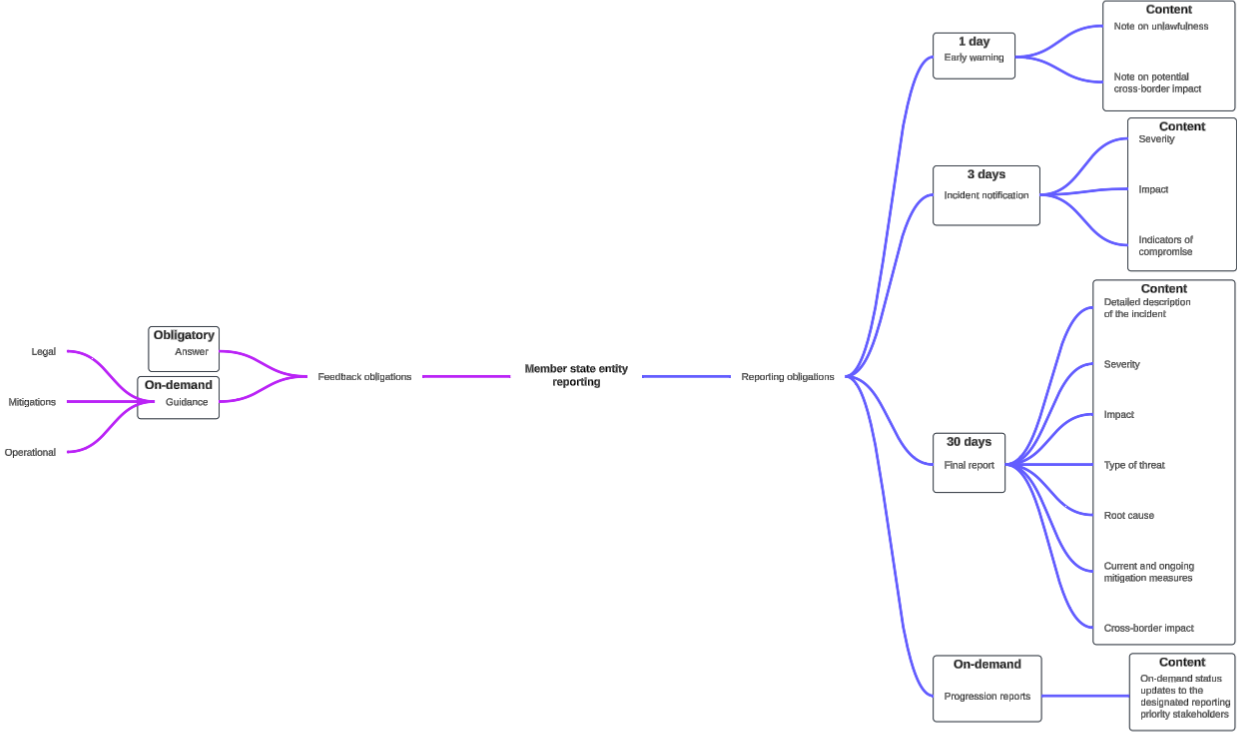}
    \caption{The Concept of Reporting}
    \label{fig:Reporting_Concept}
\end{figure}

While full implementation remains future work, we have developed a conceptual model outlining how integrated reporting might function within the system. This model (Figure~\ref{fig:Reporting_Concept}) illustrates how incident, compliance, and threat reporting could be embedded into the existing registry architecture through a unified, role-based interface.

This conceptualization offers a foundation for future development and evaluation, particularly in terms of legal traceability, structured notification channels, and real-time supervisory oversight.

\subsection{Dynamic Risk Assessments}
The functionality of the registry could be significantly enhanced through the integration of dynamic risk assessment capabilities. However, this constitutes a research area in its own right and falls beyond the scope of the current implementation. At present, the registry accounts for the predefined risk categories outlined earlier in the thesis, as specified by the NIS2 Directive \cite{NIS2}. Based on these categories, risk-related labels can be manually set to TRUE, which in turn influences the classification outcome. Nonetheless, the risk assessment process itself—along with supporting documentation—currently remains external to the registry system.

A promising direction for future work would be the development of integrated risk dashboards. Specifically, we envision two complementary components: an entity-specific risk dashboard, allowing agents to perform and document assessments tied to individual organizations; and a sector-specific risk dashboard, enabling aggregation, comparison, and visualization of risk trends across domains. These dashboards would provide a structured, traceable, and interactive interface for risk analysis—supporting more informed supervision and facilitating proactive cybersecurity governance.

\subsection{Summary}
This discussion has examined the design, implementation, and implications of a legally grounded registry system that supports NIS2 compliance through automation, traceability, and modular governance workflows. Through the application of Design Science Research, the project translated complex legal mandates into structured classification logic and semi-automated processes that align with both national supervisory needs and broader EU harmonization goals. While the registry fulfills its primary objectives, key limitations—particularly in automation boundaries, data dependencies, and risk contextualization—highlight areas for continued development. Nonetheless, the findings demonstrate that legal and technical integration is both feasible and impactful. The system offers a reusable, scalable model for member states and a foundation for future research in cybersecurity governance, cross-border interoperability, and regulatory technology innovation.

\chapter{Implications}
This research presents a comprehensive conceptualization of the NIS2 Directive’s requirements for establishing a registry of essential and important entities. It translates the directive’s legal and procedural objectives into functional workflows, deterministic algorithms, and concrete technical requirements suitable for real-world implementation. The resulting system effectively operationalizes the directive’s provisions on registration, classification, supervision, and notification—while aligning with the overarching goals of EU-wide harmonization and reduced administrative burden.

Moreover, the developed artifact provides a working model for a NIS2-compliant single-point registry. It is tailored to the Norwegian regulatory and data context but is designed with sufficient modularity and generality to be adapted by competent authorities across the EU with minimal modifications. As such, it serves as both a national solution and a transferable blueprint for wider implementation.

\section{Policy and Governance Implications}

The registry system developed in this research offers more than a technical solution; it provides a practical blueprint for implementing the NIS2 Directive at the national level. By translating legal provisions into structured, rule-based processes, it supports competent authorities in fulfilling their supervisory obligations in a harmonized and traceable manner. This contributes directly to the directive’s dual objectives of administrative efficiency and regulatory coherence across member states.

Moreover, the system reinforces harmonization not only through policy alignment but through algorithmic determinism. By applying consistent classification logic regardless of jurisdiction, it reduces interpretive variability and lays the groundwork for more unified compliance practices across the EU. As such, it holds potential value not only for Norwegian authorities but for other member states and EU institutions seeking scalable, legally grounded tools for NIS2 cybersecurity governance.

A key governance implication is that such registries can function as regulatory infrastructure. By embedding legal interpretation into software, competent authorities reduce reliance on manual discretion and increase traceability, reproducibility, and legal accountability. This can help resolve common policy implementation challenges, such as uneven national interpretation of EU directives or delayed supervisory enforcement.

Finally, the system’s modular architecture—combined with its alignment to the directive’s legal text—makes it easily transferable. While currently integrated with Norwegian data sources, its core logic can be reused by other member states with minimal adaptation. This supports the directive’s vision of harmonization not only through policy, but through interoperable, reproducible supervisory infrastructure.

\chapter{Conclusion}
NIS2 represents the de facto cybersecurity governance framework for the European Union, requiring all member states to implement mechanisms for regulatory enforcement. This research not only outlines a conceptual model for a NIS2-compliant registry, but goes further by delivering a complete registration-to-notification pipeline. Through the use of automation, deterministic algorithms, and contextualized dashboards, the system enforces a harmonized classification process while reducing the administrative burden imposed by the directive.

By providing a conceptual framework, concrete technical requirements, and a functional prototype, this work operationalizes the NIS2 Directive for the Norwegian competent authority—and with minimal modification, for other EU member states as well. In doing so, it bridges the gap between regulatory intent and technical execution.

Finally, the thesis lays the groundwork for further research and development. It introduces a sound, modular architecture that can be extended with features such as dynamic risk assessment, integrated reporting, and cross-directive supervision, offering a foundation for continued innovation in cybersecurity governance.

% References

% Use  reference style (Chicago, Harvard, APA)
\addcontentsline{toc}{chapter}{References}
 \bibliographystyle{apalike}

\bibliography{references}

% Appendix (if needed)
\appendix
\chapter{Automatic Benchmarking}

\begin{figure}[H]
    \centering
    \includegraphics[width=1.0\linewidth]{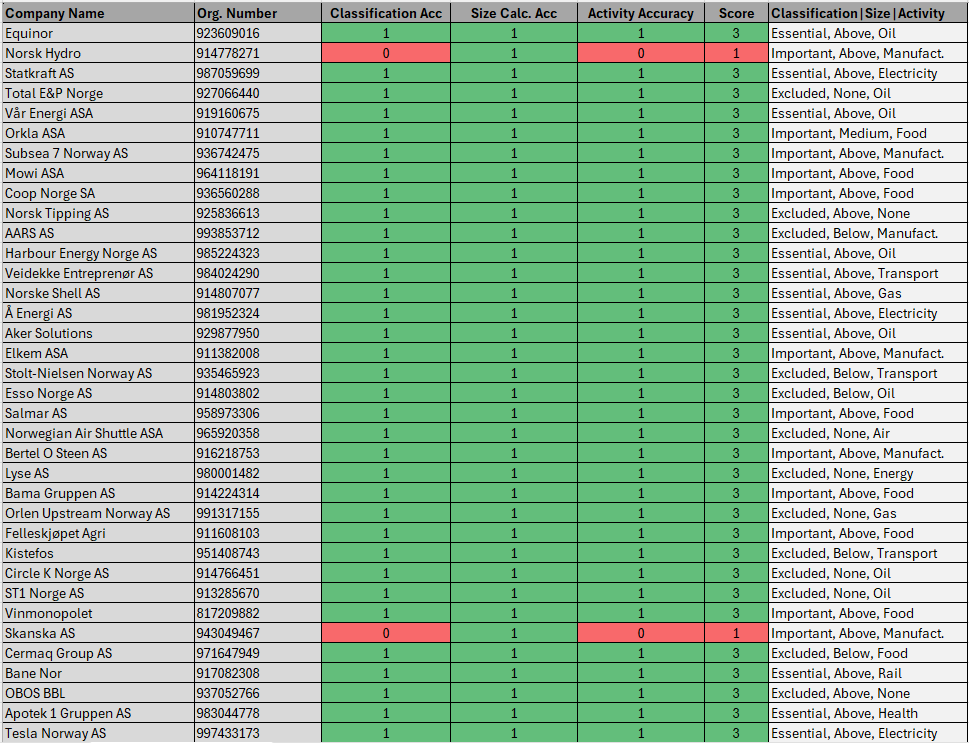}
    \caption{Automatic Benchmarking - Rows 1-37}
    \label{fig:auto_bench_one}
\end{figure}

\begin{figure}[H]
    \centering
    \includegraphics[width=1.0\linewidth]{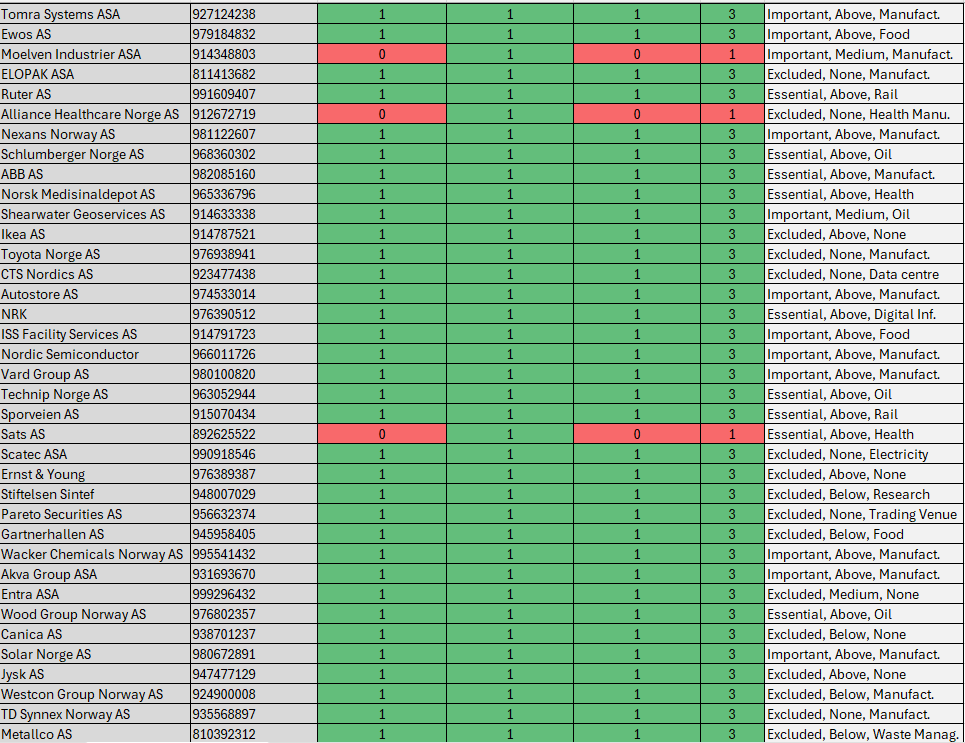}
    \caption{Automatic Benchmarking - Rows 38-74}
    \label{fig:auto_bench_two}
\end{figure}

\begin{figure}[H]
    \centering
    \includegraphics[width=1.0\linewidth]{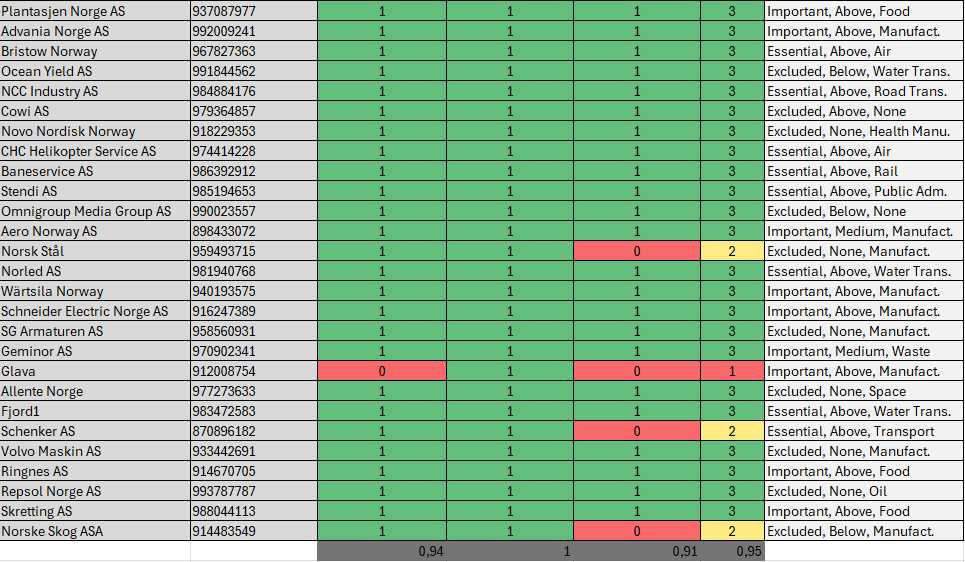}
    \caption{Automatic Benchmarking - Rows 75-101}
    \label{fig:auto_bench_three}
\end{figure}
% Insert appendixes here (if applicable)
\chapter{Application Source Code}

A complete version of the application developed as part of this thesis is provided in a separate ZIP file titled:

\begin{center}
    \texttt{NIS2\_Registry\_App.zip}
\end{center}

The archive contains:

\begin{itemize}
    \item The full application source code (including both frontend and backend)
    \item Configuration files and Docker Compose files
    \item A minimal starter \texttt{README.md} with guidance on usage and deployment
\end{itemize}

This supplementary material supports the implementation details discussed throughout the thesis and is provided to enable reproducibility and further evaluation.

\end{document}